\RequirePackage{ifpdf}
\documentclass[11pt]{JHEP3}
\pdfoutput=1
\usepackage[utf8]{inputenc}
\usepackage[babel]{csquotes}
\usepackage{amsmath}
\usepackage{amsfonts}
\usepackage{amssymb}
\usepackage{mathtools}
\usepackage{mathrsfs}
\usepackage{braket}
\usepackage{booktabs}
\usepackage{array}
\usepackage{graphicx}
\usepackage{comment}
\usepackage{aas_macros}

\newcommand{\al}{\alpha}

\newcommand{\ga}{\gamma}
\newcommand{\de}{\delta}
\newcommand{\D}{\Delta}
\newcommand{\ep}{\epsilon}

\newcommand{\e}{\eta}

\newcommand{\la}{\lambda}
\newcommand{\La}{\Lambda}
\newcommand{\Ga}{\Gamma}
\newcommand{\m}{\mu}
\newcommand{\n}{\nu}
\renewcommand{\r}{\rho}
\newcommand{\si}{\sigma}
\newcommand{\Si}{\Sigma}

\newcommand{\p}{\phi}

\renewcommand{\o}{\omega}
\renewcommand{\O}{\Omega}

\newcommand{\pa}{\partial}
\DeclarePairedDelimiter{\abs}{\lvert}{\rvert}

\newcommand{\beq}{\begin{equation}}
\newcommand{\eeq}{\end{equation}}
\newcommand{\mc}{\mathcal}

\newcommand{\mycomment}[1]{}

\newcommand{\wh}[1]{\widehat{#1}}

\newcommand{\vev}[1]{\langle #1 \rangle}

\newcommand{\ii}{\mathrm{i}}

\newcommand{\gbullet}{\!\bullet\!}
\newcommand{\wbullet}{\!\circ\!}

\newcommand{\tT}{\tilde{T}}

\newcommand{\bz}{\bar{z}}

\newcommand{\bT}{\bar{T}}

\newcommand{\whD}{{\wh{\Delta}}}

\author{Marco Billò$^a$, Vasco Gon\c calves$^{b,c}$, Edoardo Lauria$^d$, Marco Meineri$^e$\\
$^a$ Dipartimento di Fisica, Universit\`a di Torino \\
and Istituto Nazionale di Fisica Nucleare - sezione di Torino\\  
Via P. Giuria 1 I-10125 Torino,Italy\\
$^b$ Centro de Física do Porto, Departamento de Física e Astronomia
Faculdade de Ciências da Universidade do Porto\\
Rua do Campo Alegre 687, 4169–007 Porto, Portugall\\
$^c$ ICTP South American Institute for Fundamental Research
Instituto de Física Teórica, UNESP - Univ. Estadual Paulista
Rua Dr. Bento T. Ferraz 271, 01140-070, São Paulo, SP, Brasil\\
$^d$ KU Leuven, Institute for Theoretical Physics, Celestijnenlaan 200D, B-3001
Leuven, Belgium\\
$^e$ Perimeter Institute for Theoretical Physics, Waterloo, Ontario, N2L 2Y5, Canada\\
and Scuola Normale Superiore, Piazza dei Cavalieri 7 I-56126 Pisa, Italy \\
and Istituto Nazionale di Fisica Nucleare - sezione di Pisa\\
\vspace{0.35cm}
{\tiny E-mail: \email{billo@to.infn.it, vasco.dfg@gmail.com, edo.lauria@gmail.com, marco.meineri@sns.it}}
}

\bigskip
\abstract{We discuss consequences of the breaking of conformal symmetry by a flat or spherical extended operator. We adapt the embedding formalism to the study of correlation functions of symmetric traceless tensors in the presence of the defect. Two-point functions of a bulk and a defect primary are fixed by conformal invariance up to a set of OPE coefficients, and we identify the allowed tensor structures. A correlator of two bulk primaries depends on two cross-ratios, and we study its conformal block decomposition in the case of external scalars. The Casimir equation in the defect channel reduces to a hypergeometric equation, while the bulk channel blocks are recursively determined in the light-cone limit. In the special case of a defect of codimension two, we map the Casimir equation in the bulk channel to the one of a four-point function without defect. Finally, we analyze the contact terms of the stress-tensor with the extended operator, and we deduce constraints on the CFT data. In two dimensions, we relate the displacement operator, which appears among the contact terms, to the reflection coefficient of a conformal interface, and we find unitarity bounds for the latter.
}

\title{Defects in conformal field theory.}


\begin{document}

\section[Introduction]{Introduction}
\label{sec:intro}

Conformal symmetry imposes powerful constraints on the dynamics of a theory and greatly reduces the number of model dependent physical quantities. This is a blessing, and at the same time a downside, since very symmetric theories are less rich in structure. However, conformal field theories (CFTs) are redeemed by the notable fact that they describe a wealth of physical phenomena with great accuracy, thanks to the robust mechanism of symmetry enhancement which goes under the name of renormalization group. More surprisingly, they even encode the secrets of quantum gravity \cite{Maldacena:1997re}. 

Due to the non-linear action of the special conformal generators on flat space coordinates, the constraints imposed by the conformal group on correlation functions 
can be annoyingly complicated, especially in the case of operators carrying spin. 
Luckily, the action of all generators can be linearized by embedding the physical space in a bigger one \cite{Dirac:1936fq}. The physical $d$ dimensional space is embedded in the light cone\footnote{We focus on Euclidean signature, the extension being obvious.} of $\mathbb{R}^{d+1,1}$, and the linear action is given by the usual Lorentz transformations in $(d+2)$ dimensions. 
Spinning operators in $(d+2)$ dimensions have more degrees of freedom than their lower dimensional counterparts, so this redundancy needs to be resolved by a gauge choice. The machinery to do this was elucidated in \cite{Costa:2011mg} (see also \cite{Weinberg:2010fx}), where an algorithmic way of constructing conformally covariant correlators of symmetric traceless tensors (STT) was set up. 

In the present paper we adapt the embedding formalism to the presence of an extended operator, which preserves a large subgroup of the conformal symmetry of the homogeneous vacuum. Technically, we consider correlation functions of local operators constrained by $SO(p+1,1)\times SO(q)$ symmetry, with $p+q=d$. This is the residual symmetry preserved by a flat or spherical $p$-dimensional extended operator placed in $\mathbb{R}^d$, which is usually referred to as a conformal defect.
To fix ideas, in what follows we shall mostly stick to the picture of a flat defect, but we explicitly discuss the spherical case in subsection \ref{subsec:spherical}. 
The discussion is easily adapted to defect CFTs in conformally flat spaces.

Besides the symmetry breaking pattern, a specific conformal defect is defined by the CFT data that completely specify the correlation functions of local operators. 
The fusion of primary operators in the bulk is clearly unaffected by the defect, and controlled by the usual bulk Operator Product Expansion (OPE), of the schematic form\footnote{We gather our set of conventions in appendix \ref{notations}.}
\beq
O_1(x_1) O_2(x_2) \sim c_{123}\, \abs{x_{12}^\m}^{\D_3-\D_1-\D_2}\, O_3(x_2)+\dots
\label{bulkopeloose}
\eeq
However, knowledge of the set of scale dimensions $\D_i$ and coefficients $c_{ijk}$ is not sufficient to compute the correlation functions. 
The defect possesses local excitations, the defect operators $\wh{O}_i$,
whose conformal weights $\wh{\D}_i$ are not related by symmetry to the bulk ones.
When a bulk excitation is brought close to the extended operator, it becomes indistinguishable from a defect excitation, and the process is captured by a new OPE, with new bulk-to-defect OPE coefficients:
\beq
O(x^a,x^i) \sim b_{O\wh{O}}\, \abs{x^i}^{\wh{\D}_2-\D_1} \wh{O}(x^a) +\dots
\label{defopeloose}
\eeq
Here we denoted by $x^a$ the coordinates parallel to the flat extended operator, and by $x^i$ the orthogonal ones. The defect OPE \eqref{defopeloose} converges within correlation functions like its bulk counterpart. Furthermore, the contribution of descendants of a defect primary $\wh{O}$, i.e. its derivatives in directions parallel to the defect, is fixed by conformal symmetry. Precisely as the bulk OPE coefficients $c_{ijk}$ appear in the three-point functions of a homogeneous CFT, the coefficients $b_{O\wh{O}}$ determine the two-point function of a bulk and a defect operator. The existence of this non-trivial coupling between primaries of different scale dimensions is the trademark of a non trivial defect CFT. Its functional form is easily written down in the scalar case:
\beq
\braket{O(x)\wh{O}(0)} = b_{O\wh{O}}\, \abs{x^i}^{\wh{\D}_2-\D_1} 
\abs{x^\m}^{-2\wh{\D}_2},
\label{bulkdefscalars}
\eeq
while it is one of the task of this paper to treat the case of spinning primaries. Among the bulk-to-defect OPE coefficients, the one of the identity plays a special r\^ole, as it allows bulk operators to acquire an expectation value:
\beq
\braket{O(x)} = a_O \abs{x^i}^{-\D_1}.
\eeq 
Here, following the literature, we employ the notation $a_O$ for $b_{O\wh{\mathbf{1}}}$.

Finally, defect operators can be fused as well, so that one last OPE exists, which can be written just by adding hats to eq. \eqref{bulkopeloose}. When considering correlation functions of defect operators, we are faced with an ordinary conformal field theory in $p$ dimensions, with some specific features. The rotational symmetry around the defect is a global symmetry from the point of view of the defect theory, so that operators fall into representations of the $so(q)$ algebra. Moreover, unless there is a decoupled sector on the defect, there is no conserved defect stress-tensor. Indeed, energy is expected to be exchanged with the bulk, so that only the global stress-tensor is conserved. From a more formal point of view, if a separate defect current existed, the associated charge would translate the defect operators that couple to it without affecting bulk insertions. But such a symmetry is compatible with eq. \eqref{bulkdefscalars} only if $b_{12}$ vanishes. Besides the defect stress-tensor, which encodes the response of the defect to a change in its intrinsic geometry, other local operators might be present, which are associated to variations of the extrinsic geometry. Finally, there is always one protected defect primary for every bulk conserved current which is broken by the defect. In particular, the breaking of translational invariance induces a primary which is present in any local theory: the \emph{displacement} operator. This primary appears as a delta function contribution to the divergence of the stress-tensor - see eq. \eqref{newward} - that is, it measures its discontinuity across the defect. The displacement vanishes for trivial and topological defects. More generally, since the normalization of the displacement operator is fixed in terms of the stress-tensor, the coefficient of its two-point function is a physical quantity\footnote{In fact, analogously to the central charge, the Zamolodchikov norm of the displacement is the coefficient of an anomaly in four dimensions \cite{EEconfdef}. We will not treat the problem of defect anomalies in this paper. For recent work on the subject, see also \cite{Jensen:2015swa,Solodukhin:2015eca}.}, which is parametrically small when a continuous family of defects connected to the trivial one is considered.

These remarks conclude the overview of the CFT data attached to a generic defect\footnote{Interfaces make a small exception, for they require the set of data to be slightly enlarged \cite{Gliozzi:2015qsa}.}. Any correlation function of local operators can be reduced to a sum of bulk-to-defect couplings by repeatedly fusing bulk and defect operators. Yet the coefficients $a_i,\,b_{ij},\,c_{ijk},\ \hat{c}_{ijk}$ are not independent: they obey crossing symmetry constraints, one instance of which we will consider in section \ref{sec:blocks}.

Our aim is to put together a toolbox for analyzing a generic Defect CFT; this should
be useful to tackle specific examples. 
In fact conformal defects appear in a variety of situations of phenomenological and theoretical interest. The prototypical example is provided by boundaries and interfaces, whose conformal data have already been characterized in general dimensions%
\footnote{Our discussion of the Ward identities is based on the one 
of \cite{McAvity:1993ue}, extending it to defects of codimension greater than one.}
in \cite{McAvity:1993ue,McAvity:1995zd}. 
Two-dimensional boundary and interface CFTs are as usual under far better control - see \cite{Cardy:1984bb,Cardy:1989ir,Cardy:1991tv,Behrend:1999bn,Bachas:2001vj,Frohlich:2006ch} and many others - but higher dimensional conformal boundaries have been also extensively studied both in statistical and in high energy  physics. With the only purpose of illustration, let us mention a few examples. Surface critical exponents of the $\p^4$ theory have been computed numerically \cite{2011PhRvB..83m4425H} and perturbatively \cite{Diehl:1981}, and recently CFT predictions for correlation functions were tested on the lattice using the 3d Ising model with a spherical boundary \cite{Cosme:2015cxa}. The enhancement from scale to conformal symmetry in the case of a boundary was considered in \cite{Nakayama:2012ed}. In high energy physics, boundaries and interfaces can be engineered holographically \cite{Karch:2000gx, DeWolfe:2001pq, Erdmenger:2002ex,Aharony:2003qf}, while the systematic exploration of superconformal boundary conditions in $\mathcal{N}=4$ SYM was carried out in \cite{Gaiotto:2008sa}. Recently, the study of the CFT data associated to the D3-D5 brane system was initiated in \cite{deLeeuw:2015hxa}, while the spectrum of defect operators on the NS5-like interface was considered in \cite{Rapcak:2015lhn}.
Defects of higher codimensions are equally common in high and low energy physics. Wilson and 't Hooft operators are pre-eminent examples \cite{Kapustin:2005py}, but surface operators also play an important role in gauge theories: in fact, order \cite{Buchbinder:2007ar} and disorder \cite{Gukov:2006jk} two-dimensional operators, analogous to Wilson and 't Hooft lines respectively, can be constructed in four dimensions. Again, lower dimensional defects also arise at brane intersections - see for instance \cite{Constable:2002xt}. Low energy examples of defects include vortices \cite{Dias:2013bwa}, magnetic-like impurities in spin systems \cite{Billo:2013jda}, localized particles acting as sources for the order parameter of some bosonic system \cite{Allais:2014fqa}, higher dimensional descriptions of theories with long range interactions  \cite{Paulos:2015jfa} etc. Finally, the defect CFT language can be fruitfully applied in the domain of quantum Entanglement \cite{EEconfdef}.

Besides specific examples, one more motivation for our work is provided by the recent progress in exploring the space of CFTs using the conformal bootstrap \cite{Rattazzi:2008pe,ElShowk:2012ht,Gliozzi:2013ysa,El-Showk:2014dwa,Kos:2014bka,Komargodski:2012ek,Fitzpatrick:2012yx}. The bootstrap program has recently been extended to conformal defects \cite{Liendo:2012hy,Gaiotto:2013nva,Gliozzi:2015qsa}, but not all the tools required to tackle the problem in the most general situation have been developed. We make some steps in this direction by providing information on the conformal blocks for a two-point function of scalar operators. Specifically, we solve the Casimir equation corresponding to the defect OPE eq. \eqref{defopeloose} exactly, and the one corresponding to the bulk OPE eq. \eqref{bulkopeloose} perturbatively in the light-cone limit. Moreover, in the specific case of a codimension two defect, and for equal external operators, we map the bulk Casimir equation into the one for the four-point function without any defect, so that all the known solutions to the latter translate into solutions to the former. A thorough analysis of the bulk Casimir equation, however, lies outside the scope of this work, and is left to a forthcoming paper \cite{defblocks}.

The paper is organized as follows. Section \ref{sec:toolbox} is devoted to a review of the embedding formalism and to its application to the defect set-up. In particular, the rules for projecting correlators to physical space are explained, with a few examples. In section \ref{sec:correlators}, we derive the tensor structures that can appear in correlation functions of spinning operators. The reader who is uninterested in the details can safely skip them, and directly apply the rules explained in the previous section to obtain the correlators in physical space. The crossing symmetry constraint for a scalar two-point function is treated in section \ref{sec:blocks}. Finally, section \ref{sec:ward} is dedicated to the Ward identities obeyed by the stress-tensor in the presence of a defect. We also derive constraints imposed by those identities on the CFT data, and illustrate the procedure in a few examples. 
Some technical details and some useful material are collected in the Appendices.

\section[Tensors as polynomials and CFTs on the light-cone]{Tensors as polynomials and CFTs on the light-cone.}
\label{sec:toolbox}

In this section, we will first review the main properties of the embedding space formalism, and then adapt it to the presence of a defect. 

The embedding space formalism is especially useful in studying correlation functions of operators with spin, so let us start by recalling a first useful device to deal with the index structures. In this paper, we will be concerned with operators that are symmetric, traceless tensors (STT).\footnote{The embedding space machinery was set up for antisymmetric tensors in \cite{Costa:2014rya} and for fields living in AdS in \cite{Costa:2014kfa}.} Such tensors can be encoded in polynomials by using an auxiliary vector $z^\mu$:
\begin{align}
F_{\m_1\dots \m_J}(x) \rightarrow F_J(x,z)\equiv z^{\mu_1}\dots z^{\mu_J} F_{\m_1\dots \m_J}( x),
\label{encodingphys}
\qquad z^2=0.
\end{align}
The null condition on the auxiliary vectors is there to enforce tracelessness of the tensor $F$.
The correspondence is one to one, and the
index structure can be recovered employing the Todorov differential operator \cite{Dobrev:1975ru}:
\begin{align}
D_{\mu}=\bigg(\frac{d-2}{2}+z\cdot \frac{\partial}{\partial z}\bigg)\frac{\partial}{\partial z^{\mu}}-\frac{1}{2}z_{\mu}\frac{\partial^2}{\partial z\cdot \partial z}\label{eq:TodorovOp}. 
\end{align}
Notice that this operator is  interior to the condition $z\cdot z=0$.
For example we can free one index by applying (\ref{eq:TodorovOp}) once
\begin{align}
F_{\mu_1\mu_2\dots \mu_J}z^{\mu_2}\dots z^{\mu_J}=\frac{D_{\mu_1}F_J(x,z)}{J\big(\frac{d}{2}+J-2\big)}
\end{align}
or we can free all indices by applying $J$ times
\begin{align}
F_{\mu_1\mu_2\dots \mu_J} = \frac{D_{\mu_1}\dots D_{\mu_J}F_J(x,z)}{J!\big(\frac{d-2}{2}\big)_J} \label{eq:OpenningIndexes}.
\end{align}
We are now ready for a lightning review of the embedding formalism. A simple observation motivates it: the conformal algebra in $d$ dimensions coincides with the group of rotations in $(d+2)$ dimensions. In fact, it is possible to embed the former space into the latter in such a way that the generators of $SO(d+1,1)$ act on the embedded $\mathbb{R}^d$ precisely as conformal transformations \cite{Dirac:1936fq,Costa:2011mg,Weinberg:2010fx}. The advantage is manifest, since Lorentz transformations act linearly on points. The embedding works as follows. Points in physical space are mapped to the light-cone of the $(d+2)$-dimensional Minkowski space; note that
the light-cone is an invariant subspace under the action of the Lorentz group. We shall find useful to pick light-cone coordinates in $\mathbb{R}^{d+1,1}$:
\begin{equation}
P\cdot P = \eta_{AB}P^{A}P^{B} = -P^{+}P^{-}+\de_{\mu\nu}P^{\mu}P^{\nu}.
\end{equation}
We still need to get rid of a dimension, and this can be obtained by declaring the light-cone to be projective, that is, by identifying points up to a rescaling: $P\sim\la P$, $\la\in \mathbb{R}^+$. This gauge freedom can then be fixed by choosing a section such that the induced metric be the Euclidean one. To this end, a point in $x\in \mathbb{R}^d$ is mapped to a null point $P_x$ in $\mathbb{R}^{d+1,1}$ in the so called Poincar\'e section:
\begin{align}
x\rightarrow P^{M}_x=(P^{+},P^{-},P^{\mu})= (1,x^2,x^{\mu})\,. \label{eq:LightconeParametrization}
\end{align} 
A generic element $g\in SO(d+1,1)$ does not fix the section \eqref{eq:LightconeParametrization}, but
one can define an action $\tilde g$ on the section by rescaling back the point: 
writing 
\beq
g P^M_x = g(x) (1,x'^2,x'^{\mu})
\eeq
then we have $\tilde g x = x'$. 
It turns out that $\tilde{g}$ is precisely a conformal transformation.

Any field, $F^{\mu_1,\dots\,\mu_J}(x)$,  of spin $J$  in  physical space can be obtained from a field, $F^{M_1,\dots M_J}(P)$, by restricting the latter to live on $P_x$ defined in (\ref{eq:LightconeParametrization}). The two operators are simply related by a pull-back: 
\begin{equation}\label{pullback}
F_{\m_1\dots \m_J}( x)=\frac{\partial P^{M_1}_x}{\partial x^{\m_1}}\dots \frac{\partial P^{M_J}_x}{\partial x^{\m_J}}F_{M_1\dots M_J}( P_x),
\end{equation}
where, on the Poincar\'e section:
\begin{equation}
\frac{\partial P^{M}_x}{\partial x^{\nu}}=(0,2x_\nu,\delta^{\mu}_\nu).
\end{equation}
We further impose the following conditions on $F^{M_1,\dots M_J}(P)$:
\begin{itemize}
\item it is homogeneous of degree $-\Delta$, {\em {i.e. $F_{M_1\dots M_J}(\lambda P)=\lambda^{-\D}F_{M_1\dots M_J}(P),\,\,\lambda >0$}},
\item it is transverse, $P_{M_1}F^{M_1,\dots M_J}(P)=0$. 
\end{itemize}
This ensures that $F_J$ projects to a primary operator in physical space (see \cite{Weinberg:2010fx} for a derivation). 

Symmetric traceless tensors in embedding space can be again easily encoded in polynomials:
\begin{align}
F_{M_1\dots M_J}( P) \rightarrow F_J(P,Z)\equiv Z^{M_1}\dots Z^{M_1} F_{M_1\dots M_J}( P),
\qquad Z^2=0
\label{encodingemb}
\end{align}
We can impose further that $Z\cdot P = 0$ since this condition preserves the transversality of the tensor. Eqs. \eqref{encodingemb} and \eqref{encodingphys} agree, once using eq. \eqref{pullback}, if
\beq
Z=(0,2x\cdot z,z^{\mu})
\label{zbulkph}
\eeq
In particular, this satisfies $Z\cdot P=Z^2=0$, if $z^2 =0$. Therefore, a correlation function in embedding space depends on a set of pairs $(P_n,Z_n)$. The rules to project it down to physical space can be summarized as follows: 
\beq
Z_m\cdot Z_n \rightarrow\, z_m\cdot z_n,\, \qquad -2P_m\cdot P_n \rightarrow\, x_{mn}^2, \qquad 
P_m\cdot Z_n \rightarrow z_n\cdot x_{mn}\label{eq:MapEmbeddingtoPhysical}, 
\eeq
where $x_{mn}^\mu\equiv (x_m-x_n)^\mu $. Once in physical space, one can free the indices using the Todorov operator (\ref{eq:TodorovOp}).

When computing the allowed structure for correlation functions of spinning operators, it is convenient in practice to reverse the logic: one starts by writing polynomials in the variables $Z_n$, and constrains their coefficients such that the polynomials obey the required properties.  For this purpose it is convenient to rephrase the transversality condition:
\begin{align}
F_J(P,Z+\alpha P) = F_J(P,Z)\,, \ \ \ \ (\forall\,   \alpha)\label{eq:TransversalityinZ}.
\end{align} 
One more simplification follows from defining identically transverse tensors that can be used as building blocks: their contractions automatically provide the right polynomials. When the vacuum is invariant under the full conformal group - i.e. no defect is present - only one tensor is required:
\begin{align}
C_{MN}\equiv Z_{M}P_N-Z_NP_M\label{eq:ElementaryBuildingBlock}~.
\end{align}
For example, the structures appearing in the two and three point functions can be written in terms of  $C_{MN}$
\begin{align}
&C_{m\,MN}C_{n}^{MN}=H_{mn}= -2\big[(Z_m\cdot Z_n)(P_m\cdot P_n)-(P_m\cdot Z_n)(P_n\cdot Z_m)\big]\,, \\ 
&V_{m,nl}= \frac{C_{m,MN}P_{n}^{M}P_{l}^{N}}{P_m\cdot P_n P_m\cdot P_l} = \frac{(Z_m\cdot P_n)(P_l\cdot P_m)-(Z_m\cdot P_l)( P_n\cdot P_m)}{(P_n\cdot P_l)}.
\end{align}
These are transverse by construction. The same logic can be applied to higher point correlation functions. 

Let us finally comment on the case in which one of the operators in a correlation function is a conserved tensor. The conservation condition in physical space is of course
\begin{align}
\partial^{\mu}D_{\mu}T(x,z)=0
\label{conservationphys}
\end{align}
where we have used the Todorov operator (\ref{eq:TodorovOp}) to open one index. 
As it is well known, this condition is only consistent with the conformal algebra if $T$ has dimension $\Delta=d-2+J$. Furthermore, in a unitary theory the reverse is true as well: a tensor with the above scale dimension is conserved. Conservation of the tensor also implies constraints on correlation functions, which are again easier to analyze in embedding space. To this end, eq. \eqref{conservationphys} must be uplifted to the light-cone. This is easily done:
\begin{align}
\partial^{M}D_{M}\widetilde{T}(X,Z)=0\label{eq:ConservationEmbeddingNoDefect}.
\end{align}
Here $D_{M}$ has the same expression as (\ref{eq:TodorovOp}), but with $z$ replaced by $Z$. 
The tensor $\widetilde{T}(P,Z)$ is obtained by uplifting $T(x,z)$ and imposing $Z^2=Z\cdot P=0$. It is important to impose  $Z\cdot P=0$, because $\partial^{M}D_{M}$ 
does not preserve this condition - for more details see the discussion in section $5$ of \cite{Costa:2011mg}.

In the remaining of the section, we introduce the necessary modifications to the formalism in order to deal with situations in which the conformal group is broken by the presence of a defect in the vacuum.

\subsection{Defect CFTs on the light-cone}
\label{subsec:tooldefect}
We now place a $p$-dimensional defect $\mc{O}_\mathcal{D}$ in the vacuum of the CFT, by which we mean that correlation functions are measured in the presence of this extended operator, whose expectation value is divided out. That is, a correlator with $n$ bulk insertions and $m$ defect insertions is defined as follows:
\begin{equation}\label{genericcorr}
\braket{O_1(x_1)\dots O_n(x_n)\wh{O}_1({x}_1^a)\dots\wh{O}_m({x}_m^a)}\equiv
\frac{1}{\braket{\mc{O}_\mathcal{D}}}_0\braket{O_1(x_1)\dots O_n(x_n)\wh{O}_1({x}_1^a)\dots\wh{O}_m({x}_m^a)\,\mc{O}_\mathcal{D}}_0,
\end{equation} 
where the subscript $0$ denotes expectation values taken in the conformal invariant vacuum. As mentioned in the introduction, the reader can keep in mind the example of a flat defect: in subsection \ref{subsec:spherical} we show explicitly how to deal with a spherical defect.
Let us denote with $q$ the codimension of the defect, so that $p+q=d$. The generators which belong to an $so(p+1,1)\times so(q)$ sub-algebra of $so(d+1,1)$ still annihilate the vacuum. In the picture in which the defect is flat these are just conformal transformations on the defect and rotations around it. In the special case of a codimension one defect, {\em i.e.} a CFT with a boundary or an interface, the embedding formalism has been set up in \cite{Liendo:2012hy}.

Defect operators carry both $SO(q)$ and $SO(p)$ quantum numbers. We call them transverse and parallel spin, and denote them with $s$ and $j$ respectively. Clearly, one can still encode spinning defect operators into polynomials. This time, two auxiliary variables $w^i$ and $z^a$ are required, associated with transverse and parallel spin respectively. Again we restrict ourselves to symmetric traceless representations of both $SO(q)$ and $SO(p)$, thus imposing $w^i w_i=0$ and $z^az_a=0$. When recovering tensors from the polynomials, one needs to remove the two kind of polarization vectors by use of the appropriate Todorov operators, that is, respectively,
\begin{align}
D_a &=\bigg(\frac{p-2}{2}+z^b \frac{\partial}{\partial z^b}\bigg)\frac{\partial}{\partial z^a}-\frac{1}{2}z_a\frac{\partial^2}{\partial z^b \partial z_b},\label{eq:TodorovPar} \\ 
D_i &=\bigg(\frac{q-2}{2}+w^j \frac{\partial}{\partial w^j}\bigg)\frac{\partial}{\partial w^i}-\frac{1}{2}w_i\frac{\partial^2}{\partial w^j \partial w_j}\label{eq:TodorovOrt}. 
\end{align}
From the point of view of the defect theory, the transverse spin is the charge under an internal symmetry, but of course both symmetries arise from the Euclidean group in the ambient space.  In the correlator of a bulk and a defect operator, the allowed tensor structures couple indeed both transverse and parallel spins to the bulk Lorentz indices, in every way that preserves the stability group of the defect. 

In the embedding space, let us split the coordinates in the two sets which are acted upon respectively by $SO(p+1,1)$ and $SO(q)$. We loosely call the first set \enquote{parallel} directions and denote them $A,B,\dots$, while \enquote{orthogonal} directions are labelled $I,J,\dots$:
\begin{align}
M=(A,I), \qquad A=1,\dots, p+2\,,\ \ I=1,\dots, q\,. \label{eq:IndicesConventions}
\end{align}

Since the symmetry is still linearly realized in embedding space, scalar quantities are simply built out of two scalar products instead of one:
\begin{align}
P\gbullet Q = P^A\eta_{AB} Q^B \qquad\quad P\wbullet Q= P^I\de_{IJ} Q^J.
\end{align}
There is of course also the possibility of using the Levi-Civita tensor density, which is relevant for correlators of parity odd primaries. We comment on this in subsection \ref{subsec:odd}.\footnote{There is at least another method to set up the formalism, which was suggested to us by Joao Penedones. The defect is specified by a $q$-form $V_{M_1\dots M_q}$. Allowed tensor structures are produced via contraction of $V$, or of the Hodge dual, with the position and polarization vectors. One can take $V^{M_1,\dots, M_q}\propto n_{1}^{[M_1}\dots n_{q}^{M_q]}$, $n^M_I$ being vectors normal to the embedded defect. This approach is equivalent to the one chosen in this paper. Indeed, in the coordinate system \eqref{eq:IndicesConventions} $n^M_I=\de^M_I$ and $V_{M_1\dots M_q}\propto \de_{M_1}^{I_1}\dots \de_{M_q}^{I_q} \ep_{I_1\dots I_q}$, $\ep$ being the Levi-Civita symbol. Furthermore, a product of an even number of copies of $V$ can be expressed in terms of the the orthogonal scalar product $\wbullet\,$, thus proving the equivalence.}
Bulk insertions still obey the conditions $P^2,\, Z^2\,, Z\cdot P=0$. 
This implies that, for a single insertion,  only  a subset of scalar products
is independent, since
\begin{align}
\label{spni}
P\gbullet P = -P\wbullet P\,, \ \ Z\gbullet Z = - Z\wbullet Z\,, \ \ Z\gbullet P = -Z \wbullet P.
\end{align}

Under the symmetry breaking pattern,
the transverse tensor \eqref{eq:ElementaryBuildingBlock} breaks up in three pieces: $C^{AB},C^{IJ},C^{AJ}$. It turns out that only the last one, 
\begin{align}
C^{A\,I}=P^AZ^I-P^IZ^A,\label{eq:BasicTransverseDefect}
\end{align}
is necessary when dealing with bulk insertions.
Indeed, the other structures can be written as linear combinations of $C^{AI}$:
\begin{align}
&C_{AB}Q^{A}R^{B}=\frac{P\gbullet R}{P\wbullet G}C_{AI}Q^{A}G^{I}-\frac{P\gbullet Q}{P\wbullet G} C_{AI}R^{A}G^{I},\\
&C^{IJ}Q^{I}R^{J} = \frac{P\wbullet Q}{P\gbullet G}C_{AI}G^{A}R^{I} -\frac{P\wbullet R}{P\gbullet G}C_{AI}G^{A}Q^{I},
\end{align}
for generic vectors $Q,R$ and $G$ - in particular, one can always choose $G^M=P^M$. The tensor \eqref{eq:BasicTransverseDefect} also obeys the following identity:
\beq
C^{AI} C_{BI} C^{BJ} = \frac{1}{2} (C^{BI} C_{BI}) C^{AJ},
\label{3t}
\eeq
so we never need to concatenate more than two of these structures. 

Defect operators live on a $(p+1)$-dimensional light-cone within of the full $(d+1)$-dimensional one, and again they are encoded into polynomials of the two variables $W^I$ and $Z^A$. They are subject to the usual transversality rule, so that parallel indices satisfy $Z\gbullet Z=Z\gbullet P=0$. In particular, the polarization $Z^A$ should appear in correlation functions only through the structure \eqref{eq:ElementaryBuildingBlock}, restricted to the parallel indices.

Let us finally briefly comment on the issue of conservation in the presence of a defect.
One possibility to study consequences of conservation on a correlator is to project the embedding space expression to physical space, open the indices with the Todorov operator and then take a derivative. This is completely harmless, but sometimes slightly inconvenient. The other possibility is to work directly in the embedding space. In this case, conservation corresponds to 
\begin{align}
\partial^{M}D_{M}\overline{T}(X,Z)=0\label{eq:ConservationEmbeddingDefect}.
\end{align}
We use a different symbol with respect to (\ref{eq:ConservationEmbeddingNoDefect}), because $\overline{T}(X,Z)$ 
is obtained from $T(X,Z)$ imposing $Z\cdot P,\ Z\cdot Z=0$ but in a specific way, namely, using eqs. (\ref{spni}) to replace
everywhere
\begin{align}
\label{rplZP}
Z\gbullet P \rightarrow -Z\wbullet P,\, \ \ \ \ \  Z\gbullet Z\rightarrow -Z\wbullet Z .
\end{align}
The reason for this is that the operator  $\partial^{M}D_{M}$ is not interior to the 
conditions $Z\gbullet P=-Z \wbullet P$ and $Z\gbullet Z=-Z \wbullet Z$
and, for any expression $g$, gives different results when applied to the l.h.s. or r.h.s. of 
\begin{align}
g(Z\wbullet Z,Z\wbullet P,\dots)=g(-Z\gbullet Z,-Z\gbullet P,\dots);
\end{align}
we thus have to make a choice. 
The easiest way to establish the correct one is to notice that $Z\wbullet Z$ and $Z\wbullet P$ are mapped to $z\wbullet z$ and $z\wbullet x$ in physical space when the defect is flat - see subsection \ref{subsec:flat} - and that $\partial^{M}D_{M}$ acting on $Z\wbullet Z$ and $Z\wbullet P$ gives the same result as $\partial^{\mu}D_{\mu}$ acting on $z\wbullet z$ and $z\wbullet x$. Clearly, there is no ambiguity in physical space. The correct prescription is thus to use $Z\wbullet Z$ and $Z\wbullet P$ instead of $Z\gbullet P$ and $Z\gbullet Z$, \emph{i.e.} to apply eq. (\ref{rplZP}). Since the embedding formalism is insensitive to whether the defect is flat or spherical - see subsection \ref{subsec:spherical} - the same prescription works in the latter case as well.
\subsection{Projection to physical space: flat defect}
\label{subsec:flat}
The embedding space is a useful tool, but sometimes one is also interested in the result in physical space.
A simple rule to project down is to pick the polynomial expressions in the embedding space, project to physical space using the Poincar\'e section \eqref{eq:LightconeParametrization} and then open the indices with the appropriate Todorov operator. A defect $\mc{O}_\mc{D}$ extended on a flat sub-manifold $\mc{D}$ is embedded in the Poincar\'e section as follows:
\beq
P^M \in \mc{D}:\quad P^A=(1,x^2,x^a),\quad P^I=0.
\label{embedflat}
\eeq
Polarization vectors of bulk operators are evaluated according to \eqref{zbulkph}, and a similar rule holds for defect operators:
\beq
W^I=w^i,\quad Z^A=(0,2 x^a z_a, z^a).
\eeq
The projection to real space of orthogonal scalar products $(P\wbullet Q)$ is trivial, while for generic vectors $P_m,\ P_n$ and polarizations $Z_m,\ Z_n$ the rule is
\beq
-2P_m\gbullet P_n=|x_{mn}^a|^2+|x_m^i|^2+|x_n^i|^2,\, \qquad
 P_m\gbullet Z_n=x_{mn}^a z_n^a-z_n^i x_n^i.
 \label{eq:SimpleRulesProjectDefect}
\eeq

Let us now present a few examples. They can be obtained from the embedding space correlators derived in section \ref{sec:correlators}. Let us first consider the generic two-point function of a defect operator. This is a trivial example: we report it because it corresponds to a choice of normalization. The correlator in embedding space appears in eq. \eqref{defecttwopoint}. Projection to physical space yields
\begin{equation}
\langle\widehat{O}^{i_1...i_s}_\whD(x_1^a)\widehat{O}^{j_1...j_s}_\whD(x_2^a)\rangle 
=\frac{\mathcal{P}^{i_1...i_s; j_1...j_s}}{(x_{12}^2)^{\widehat{\Delta}}},
\label{2pdefectph}
\end{equation}
where $\mathcal{P}$ is the projector onto symmetric and traceless tensors, as defined in \cite{Costa:2011mg} in terms of the Todorov operators \eqref{eq:TodorovOrt}:
\begin{equation}\label{sttprojector}
\mathcal{P}^{i_1...i_s; j_1...j_s}\equiv \frac{1}{s!\left(\frac{q}{2}-1\right)_s}D_{i_1}\dots D_{i_s}w_{j_1}\dots w_{j_s}.
\end{equation}

Let us consider next the one-point function of a $J=2$ bulk primary $O_{\Delta,2}(x)$:

\begin{align}\label{tensonept}
\braket{O^{ab}_\Delta(x)} = \frac{q-1}{d}\frac{a_O}{\abs{x^i}^{\Delta}} \delta^{ab},\, \qquad 
\braket{O^{ij}_\Delta(x)} = -\frac{a_O}{\abs{x^i}^{\Delta}}\left(\frac{p+1}{d}\delta^{ij}-n^in^j \right),
\end{align}
where we introduced the versor $n^i\equiv x^i/|x^i|$ and we wrote $a_O$ for 
$a_{O_{\Delta,2}}$ for simplicity. Notice that this correlator is compatible with conservation, so that in particular the stress-tensor can acquire expectation value.

Among defect operators, a special role is played by a primary of transverse spin $s=1$ and null parallel spin. This is the displacement operator, which we describe in some detail in section \ref{sec:ward}. Hence we choose to write here the correlator of a defect primary $\wh{O}_{\whD,0,1}$ with these quantum numbers with bulk primaries of spin $J=1$ and $J=2$. The two-point function with the vector reads
\beq\label{bulkdefc0}
\!\!\!\!\braket{O_\Delta^{a}(x_1)\widehat{O}^i_\whD(x_2^b)}= \frac{-b_{O\wh{O}}}{(x_{12}^2)^{\widehat{\Delta}}\abs{x_1^i}^{\Delta-\widehat{\Delta}}} \frac{{x}_1^i x_{12}^a}{x_{12}^2} ,\qquad
\braket{O^{j}_\Delta(x_1)\widehat{O}^i_\whD(x_2^b)}= \frac{-b_{O\wh{O}}}{(x_{12}^2)^{\widehat{\Delta}}\abs{x_1^i}^{\Delta-\widehat{\Delta}}} \frac{{x}_1^i{x}_1^j}{x_{12}^2}.
\eeq
The correlator with a rank-two tensor $O_{\Delta,2}$ is a bit more lengthy, but straightforward to obtain:\footnote{The notation used in the following section is slightly different from this one. The coefficients $b_{n_1\dots n_4}$ in (\ref{eq:Structures2PtBulkdefect}) correspond to 
\begin{equation}
b_{1,2,0,0}=b_{O\wh{O}}^1\,, \quad b_{1,0,0,1}=b_{O\wh{O}}^2\,, \quad b_{0,1,1,0}=b_{O\wh{O}}^3.
\end{equation}}
\begin{align}\label{bulkdefc}
\braket{O^{ab}_\Delta(x_1)\widehat{O}_\whD^i(x_2^b)}&= \frac{n_1^i/d}{(x_{12}^2)^{\widehat{\Delta}}\abs{x_1^i}^{\Delta-\widehat{\Delta}}} \left\{b^1_{O\wh{O}}\left(\frac{4d \abs{x_1^i}^2 x_{12}^a x_{12}^b}{x_{12}^4}-\delta^{ab}\right)
+b^2_{O\wh{O}}(q-1)\delta^{ab}\right\}\notag\\
\braket{O^{ja}_\Delta(x_1)\widehat{O}^i_\whD(x_2^b)}&=\frac{1}{(x_{12}^2)^{\widehat{\Delta}}\abs{x_1}^{\Delta-\widehat{\Delta}}}\frac{-x_{12}^a \abs{x_1^i}}{x_{12}^2}\left\{ 2b^1_{O\wh{O}}\, n^i_1 n_1^j \left(1-\frac{2\abs{x_1^i}^2}{x_{12}^2}\right)
+b^3_{O\wh{O}}(\delta^{ij}-n^i_1 n_1^j) \right\},\notag\\
\braket{O^{jk}_\Delta(x_1)\widehat{O}^i_\whD(x_2^b)}&=\frac{1}{(x_{12}^2)^{\widehat{\Delta}}\abs{x_1^i}^{\Delta-\widehat{\Delta}}} \left\{b^1_{O\wh{O}}\,n_1^i\left[ 
n_1^jn_1^k\frac{\left(x_{12}^2-2 \abs{x_1^i}^2\right)^2}{x_{12}^4}
-\frac{1}{d}\delta^{jk}\right]\right.\notag\\
&+\left.b^2_{O\wh{O}}\,n_1^i \left(n^j_1n^k_1-\frac{p+1}{d}\delta^{jk}\right) \right.\notag\\
&+\left.b^3_{O\wh{O}} \left(1-\frac{2\abs{x_1^i}^2}{x_{12}^2}\right)\left(\frac{(\delta^{ik}n_1^j+\delta^{ij}n_1^k)}{2}-n_1^i n_1^j n_1^k \right) \right \}.
\end{align}

\subsection{Projection to physical space: spherical defect}
\label{subsec:spherical}

\begin{figure}[t]
\begin{center}
\includegraphics[scale=.55]{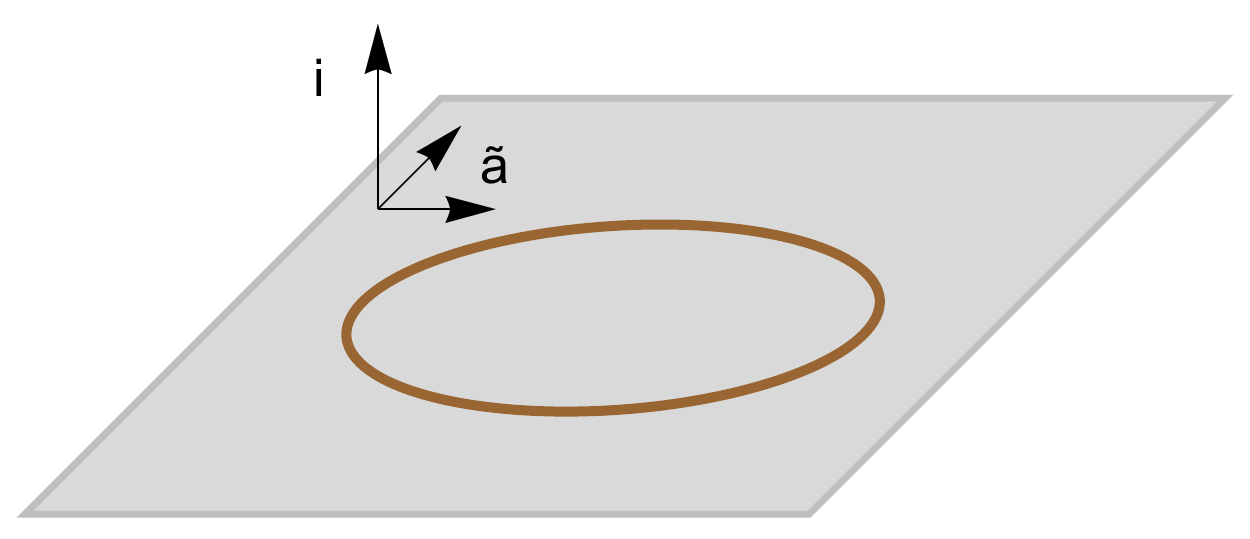}
\end{center}
\caption{The picture illustrates the choice of coordinates in this subsection: the spherical defect is drawn in brown and is placed on the plane spanned by the coordinates $x^{\tilde{a}}$.}\label{fig:sphericaldefect}
\end{figure}

A spherical defect is conformally equivalent to a flat one, therefore correlators in the presence of the former can be obtained from the homologous ones via a special conformal transformation - or simply an inversion. On the light-cone, such a transformation changes the embedding of the defect into the Poincar\'e section. On the other hand, the rules explained in subsection \ref{subsec:tooldefect} only care about the $so(p+1,1)\times so(q)$ symmetry pattern, which stays unchanged.\footnote{The embedding of the stability subgroup into $SO(d+1,1)$ gets conjugated by the same special conformal transformation.} As a consequence, correlators in embedding space encode both the flat and the spherical cases, the only difference lying in the choice of parallel and orthogonal coordinates, which we now describe. Without loss of generality, we consider a spherical $p$-dimensional defect of unit radius. Then, we abandon the light-cone coordinates in embedding space and use Cartesian coordinates instead:
\beq
P_x^{M}=(P^{0},P^{1},\dots,P^{d},P^{d+1})=\bigg(\frac{x^2+1}{2},x^{\mu},\frac{1-x^2}{2}\bigg),
\label{embedcart}
\eeq
the first entry being the time-like one. The relation between light-cone and Cartesian coordinates is $P^{\pm}=P^0\pm P^{d+1}$. We parametrize the $p$-sphere with stereographic coordinates $\si^a$ and center it at the origin of the $d$-dimensional space. A point on the defect turns out to be embedded as follows:
\beq
P^M \in \mc{D}:\ \ \, P^M(\si) = \bigg(1, \frac{2 \si^a}{\si^2+1},\frac{1-\si^2}{\si^2+1}, \underbrace{0,\dots, 0}_{\textup{q times}} \bigg) = 
\frac{2}{\si^2+1}\bigg(\frac{\si^2+1}{2},\si^a,\frac{1-\si^2}{2},0,\dots,0\bigg).
\label{embedsphere}
\eeq
By comparing the second equality with the embedding of the flat defect eq. \eqref{embedflat}, we see that the two defects are related by a rotation in the plane $(P^{p+1},P^{d+1})$, up to the conformal factor needed to bring us back to the Poincar\'e section. The action of the two factors of the stability subgroup is clear from eq. \eqref{embedsphere}, so that parallel and orthogonal coordinates are as follows:
\beq
P^A=(P^0,P^1,\dots, P^{p+1})\,, \qquad P^I=(P^{p+2},\dots, P^{d+1}).
\label{parortsphere}
\eeq
This is already enough to project the one-point functions to real space. It is sufficient to plug this choice of indices into eqs. \eqref{onepointboh}, \eqref{onepointstruc} and evaluate the correlator on the Poincar\'e section \eqref{embedcart}. Before giving an example, we introduce one more bit of notation: we use a tilde for the $(p+1)$ directions in which the $p$-sphere is embedded - see figure \ref{fig:sphericaldefect} 
Correspondingly, the index $i$ now only runs over the $(q-1)$ directions of the orthogonal subspace:
\beq
x^\m=(x^{\tilde{a}},x^i)\,,\qquad \tilde{a}=1,\dots, p+1\,,\ \ i=1,\dots,q-1.
\label{defatilde}
\eeq
We also use the radial coordinate $r= \left| x^{\tilde{a}}\right|$. The defect is placed in $r=1$. Let us now consider the expectation value of a spin two primary. The projection to physical space is done using \eqref{embedcart} and the polarization vector (\ref{zbulkph}) in cartesian coordinates reads
\begin{align}
Z= (x\cdot z,z^{\mu},-x\cdot z).
\label{Zcartesian}
\end{align}
Combining eqs. \eqref{embedcart} and \eqref{Zcartesian} we get the following projections:
\begin{gather}
P_m\wbullet P_n = x_m^{i}x_n^i+\frac{(1-x_m^2)(1-x_n^2)}{4},\, \ \ \ 
P_m\wbullet Z_n = x_m^{i}z_n^{i}+\frac{(x_m^2-1)}{2}\,z_n\cdot x_n,\notag\\
 Z_m\wbullet Z_n = z_m^{i}z_n^{i} + x_m\!\cdot\! z_m\, x_n\!\cdot\! z_n,
\label{projsphere}
\end{gather}
and similarly for the parallel scalar products.
Let us now consider the one-point function of the stress-tensor. The embedding space formula is eq. \eqref{onepointboh}. We project to physical space via eq. \eqref{projsphere}, we remove the polarization vectors with (\ref{eq:TodorovOp}), and we obtain
\begin{align}
&\langle T_{ij} \rangle  =a_{\mathcal{T}}\frac{\frac{d+1-q}{4d}(r^4+2r^2(|x^{i}|^2-1)+(1+|x^{i}|^2)^2)\delta_{ij}-r^2x_{i}x_{j}}{\left(\frac{(x^2-1)^2}{4}+|x^{i}|^2\right)^{d/2+1}},\\
&\langle T_{\tilde{a}i} \rangle = a_{\mathcal{T}}x_{\tilde{a}}x_{i}\frac{1-r^2+|x^{i}|^2}{2\left(\frac{\left(x^2-1\right)^2}{4}+|x^{i}|^2\right)^{d/2+1}} \nonumber\\
& \langle T_{\tilde{a}\tilde{b}} \rangle  = a_{\mathcal{T}}\frac{\frac{1-q}{4d}(r^4+2r^2(|x^i|^2-1)+(1+|x^{i}|^2)^2)\delta_{\tilde{a}\tilde{b}}+r^2x_{\tilde{a}}x_{\tilde{b}}}{\left(\frac{(x^2-1)^2}{4}+|x^{i}|^2\right)^{d/2+1}}.
\end{align}

As for the defect primaries, their parallel indices can be pulled-back to physical space by means of the Jacobian $\pa P^A(\si)/\pa \si^a$ of the map \eqref{embedsphere} - which is the rule for any spinning operator of a CFT on a sphere. Then, the auxiliary variable $Z^A$ is as usual determined in terms of its real space counterpart. With the choice of stereographic coordinates we get
\beq
Z^A= z^b\, \frac{\pa P^A(\si)}{\pa \si^b} = 
\bigg(0\,,\ 2\frac{z^a}{\si^2+1}-4\frac{z^b \si_b\, x^a}{(\si^2+1)^2}\,,\ 
-8\frac{z^b\si_b}{(\si^2+1)^2}\bigg).
\eeq
The $so(q)$ global symmetry now rotates vectors in the normal bundle to the sphere, and it is natural to choose the following basis of orthonormal vectors $n^\m_I$:
\beq
n^\m_i = \de^\m_i\,, \qquad
n^\m_r=-\de^\m_{\tilde{a}}\,\frac{x^{\tilde{a}}}{r}=-\de^\m_{\tilde{a}}\,x^{\tilde{a}}.
\label{normalsphere}
\eeq
We chose the radial vector to point inward. The reason for this will be clear in a moment.
Notice that there is an apparent clash of notation, between the index $I=(i,r)$ of $n^\m_I$ and the one in eq. \eqref{parortsphere}. The clash is, indeed, only apparent. To see this, let us consider the projection to physical space of a defect primary $\wh{O}^I(P(\si))$. The index has to be pulled back along the Poincar\'e section eq. \eqref{embedcart}:
\beq
\wh{O}_\m(\si)=\frac{\pa P_x^I}{\pa x^\m} \wh{O}_I(P(\si)).
\eeq
We know from the case of a trivial defect that $\wh{O}_\m(\si)$ transforms as a vector under the $so(q)$ factor of the conformal algebra. Furthermore, it is easy to verify that this operator has only components normal to the $p$-sphere. Therefore we can further recover an operator $\wh{O}_I(\si)$ in physical space via contraction with $n^\m_I$. However, we now notice that
\beq
\left. n^\m_J \frac{\pa P_x^I}{\pa x^\m} \right|_{P_x=P(\si)} = \de^I_J\,,
\eeq
which trivializes the projection with the identification of the $(d+1)$-th direction on the light-cone with the radial one in physical space. The last equation also provides the rule for projecting down the transverse polarization vectors:
\beq
W^I=(w^i,w^r)\,.
\label{wprojsphere}
\eeq
The choice of an inward pointing radial vector in eq. \eqref{normalsphere} has been made in order to avoid a minus sign in the last entry of eq. \eqref{wprojsphere}.
At this point, we are able to write down the expression in physical space of any correlation function. For instance, any bulk-to-defect two-point function involving a defect operator with $j=0$ is obtained by making the following substitutions in (\ref{eq:Structures2PtBulkdefect}):
\begin{align}
&P_1\wbullet W = x^{i}_1w^{i}+\frac{1-x^2_1}{2}w^{r},\, \ \ \ 
P_1\gbullet P_2 = x_1^{\tilde a} x_2^{\tilde a}  -\frac{1+x_1^2}{2}, \nonumber\\
& P_2\gbullet Z_1 = z^{\tilde a} x_2^{\tilde a} - z\cdot x_1\,, \ \ 
W_2\wbullet Z_1 = w^{i}z^{i}-x_1\!\cdot\! z\, w^{r}.
\end{align}
where we labelled the physical point corresponding to $P_2$ by means of its $p+1$ cartesian coordinates $x_2^{\tilde a}$, see eq. (\ref{defatilde}). Indeed, while the parametrization \eqref{embedsphere} makes it clear that operators on a spherical defect obey the same rules as operators in a CFT on a sphere, formulae may look nicer with this different choice.
Removing the polarization vectors is straightforward. The bulk polarizations can be removed using eq. (\ref{eq:TodorovOp}) as in the case of the one-point function, while the defect polarizations are removed by means of eq.s  (\ref{eq:TodorovPar}) and (\ref{eq:TodorovOrt}). As a simple example, let us present
the correlator of a bulk scalar primary and the displacement operator:
\begin{align}
&\langle O_{\D,0}(x_1)\textup{D}_{i}(x_2) \rangle = b_{O\textup{D}} \frac{x_{1,i}}{\left(\frac{(1-x^2_1)^2}{4}+|x^{i}_1|^2\right)^{\frac{\Delta_1-p}{2}}\left(x_1^{\tilde a} x_2^{\tilde a} -\frac{1+x_1^2}{2}\right)^{p+1}}\,,
\nonumber\\ 
&\langle O_{\D,0}(x_1)\textup{D}_{r}(x_2) \rangle = b_{O\textup{D}} \frac{(1 - x_1^2)/2}{\left(\frac{(1-x^2_1)^2}{4}+|x^{i}_1|^2\right)^{\frac{\Delta_1-p}{2}}\left(x_1^{\tilde a} x_2^{\tilde a} -\frac{1+x_1^2}{2}\right)^{p+1}}.
\end{align}

\section[Correlation functions in a Defect CFT]{Correlation functions in a Defect CFT.}
\label{sec:correlators}

In the previous section we established the rules of the game: we now would like to play and construct the tensor structures appearing in correlation functions of a defect CFT. It makes sense to start from the correlators which are fixed by symmetry up to numerical coefficients - the latter being the CFT data associated to the defect.

Correlation functions with only defect insertions obey the constraints of a $p$ dimensional CFT with a global symmetry. We will not have anything to say about this, besides the choice of normalization of a defect primary charged under the global symmetry:
\beq
\braket{\wh{O}_{\widehat{\Delta},0,s}(P_1,W_1)\wh{O}_{\widehat{\Delta},0,s}(P_2,W_2)}=\frac{(W_1\wbullet W_2)^s}{(-2P_1\gbullet P_2)^\D}.
\label{defecttwopoint}
\eeq

As mentioned in the introduction, the distinguishing feature of a conformal defect is the presence of bulk-to-defect couplings. The prototype of such interaction is the two-point function of a bulk and a defect operator, eq. \eqref{eq:BulkToDefectTwopoint}.
This correlator is fixed by conformal invariance, up to a finite number of coefficients. The set of two-point functions of a bulk primary with all defect primaries fixes its defect OPE. The simplest among such couplings is the expectation value of the bulk operator itself, aka the coupling with the identity on the defect. This has appeared in various places, and we derive it in the next subsection as a simple warm up.
The first correlator which includes dependence on cross-ratios is the two-point function of bulk primaries. In section \ref{sec:bulk2point} we provide the elementary building blocks for the tensor structures in this case, and we briefly comment on the choice of cross-ratios. Finally, subsection \ref{subsec:odd} is dedicated to parity odd structures.
 
\subsection{One-point function}
\label{subsec:onepoint}
The structure of the one-point function of a primary in the presence of a defect is easily constructed by means of the transverse tensor structure $C^{AI}$. Scale invariance implies that the one point function has the form 
\begin{align}
\langle O_{\D,J}(P,Z) \rangle = a_{{O}}\frac{ Q_J(P,Z)}{( P\wbullet P)^{\frac{\Delta}{2}}} \label{onepointboh}
\end{align}
where $Q_J(P,Z)$ is a homogeneous polynomial - whose normalization we will fix shortly - of degree $J$ in $Z$. 
It must moreover have degree zero in $P$ and be transverse, {\em i.e.}, it must satisfy
 $Q_J(P,Z+~\alpha P)=~Q_J(P,Z) $. The unique function with the aforementioned properties is
\begin{align}
Q_J=\left(\frac{C^{AI}C_{AI}}{2 P \wbullet P }\right)^{\frac{J}{2}} =
\left( \frac{(P\wbullet Z)^2}{P \wbullet P } - Z\wbullet Z\right)^{\frac{J}{2}}. 
\label{onepointstruc}
\end{align}
The proof just follows from the fact that we are only allowed to use the $C^{AI}$ building block, together with the identity \eqref{3t}.
Clearly, this implies that only even spin operators acquire an expectation value in a parity preserving theory (but pseudo-tensors make an exception, see subsection \ref{subsec:odd}). Furthermore, when the codimension is one the polynomial becomes trivial, which means that only scalars have non-vanishing one-point functions \cite{Liendo:2012hy}.
As a final remark, let us notice that the structure of the one-point function is compatible with conservation. Indeed, the condition
\begin{align}
\partial_{M}D^{M}\frac{ Q_{J}}{(P\wbullet P)^{\frac{\Delta}{2}}}=J(q+J-3)(d-\Delta+J-2)\, \frac{P\wbullet Z\,\mbox{ }Q_{J-2}}{2(P\wbullet P)^{\frac{\Delta+2}{2}}}=0
\end{align}
is satisfied if $\Delta=d-2+J$.

\subsection{Bulk-to-defect two-point function}
\label{subsec:btodefcorr}
A bulk-to-defect two-point function is a function of five variables:
\beq
\braket{O_{\Delta,J}(P_1,Z_1)}\widehat{O}_{\widehat{\Delta},j,s}(P_2,Z_2,W_2)\label{eq:BulkToDefectTwopoint}.
\eeq
As explained in subsection \ref{subsec:tooldefect}, $Z_2$ should appear in correlation functions only through the building block
\beq
C_2^{AB}=P_2^A Z_2^B-P_2^B Z_2^A.
\label{BasicPar}
\eeq
This elementary structure can be contracted with $P_1$, or with $Z_1$ through the building block $C_1^{AI}$ defined in \eqref{eq:BasicTransverseDefect}, while the other options lead to ``pure gauge'' terms, i.e., terms
that vanish upon enforcing $P^2=Z^2=P\cdot Z=0$.
In turn, terms of type $C_1^{AI}$ can be linked in chains of contractions, but luckily these chains do not become too long, thanks to the identity \eqref{3t}. When writing all the contractions which involve $C_2^{AB}$, one can remain loyal to the rule of exclusively using $C_1^{AI}$ in order to introduce the vector $Z_1$ in the game. However, one soon realizes that all the structures factorize. The factor that contains $Z_2$ is always given by
\beq
Q^0_{BD}= \frac{C_1^{AB} C_{2,AB}}{2P_1\gbullet P_2} = \frac{P_1\gbullet P_2\mbox{ } Z_1 \gbullet Z_2-P_2\gbullet Z_1\mbox{ } Z_2 \gbullet P_1}{P_1\gbullet P_2}.
\label{wstructure}
\eeq
We conclude that the most general tensor structure is given by the product of \eqref{wstructure} with other factors built out of at most two copies of $C_1^{AI}$, contracted with $P_1$, $P_2$ and $W_2$. The independent ones among these remaining structures are
\begin{align}
&Q_{BD}^{1}=\frac{P_1\wbullet W_2}{(P_1\wbullet P_1)^{1/2}},\qquad
Q_{BD}^{2}=\frac{P_1\wbullet Z_1\, P_1\gbullet P_2-P_2\gbullet Z_1\, P_1\wbullet P_1}{(P_1\wbullet P_1)^{1/2} (P_1\gbullet P_2)},\nonumber\\
&Q_{BD}^3= \frac{W_2\wbullet Z_1\, P_1\wbullet P_1-P_1\wbullet W_2\, P_1\wbullet Z_1}{P_1\wbullet P_1},\qquad Q_{BD}^4= Q_2(P_1,Z_1),
\label{bdefectgen}
\end{align}
where $Q_2$ was defined in eq. 
\eqref{onepointstruc}.
Thus, a generic bulk-to-defect two point function (\ref{eq:BulkToDefectTwopoint}) is given by  
\begin{align}
\braket{O_{\Delta,J}(P_1,Z_1)\widehat{O}_{\widehat{\Delta},j,s}(P_2,Z_2,W_2)}= (Q_{BD}^0)^{j}\sum_{\{n_i\}}b_{n_1\dots n_4}\frac{ 
 \prod_{k=1}^4 (Q_{BD}^{k})^{n_k}}{(-2P_1\gbullet P_2)^{\widehat{\Delta}}(P_1\wbullet P_1)^{\frac{\Delta-\widehat{\Delta}}{2}}}\,,
 \label{eq:Structures2PtBulkdefect}
\end{align}
where the sum runs over integers $n_i$ satisfying the condition $n_1+n_3=s$ and $n_2+n_3+2n_4=J-j$. The number of structures is given by
\begin{align}
N_{s,j;J}=\sum_{k=0}^{\textrm{Min}(s,J-j)}\left(1+\left\lfloor \frac{J-k-j}{2}\right\rfloor\right)\label{eq:CountingStructuresFormula}.
\end{align}
Notice that this only makes sense for $J\geq j$, which is easily understood from the leading order OPE. 

The $N_{s,j;J}$ structures do not correspond to as many independent coefficients when one of the primaries is conserved. 
Let us consider, as an example, the correlator of an $s=1,\,j=0$ defect primary and a conserved $J=2$ bulk primary. In this case the general results \eqref{bdefectgen}, \eqref{eq:Structures2PtBulkdefect} give:
\begin{align}
\braket{O_{\Delta,2}(P_1,Z_1)\widehat{O}_{\widehat{\Delta},0,1}(P_2,Z_2,W_2)} = \frac{ b_{1,2,0,0}Q_{BD}^{1}(Q_{BD}^{2})^2 + b_{0,1,1,0}Q_{BD}^{2}Q_{BD}^{3}+b_{1,0,0,1}Q_{BD}^{1}Q_{BD}^{4}
 }{(-2P_1\gbullet P_2)^{{\widehat{\Delta}}}(P_1\wbullet P_1)^{\frac{\Delta-{\widehat{\Delta}}}{2}}}\label{eq:Twopointfunctiondisplastresstensor}.
\end{align}
The projection to real euclidean space can be found in \eqref{bulkdefc}.
By imposing conservation in the form of eq. (\ref{eq:ConservationEmbeddingDefect}), that is
\begin{align}
&\partial_{M}D_{Z_1}^{M}\braket{O_{\Delta,2}(P_1,Z_1)\widehat{O}_{\widehat{\Delta},0,1}(P_2,Z_2,W_2)}=0,
\end{align}
we get the following constraint:
\begin{align}
&2 b_{1,2,0,0} (\widehat{\Delta}+d (p-\widehat{\Delta}))+(q-1) \left(b_{0,1,1,0} d-2 b_{1,0,0,1} \widehat{\Delta}\right) = 0,\,\nonumber\\
&  b_{0,1,1,0} d (\widehat{\Delta}-p)-2 b_{1,0,0,1} (p+1)-2 b_{1,2,0,0} = 0,\label{eq:conservation}
\end{align}
where we have used $\Delta=d$.
For generic values of $\widehat{\Delta}\,,p$ and $d$ this implies that there is just one independent coefficient. However, if $\widehat{\Delta}=p+1$ the rank decreases and two independent coefficients remain. This happens when $\widehat{O}$ is the displacement operator - see section \ref{sec:ward}.

\subsection{Two-point function of bulk primaries}
\label{sec:bulk2point}
In this section we analyze the structure of two-point functions of operators with spin in the bulk. The main novelty compared to the bulk-to-defect two-point function is that conformal symmetry is not powerful enough to fix completely the dependence on the positions. There are two cross-ratios, which we may choose as follows:
\beq
\xi=-\frac{2P_1\cdot P_2}{(P_1\wbullet P_1)^{\frac{1}{2}}(P_2\wbullet P_2)^{\frac{1}{2}}}\,, \qquad \cos\phi= \frac{P_1\wbullet P_2}{(P_1\wbullet P_1)^{\frac{1}{2}}(P_2\wbullet P_2)^{\frac{1}{2}}}.
\label{crossxiphi}
\eeq
\begin{figure}[t]
\begin{center}
\includegraphics[scale=.45]{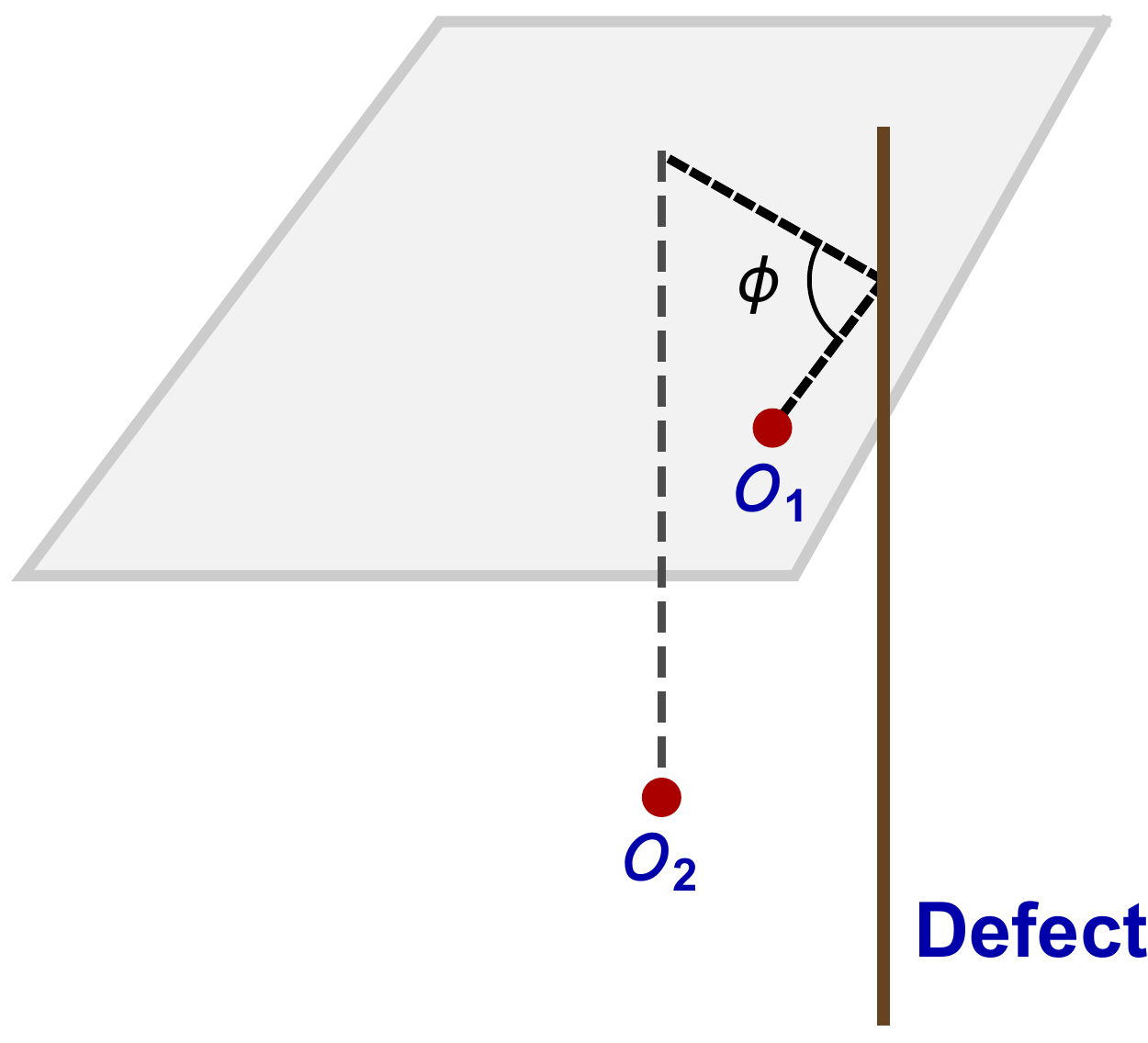}
\includegraphics[scale=.4]{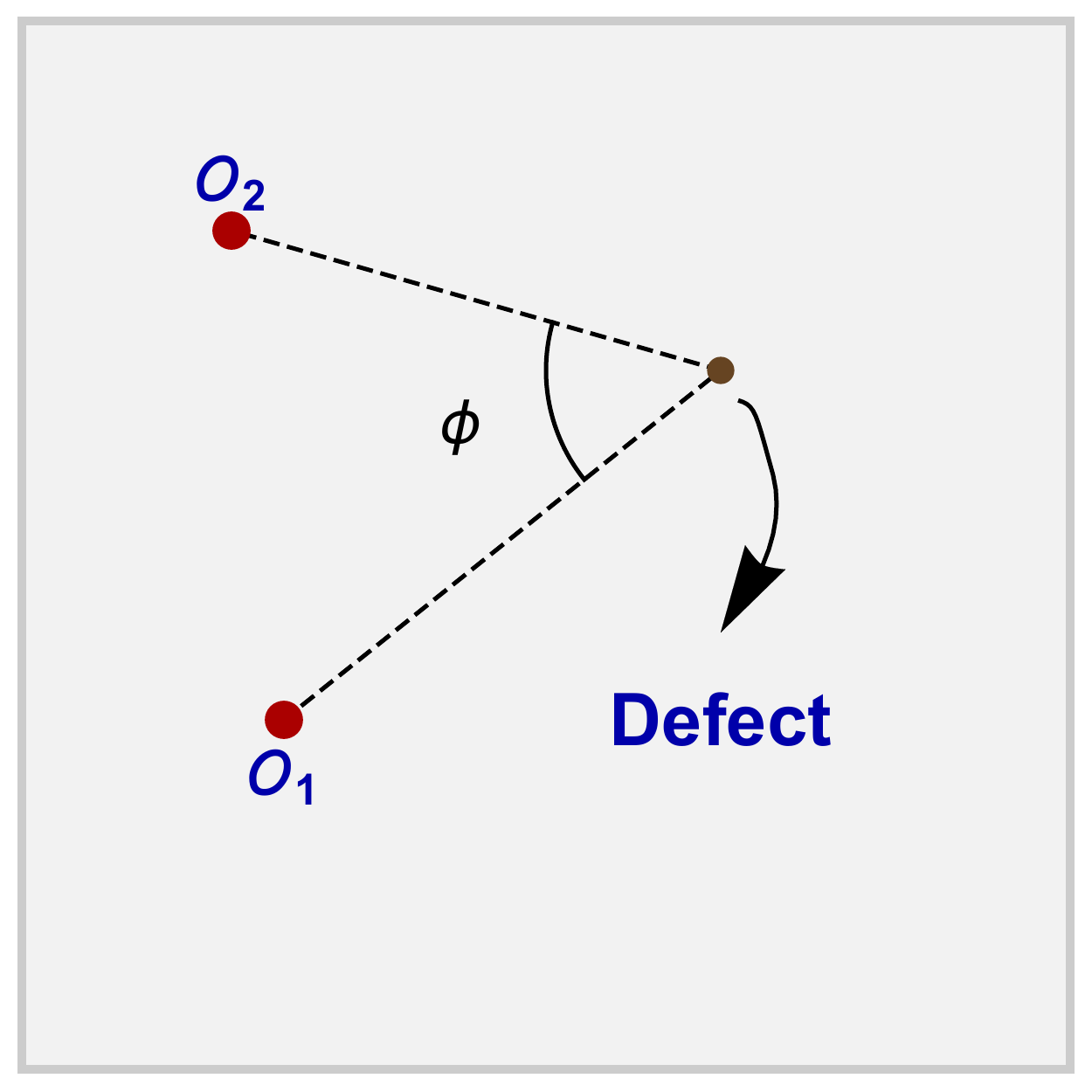}
\end{center}
\caption{The angle $\phi$ is formed by the projections of the vectors $P_1$ and $P_2$ onto the $q$-dimensional space orthogonal to the defect. The left figure gives a perspective view of this angle, while the one on the right gives the top view. The defect is represented here by a brown line (or brown point in the top view).  }\label{fig:anglephi}
\end{figure}
Let us pause to make a few comments on this choice. The first cross-ratio vanishes in the bulk OPE limit and diverges in the defect OPE one. The angle $\p$ is defined in fig. \ref{fig:anglephi}. 
This angle is not defined in the codimension one case, where the number of cross-ratios reduces to one.\footnote{Notice that $\xi=4\xi_\textup{\cite{McAvity:1995zd}}$, the latter being the cross ratio defined in \cite{McAvity:1995zd}. Their convention is motivated by the natural appearance of this factor in specific examples of boundary CFTs, as it is easy to see using the method of images. The same simplifications seem not to occur in examples 
with greater codimension $q$.} The cross-ratios \eqref{crossxiphi} are especially suitable for describing the bulk OPE in Lorentzian signature, where $\xi\to 0$ while $\cos \phi$ may remain constant. On the contrary, since $\cos \phi$ goes to one in the Euclidean OPE limit, it is not a useful variable in this case, and may be traded for instance for the following one:
\beq
\zeta = \frac{1-\cos \phi}{\xi}.
\label{crosszeta}
\eeq
Indeed, in the Euclidean OPE limit the two points approach each other along some direction $v$, or in other words $x_2=x_1+\epsilon \,v$. In the small $\epsilon$ limit the cross ratios \eqref{crossxiphi} behave as $\xi\approx O(\epsilon^2)$ and $\cos \phi \approx 1+  O(\epsilon^2)$, so that \eqref{crosszeta} stays fixed in the limit. This choice is by no means unique. We leave a thorough analysis for a future work,\footnote{V. Gon\c calves, E. Lauria, M. Meineri, E. Trevisani, \emph{work in progress}.} in which the choice of the most appropriate cross-ratio will be further discussed, along the lines of \cite{Hogervorst:2013sma}.

Finally, in the defect OPE limit, it is convenient to substitute $\xi$ with the following cross-ratio:
\beq
\chi=-\frac{2P_1\gbullet P_2}{(P_1\wbullet P_1)^{\frac{1}{2}}(P_2\wbullet P_2)^{\frac{1}{2}}},
\label{crosschi}
\eeq
which has the property of being invariant under $SO(q)$ transformations applied only to one of the two coordinates. The same applies to $\cos \phi$, now with respect to $SO(p+1,1)$ transformations. This property is useful in solving the defect Casimir equation \eqref{def_eigval}.

Let us now turn to the allowed structures in the correlation function. Recall that the polarization vectors $Z_1$ and $Z_2$ appear through the transverse structures (\ref{eq:BasicTransverseDefect}). These can be concatenated and contracted with $P_1$ and $P_2$. In the end, one can write the two-point function as follows:
\beq
\langle {O}_{\Delta_1,J_1}(P_1,Z_1){O}_{\Delta_2,J_2}(P_2,Z_2) \rangle =\sum_{\{n_i\}} \frac{\prod_{k=1}^8(Q_{BB}^k)^{n_k}f_{n_1\dots n_8}(\chi,\phi)}{(P_1\wbullet P_1)^{\frac{\Delta_1}{2}}(P_2\wbullet P_2)^{\frac{\Delta_2}{2}}}.\,\\
\eeq
The indices $n_i$ are subject to the constraints $n_1+n_2+n_5+n_6+2n_7=J_1$ and $n_3+n_4+n_5+n_6+2n_8=J_2$, which impose that the two point function is homogeneous of degrees $J_1$ and $J_2$ in the polarization vectors $Z_1$ and $Z_2$. Here are the building blocks $Q_{BB}^k$:
\begin{align}
&Q_{BB}^1=\frac{C^{AI}_1 P_{1A}P_{2I}}{(P_1\wbullet P_1)\, (P_2\wbullet P_2)^{1/2}},\, \ \ \ Q_{BB}^2= \frac{C^{AI}_1 P_{2A}P_{2I}}{(P_2\wbullet P_2)\, (P_1\wbullet P_1)^{1/2}}\,, \ \ Q_{BB}^3=\frac{C^{AI}_2 P_{1A}P_{2I}}{(P_2\wbullet P_2)\, (P_1\wbullet P_1)^{1/2}},\nonumber\\
&Q_{BB}^4=\frac{C^{AI}_2 P_{1A}P_{1I}}{(P_2\wbullet P_2)^{1/2} (P_1\wbullet P_1)},\, \ \ \ Q_{BB}^5= \frac{C^{AI}_1 C_2^{BI}P_{1A} P_{2B}}{(P_2\wbullet P_2) (P_1\wbullet P_1)},\, \ \ Q_{BB}^6=\frac{C^{AI}_1C_2^{AJ} P_{2I}P_{2J}}{(P_2\wbullet P_2)^{3/2} (P_1\wbullet P_1)^{1/2}}, \nonumber \\ 
&Q_{BB}^7=Q_{2}(P_1,Z_1),\, \ \ \ Q_{BB}^8=Q_{2}(P_2,Z_2).
\end{align}
The number of allowed structures is
\begin{align}
\sum_{k=0}^{\textrm{Min}(J_1,J_2)}(k+1)\left(J_1-k+1+\big\lfloor{\textstyle{\frac{J_1-k}{2}}} \big\rfloor\right)\left(J_2-k+1+\big\lfloor{\textstyle{\frac{J_2-k}{2}}} \big\rfloor\right).
\end{align}
Again, in the case of conserved operators the number of independent coefficients is smaller. The procedure used  to obtain the relevant constraints in the previous subsection can be applied here as well.
\subsection{Parity odd correlators}
\label{subsec:odd}
Let us finally comment on correlation functions of parity odd primaries. The strategy to construct them follows the same lines as in a homogeneous CFT: one has to consider all the allowed additional structures involving the $\epsilon$-tensor \cite{Costa:2011mg}. The main difference here is that the part of the residual symmetry group connected to the identity does not relate parity transformations applied to parallel and orthogonal coordinates. As a consequence, it is possible to use two more $\ep$-tensors.

The simplest possible example is provided by the one-point function of a bulk pseudo vector field in a CFT with a codimension $q=2$ defect. Since it is a bulk operator, it transforms according to irreducible representations of $O(d+1,1)$, so we have to use the total $\epsilon$-tensor:
\begin{align}
\langle  O_{\D,1}(P,Z)\rangle = a_{O} \frac{\epsilon_{01\dots \,p+1\, IJ}Z^{I}P^{J}}{(P\wbullet P)^{\frac{\Delta+1}{2}}}\,.
\end{align}
where the first $p+2$ coordinates are fixed by the defect. 

Defect operators carry separate parallel and orthogonal parity quantum numbers. Correlation functions involving primaries which are odd under one or the other require the use of the $\epsilon$-tensors $\ep_{AB\dots}$ and $\ep_{IJ\dots}$ respectively. The orthogonal $\epsilon$-tensor has $q$ indices that can be contracted with the vectors $P_1,\, Z_1$ and $W_2$. Therefore there are solutions only up to $q=3$. When $q=2$, the only transverse contractions are $\ep_{IJ}P_1^IW_2^J$ and $\ep_{IJ}P_1^IZ_1^J$, while for $q=3$, $\ep_{IJK}P_1^IZ_1^JW_2^K$ is the only possible parity odd structure. For instance, the two-point function of a defect operator odd under inversion of an orthogonal coordinate and a bulk vector, with $q=3$, reads
\begin{align}
\langle O_{\D, 1}(P_1,Z_1) \wh{O}_{\wh{\D},0,1}(P_2,W_2)\rangle = b_{O\wh{O}} \frac{\epsilon_{IJK}W_2^{I}Z_1^{J}P_1^{K}}{(P_1\wbullet P_1)^{\frac{\D-\wh{\Delta}+1}{2}}(-2P_1\gbullet P_2)^{\wh{\D}}}\,.
\end{align}
On the other hand, the parallel $\epsilon$-tensor has $p+2$ indices that can be contracted with four vectors $P_1,\, P_2,\, Z_1$ and $Z_2$, which implies that there are solutions up to $p=2$ and the correlators can be constructed exactly in the same way as was done above.
In particular, when $p=1$ only contractions of the kind $\ep_{ABC}P_1^AP_2^B Z_n^C$, $n=1,2$ are allowed.

\section[Scalar two-point function and the conformal blocks]{Scalar two-point function and the conformal blocks.}
\label{sec:blocks}

This section is devoted to the simplest crossing equation which contains information about the bulk-to-defect couplings. 
Let us consider the correlator between two scalar bulk operators $O_{1,2}$ of dimensions $\Delta_{1,2}$, which we write again in terms of the cross-ratios defined in subsection \ref{sec:bulk2point}:
\begin{equation}\label{sc_sc_bulkfunct}
\begin{array}{ll}
\displaystyle{\langle{O}_{1}(P_1){O}_{2}(P_2)\rangle=}&\displaystyle{\frac{f_{12}(\xi,\phi)}{(P_1\wbullet P_1)^{\frac{\Delta_1}{2}}(P_2\wbullet P_2)^{\frac{\Delta_2}{2}}}}.
\end{array}
\end{equation}
The function $f_{12}(\xi,\phi)$ can be decomposed into two complete sets of conformal blocks\footnote{See \cite{Dolan:2000ut,Dolan:2011dv} for an extensive discussion of conformal blocks in CFT, and \cite{McAvity:1995zd,Liendo:2012hy} for the case of a boundary.} by plugging either the bulk or the defect OPE \eqref{OPEI}, \eqref{compBOPE} in the l.h.s. of eq. \eqref{sc_sc_bulkfunct}:
\beq
f_{12}(\xi,\phi) =\xi^{-(\D_1+\D_2)/2}\sum_{k} c_{12k}\, a_k \,f_{\Delta_k,J}(\xi,\phi) =\sum_{\widehat{O}} b_{1\widehat{O}}\, b_{2\widehat{O}} \,\widehat{f}_{\widehat{\Delta},0,s}(\xi,\phi)\,.
\label{eq:crossingsy}
\eeq
Equality of the last two expressions is an instance of the crossing symmetry constraint.
 The bulk conformal blocks $f_{\D_k,J}$ are eigenfunctions of the quadratic Casimir of the full conformal group $SO(d+1,1)$, while the defect conformal blocks $\widehat{f}_{\widehat{\Delta},j,s}$ are eigenfunctions of the quadratic Casimir of the stability subgroup $SO(p+1,1)\times SO(q)$. In what follows we study the solutions of the two associated Casimir equations. We refer to appendix \eqref{Derivation_casimir} for a derivation of the equations themselves.

\subsection{Defect channel Casimir equation}

The sum in the rightmost side of eq. \eqref{eq:crossingsy} runs over the defect primaries that appear in both the defect OPE of $O_1$ and $O_2$. As we remarked in subsection \ref{subsec:btodefcorr}, only defect scalars ($j=0$) have a chance of being present. Since the stability subgroup is a direct product, the defect channel Casimir equation factorizes correspondingly, so that each defect conformal block satisfies separately the following eigenvalues equations:
\begin{align}
(\mathcal{L}^2 +\widehat{C}_{\widehat{\Delta},0})\frac{\widehat{f}_{\widehat{\Delta},0,s}(\xi,\phi)}{(P_1\wbullet P_1)^{\frac{\Delta_1}{2}}(P_2\wbullet P_2)^{\frac{\Delta_2}{2}}}  = 0,\notag\\
(\mathcal{S}^2 +\widehat{C}_{0,s})\frac{\widehat{f}_{\widehat{\Delta},0,s}(\xi,\phi)}{(P_1\wbullet P_1)^{\frac{\Delta_1}{2}}(P_2\wbullet P_2)^{\frac{\Delta_2}{2}}}  = 0,
\label{def_eigval}
\end{align}
where $\widehat{C}_{\widehat{{\Delta}},s}=\widehat{\Delta}(\widehat{\Delta}-p)+s(s+q-2)$ and the differential operators $\mathcal{L}\equiv \frac{1}{2} (\mathcal{J}_{AB})^2$,  $\mathcal{S}\equiv~\frac{1}{2}~(\mathcal{J}_{IJ})^2$ are defined through the generators
\begin{equation}
\displaystyle{\mathcal{J}_{MN}=P_M\frac{\partial}{\partial P^N}-P_N\frac{\partial}{\partial P^M}}.
\label{generator}
\end{equation}
In eq. \eqref{def_eigval}, the operators $\mc{L}^2$ and $\mc{S}^2$ act only on one of the points, say $P_1$.  Eqs. \eqref{def_eigval} immediately translate into differential equations for $\widehat{f}_{\widehat{\Delta},0,s}(\xi,\phi)$:
\begin{align}\label{casimdef}
\mathcal{D}_{\textrm{def}}^{L^2}\mbox{ }\widehat{f}_{\widehat{\Delta},0,s}(\xi,\phi)=0,\nonumber\\
\mathcal{D}_{\textrm{def}}^{S^2}\mbox{ }\widehat{f}_{\widehat{\Delta},0,s}(\xi,\phi)=0.
\end{align}
The differential operators are most conveniently expressed by trading $\xi$ for the variable $\chi$, which was defined in \eqref{crosschi}:
\begin{equation}
\begin{array}{ll}
\displaystyle{\mathcal{D}^{S^2}_{\textrm{def}} \equiv 4\cos\phi(1-\cos\phi) \frac{\partial^2}{\partial \cos\phi^2}+2(1-q\cos\phi) \frac{\partial}{\partial\cos\phi}+\widehat{C}_{0,s}},\\
\displaystyle{\mathcal{D}^{L^2}_{\textrm{def}} \equiv (4-\chi^2) \frac{\partial^2}{\partial \chi^2} - (p+1)\chi \frac{\partial}{\partial \chi}+\widehat{C}_{\Delta,0}}.
\end{array}
\end{equation}
The complete solution of the system (\ref{casimdef}) is then:
\begin{multline}\label{def_block}
\widehat{f}_{\widehat{\Delta},0,s}(\chi,\phi)=\alpha_{s;q}\,\chi^{-\widehat{\Delta}} \, _2F_1\left(\frac{q}{2}+\frac{s}{2}-1,-\frac{s}{2};\frac{q}{2}-\frac{1}{2};\sin^2\phi\right)\\
\times\, _2F_1\left(\frac{\widehat{\Delta} }{2}+\frac{1}{2},\frac{\widehat{\Delta} }{2};\widehat{\Delta} +1-\frac{p}{2};\frac{4}{\chi^2}\right),
\end{multline}
where $\alpha_{s;q}=2^{-s}\frac{\Gamma\left(q+s-2\right)}{\Gamma\left(\frac{q}{2}+s-1\right)}\frac{\Gamma\left(\frac{q}{2}-1\right)}{\Gamma\left(q-2\right)}$. The normalization has been chosen such that, given a leading contribution to the defect OPE of the kind:
\begin{equation}
O_{1} (x)=b_{1\widehat{O}}|x^i|^{-\Delta_1+\widehat{\Delta}-s}\,x_{i_1}...x_{i_s}\widehat{O}^{i_1...i_s}(\hat{x})+...,
\end{equation}
the contribution of $\widehat{O}$ to the two-point function \eqref{sc_sc_bulkfunct} is as shown in eq. \eqref{eq:crossingsy}.
Also, recall that the normalization of the defect-defect correlator is fixed by \eqref{defecttwopoint} (see also eq. \eqref{2pdefectph}).
Finally, note that the transverse factor in the conformal block \eqref{def_block} correctly reduces to a Gegenbauer polynomial for integer $s$:
\begin{equation}
{}_2F_1\left(\frac{q}{2}+\frac{s}{2}-1,-\frac{s}{2};\frac{q}{2}-\frac{1}{2};\sin^2\phi\right)=\frac{\Gamma(s+1)\Gamma(q-2)}{\Gamma(q+s-2)}\,C_s^{(\frac{q}{2}-1)}\!\left(\cos\phi\right).
\end{equation}

\subsection{Bulk channel Casimir equation}
The sum over bulk operators in the second equality of eq. \eqref{eq:crossingsy} runs over all primaries admitted in $O\times O$ with non-vanishing one point functions. In particular, as follows from subsection \ref{subsec:onepoint}, the sum can be restricted to even spins $J$. 
Each bulk conformal block is an eigenfunction of the $SO(d+1,1)$ Casimir operator $\mathcal{J}^2$ with eigenvalue $C_{\Delta_k,J}=\Delta_k(\Delta_k-d)+J(J+d-2)$:
\begin{align}\label{bulk_eigval}
(\mathcal{J}^2 +C_{\Delta_k,J})\frac{ \xi^{-(\D_1+\D_2)/2}f_{\Delta_k,J}(\xi,\phi)}{(P_1\wbullet P_1)^{\frac{\Delta_1}{2}}(P_2\wbullet P_2)^{\frac{\Delta_2}{2}}}  = 0,
\end{align}   
where $\mathcal{J}^2\equiv \frac{1}{2}(\mathcal{J}_{MN}^{(1)}+\mathcal{J}_{MN}^{(2)})^2$ and $\mc{J}_{MN}$ is defined in \eqref{generator}.
The differential equation for $f_{\Delta_k,J}(\xi,\phi)$ which follows from eq. \eqref{bulk_eigval},
\begin{align}\label{casim}
\mathcal{D}_{\textrm{bulk}}f_{\Delta_k,J}(\xi,\phi)=0,
\end{align}
contains the differential operator
\begin{multline}
\mathcal{D}_{\textrm{bulk}} \equiv 2\xi^2\left(2+\xi\cos\phi+2\cos^2\phi\right)\frac{\partial^2}{\partial \xi^2}+2\sin^2\phi \left(2\sin^2\phi-\xi\cos\phi\right)\frac{\partial^2}{\partial \cos\phi^2} \\
-4\xi \sin^2\phi\left(\xi +2\cos\phi\right)\frac{\partial^2}{\partial \xi\partial\cos\phi}+2\xi\left[2(1+\cos^2\phi)-(2d-\xi\cos\phi)\right]\frac{\partial}{\partial \xi}\\
 +\left[2\xi(q-2+\cos^2\phi)-4\cos\phi \sin^2\phi\right]\frac{\partial}{\partial \cos\phi}
-\left[\Delta_{12}^2 \cos \phi  \left(\cos \phi + \frac{\xi}{2} \right) -\Delta_{12}^2+2C_{\Delta_k,J} \right],
\label{bulkdiffop}
\end{multline}
where $\Delta_{12}=\Delta_{1}-\Delta_{2}$.
We will not be able to solve this differential equation in closed form in the most general case. In the next subsection, we provide a recurrence relation for the light-cone expansion of the conformal block. Then, in subsection \eqref{exact_sol_cod2}, we will consider a special case, in which the Casimir equation can be mapped to a different one, well studied in the literature.  Finally, we leave to appendix \ref{app_scalar} a solution for the scalar block in terms of an infinite series. For now, let us point out that the equation can be solved when the codimension is one ($q=1$). In this case, only scalars acquire an expectation values ($J=0$), and \eqref{casim} reduces to an hypergeometric equation in $\xi$. The solution with the correct asymptotics is \cite{McAvity:1995zd}
\begin{equation}
\begin{array}{ll}
\displaystyle{ f_{\Delta_k,0}(\xi)=\xi^{\frac{\Delta_k}{2}}{}_2F_1\left(\frac{\Delta_k+\Delta_{12}}{2},\frac{\Delta_k-\Delta_{12}}{2},\Delta_k+1-\frac{d}{2};-\frac{\xi}{4}\right)}.
\end{array}
\end{equation}
We fix the normalization of the conformal blocks in the next subsection, while discussing the collinear block - see eq. \eqref{coll}.

\subsubsection{Lightcone expansion}\label{asympt_CB} 
It is a well understood fact that the nature of the light cone limit in the Lorentzian OPE limit is different from the Euclidean one. For instance, the operators that dominate in each limit are not the same: while in the Lorentzian the operators with lowest twist contribute the most (with the twist $\tau=\Delta_k-J$ being defined as the dimension minus the spin), in the Euclidean it is the ones with lowest dimension. This is easily seen by considering the leading order OPE
\begin{align}\label{opexpr}
\!\!\!\!{O}_{1}(x_1){O}_{2}(x_2) =
\sum_k \frac{c_{12k}}{(x_{12}^2)^{\frac{\Delta_1+\Delta_2-\Delta_k}{2}}}
\left[ \frac{x_{12\mu_1}\dots x_{12\mu_J}}{(x_{12}^2)^{\frac{J}{2}} }\,
{O}_k^{\mu_1\dots \mu_J} (x_2) + {\rm descendants} \right].
\end{align}
It follows directly from the OPE that $\xi$ controls the twist of the operator being exchanged in the OPE limit. 
\begin{figure}[t]
\begin{center}
\includegraphics[scale=.4]{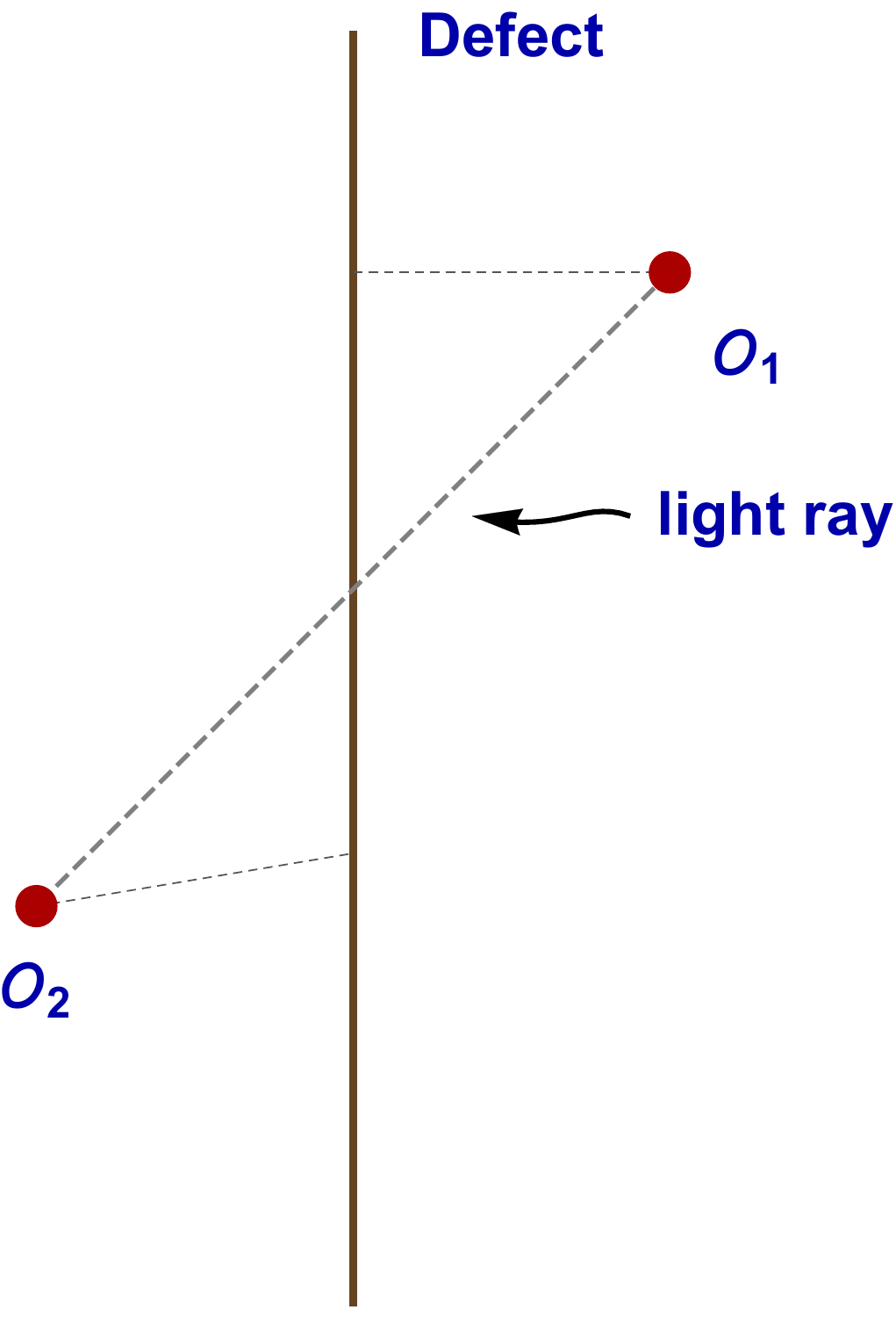}
\end{center}
\caption{The light-cone limit relevant for the bulk Casimir equation: the defect is time-like and the operators are space-like separated from it.
}\label{fig:lightcone}
\end{figure}
The conformal block resums the contribution of a conformal family. It follows from the observation above that it is possible to organize every such family into operators with the same twist, which contribute at the same order in the light-cone limit. In said limit, the operators become light-like separated without colliding - see fig. \ref{fig:lightcone}. In terms of the cross-ratios, this corresponds to sending $\xi\rightarrow 0$ while holding $\phi$ fixed.
Then, from a direct analysis of the solution of the Casimir equation for the first few orders in small $\xi$, one is led to the following ansatz:
\begin{align}\label{lightconexp}
f_{\Delta_k,J}(\xi,\phi) = \sum_{m=0}^{\infty}\sum_{k=0}^{m}c_{m,k}g_{\tau + 2m,J-2k}(\xi,\phi),\qquad g_{\tau,J}(\xi,\p)=\xi^{\tau/2} {\tilde g}_{\tau,J}(\p),
\end{align}
where $g_{\tau ,J}(\xi,\phi)$ includes the contribution of operators with twist $\tau$ and spin $J$, and we determine it below. Notice that, given a primary operator ${O}_{\mu_1,\dots,\mu_J}(x_1)$ it is possible to construct different descendants with the same twist but different spin. For example, given a primary operator with spin $J$ and twist $\tau$ we can create a descendant operator with twist $\tau+2$ either by acting with $P^2$ or by contracting one of the indices of the primary operator with $P^{\mu}$. In formula \eqref{lightconexp} this degeneracy is labelled by $k$.  For a given $m$, the number of descendants with different spin is $m+1$.

In order to constrain the functions $g_{\tau ,J}(\xi,\phi)$ and the coefficients $c_{m,k}$, it is convenient to divide the Casimir differential operator in two parts \cite{Hogervorst:2013sma}, one that keeps the degree of $\xi$ and other that does not, $\mathcal{D}_{\textrm{bulk}}=\mathcal{D}_0+\mathcal{D}_1+2C_{\Delta_k,J}$:
\begin{align}
&\mathcal{D}_{\textrm{0}} \equiv 4\xi^2(1+\cos^2\phi)\frac{\partial^2}{\partial \xi^2}+4\sin^4\phi \frac{\partial^2}{\partial \cos\phi^2}-8\xi \sin^2\phi\cos\phi\frac{\partial^2}{\partial \xi\partial\cos\phi}-4\xi(d-1-\cos^2\phi)\frac{\partial}{\partial \xi}\nonumber\\
&-4\cos\phi \sin^2\phi \frac{\partial}{\partial \cos\phi}-\Delta_{12}^2(\cos^2 \phi -1),\\
&\mathcal{D}_{\textrm{1}}\equiv 2\xi^3\cos\phi\frac{\partial^2}{\partial \xi^2}-2\xi\sin^2\phi\cos\phi \frac{\partial^2}{\partial \cos\phi^2}-4\xi^2 \sin^2\phi\frac{\partial^2}{\partial \xi\partial\cos\phi}+2\xi^2\cos \cos\phi\frac{\partial}{\partial \xi}\nonumber\\
& +2\xi(d-p-2+\cos^2\phi)\frac{\partial}{\partial \cos\phi}-\frac{\xi}{2}\Delta_{12}^2 \cos \phi .
\end{align}
By inspection of the Casimir equation, one can conclude that $g_{\tau,J}(\xi,\phi)$ is an eigenfunction of the differential operator $\mathcal{D}_{0}$ with eigenvalue $2C_{\tau,J}=(J+\tau ) (J+\tau-d  )+J (d +J-2)$. More precisely, for $m=0$ this property coincides with the leading order of the Casimir equation. At all orders, we simply check that the equation can be solved by the ansatz \eqref{lightconexp} and this choice of functions. Clearly, the solution is unique once the asymptotic behavior has been chosen.
The eigenvalue equation fixes $g_{\tau,J}(\xi,\phi)$ to be
\begin{align}\label{coll}
g_{\tau,J}(\xi,\phi) = \xi^{\frac{\tau}{2}}\sin^J \phi \,_2F_1\left(\frac{2J+\tau+\Delta_{12}}{4},\frac{2J+\tau-\Delta_{12}}{4},\frac{2J+\tau+1}{2},\sin^2\phi \right).
\end{align}
We chose the solution in such a way that $g_{\tau,J}(\xi,\phi)\sim \xi^{\frac{\tau}{2}}\sin^{J}\phi $ in the limit $\phi \rightarrow 0$, which is the asymptotics required by the Euclidean OPE limit. The normalization of $g_{\tau,J}$ has been fixed so that, given the leading order OPE eq. \eqref{opexpr}, and the normalization of the one-point function eq. \eqref{onepointboh}, the conformal block \eqref{lightconexp} contributes to the two-point function as in eq. \eqref{eq:crossingsy}, once $c_{0,0}=1$.\footnote{This matching is most easily done by noticing that in the light cone limit $x^\m$ in eq. \eqref{opexpr} becomes null. Then, comparing with the one point function (\ref{onepointboh}), we can just specify $z=x$. This immediately leads to 
\begin{align}
\langle{O}_1(x_1){O}_2(x_2) \rangle  \sim \frac{c_{12k}a_{{O}}}{(x_{12}^2)^{\frac{1}{2}\left(\Delta_1+\Delta_2-\Delta_k+J\right)}}\frac{((x_1\wbullet x_{12})^2-x_{12}\wbullet x_{12}\mbox{ } x_1\wbullet x_1)^{\frac{J}{2}}}{(x_1\wbullet x_1)^{\frac{\Delta_k+J}{2}}}\label{eq:LeadingOPE} .
\end{align}
We can compare with the collinear block by taking the limit $x_{12}\rightarrow 0$, which  corresponds to $\sin \phi \rightarrow 0 $. 
} 

The action of $\mathcal{D}_1$ on $g_{\tau,J}(\xi,\phi)$ can be expressed as a combination of the these building blocks with one more unit of twist
\begin{align}
&\mathcal{D}_1g_{\tau,J}(\xi,\phi) = b^{+}_{\tau,J}\, g_{\tau+1,J}(\xi,\phi)+b^{-}_{\tau,J}\, g_{\tau+1,J-2}(\xi,\phi),\,\\ 
&b^{-}_{\tau,J}= 8 J  (q+J-3),\quad
b^{+}_{\tau,J}=\frac{(J+\tau -1) \left(4 J^2+4 J \tau +\tau ^2-\Delta_{12}^2\right) (q-J-\tau -2)}{2 (2 J+\tau -1) (2 J+\tau +1)}.\nonumber\, \
\end{align}
Thus, the Casimir equation can be translated into a recurrence relation that the coefficients $c_{n,k}$ must satisfy:
\begin{align}
2(C_{\tau+2n,J-2k}-C_{\tau,J})c_{m,k}= b^+_{\tau+2n-2,J-2k}c_{n-1,k}+b^{-}_{\tau+2n-2,J-2k+2}c_{n-1,k-1}.
\end{align}
All coefficients $c_{n,k}$ are determined once we impose the initial conditions $c_{n,k}=0$ for $n<j$ and $c_{0,0}=1$. In particular, in the case of an exchanged scalar primary, the general term of the recursion can be recovered from eq. \eqref{scalar_res_real}. Finally, notice that this solution can be straightforwardly applied to the conformal block for the four-point function without a defect. In that case we would be solving the Casimir equation in the limit of small $u$ and fixed $v$ - see appendix A of  \cite{Costa:2012cb}.

\subsubsection{Defects of codimension two and the four-point function}\label{exact_sol_cod2}
The four-point function of scalar primaries in a homogeneous CFT has been extensively studied in the literature. Many results are available for the conformal blocks in this case \cite{Dolan:2000ut,Dolan:2011dv,Costa:2012cb,Dolan:2003hv,Costa:2011dw,Penedones:2015aga,Rychkov:2015lca}. In the special case of a codimension two defect - with equal external dimensions - all those results apply in fact to our problem. Indeed, let us consider the change of variables
\begin{align}\label{maps}
&u=-\xi e^{-i \phi}, \qquad v=e^{-2 i\phi};\\
&\xi=-\frac{u}{\sqrt{v}}, \qquad \cos\phi=\frac{1+v}{2\sqrt{v}}.
\end{align}
Let us denote by $\mathcal{D}_{\textrm{CFT}}$ the Casimir differential operator for the four-point function without defect, and let us choose pairwise equal external dimensions $\widetilde{\Delta}_{12}=\widetilde{\Delta}_{34}=0$, so that they do not appear in $\mathcal{D}_{\textrm{CFT}}$ - see for instance eqs. (2.10) and (2.11) in \cite{Dolan:2011dv}. We obtain the following relation with the operator \eqref{bulkdiffop}, when $\D_1=\Delta_2$:
\begin{align}
\mathcal{D}_{\textrm{def}}[f_{\Delta_k,J}(u,v)] = \big(\mathcal{D}_{\textrm{CFT}} f_{\Delta_k,J}(u,v)+{\mathcal{D}}_q f_{\Delta_k,J}(u,v) \big)= 0,
\end{align}
where
\begin{align}
{\mathcal{D}}_q = \frac{4u(q-2)}{1-v}\bigg(u\frac{\partial}{\partial u}+2v\frac{\partial}{\partial v}\bigg).
\end{align}
Hence, in the special case of $q=2$, the functions $ f_{\Delta_k,J}(u,v)$ solve the Casimir equation for the four point function with pairwise equal external operators, $f_{\Delta_k,J}^{4\mbox{pt}}(u,v)$:
\begin{equation}\label{map2}
f_{\Delta_k,J}(u,v)=f_{{\Delta}_k,{J}}^{4\mbox{pt}}(u,v).
\end{equation} 
To confirm that \eqref{map2} is an equality between conformal blocks, we need to match the asymptotic behavior.\footnote{Conformal blocks of the four-point function with internal quantum numbers
 $\Delta'$ and $J'$, such that $C_{\Delta_k,J}=C_{\Delta',J'}$, solve eq. \eqref{bulkdiffop} as well. The asymptotics single out the unique physical solution \eqref{map2}, with $\Delta'=\Delta_k$ and $J'=J$.} In fact, we can check that \eqref{map2} holds explicitly at leading order in the lighcone expansion \eqref{lightconexp}:
\begin{equation}
f_{\Delta_k,J}(\xi,\phi)=g_{\tau,J}(\xi,\phi)+\mathcal{O}(\xi),
\end{equation}
where $g_{\tau,J}(\xi,\phi)$ is defined in \eqref{coll}. An identity between hypergeometric functions 
\begin{multline}\label{rel_hyp}
e^{-i(a+m)\phi} {}_2F_1\left(a+m,m+\frac{a+b}{2},a+b+2m;2i\sin\phi e^{-i\phi} \right)\\
={}_2F_1\left(\frac{a+m}{2},\frac{b+m}{2},\frac{a+b+1}{2}+m;\sin^2\phi \right).
\end{multline}
allows to rewrite (\ref{coll}) in terms of $u$ and $v$
\begin{equation}\label{ll}
g_{\tau,J}(\xi,\phi)=\ii^{-\D} u^{\frac{\tau}{2}}\left(\frac{1-v}{2}\right)^{J}\mbox{ } {}_2F_1\left(\frac{\tau}{2}+J,\frac{\tau}{2}+J,\tau+2J; 1-v\right).
\end{equation} 
Here we have assumed once again $\Delta_1=\Delta_2$. The result agrees with the collinear block for the four-point function - see for instance eq. (118) in \cite{Costa:2012cb}. Hence, all formulae for the conformal blocks of a homogeneous CFT provide as many results for the bulk channel blocks of a defect CFT with codimension two. It would be interesting to have a geometric understanding of this fact. For now, let us just notice that the mapping \eqref{maps} has a chance of having some kinematic meaning only when the defect is placed in a space with Lorentzian signature.

An interesting check of eq. \eqref{map2} is related to the reality property of $f_{\Delta_k,J}^{4\mbox{pt}}$. If $\xi$ and $\p$ are real, this block should be real when evaluated on the map \eqref{maps}. Certainly, $\left(f^{4\mbox{pt}}_{\Delta_k,J}(u,v)\right)^*$ and $f^{4\mbox{pt}}_{\Delta_k,J}(u^*,v^*)$ coincide, since they satisfy the same Casimir equation with the same boundary conditions. Then one notices that, through eq. \eqref{maps}, the complex conjugation acts on $u,v$ as
\beq
{}^* : (u,v) \longrightarrow \left(\frac{u}{v},\frac{1}{v}\right).
\eeq
But this is easily recognized as the effect of exchanging points $x_1$ and $x_2$.\footnote{Recall that in the standard notation $u=\frac{x_{12}^2x_{34}^2}{x_{13}^2x_{24}^2}$, $v=\frac{x_{14}^2x_{23}^2}{x_{13}^2x_{24}^2}$.} Conformal blocks for the exchange of a primary with even spin are invariant under this crossing, when $\widetilde{\Delta}_{12}=\widetilde{\Delta}_{34}=0$. We conclude that $f^{4\mbox{pt}}_{\Delta_k,J}(u^*,v^*)=f^{4\mbox{pt}}_{\Delta_k,J}(u,v)$ precisely in the case of interest.

\section[Ward identities and the displacement operator]{Ward identities and the displacement operator.}
\label{sec:ward}
This section is devoted to the Ward identities involving the stress-tensor, in the presence of a flat defect, both in the Poincar\'e (or rather, Euclidean) invariant and in the fully conformal cases. Throughout the section, we ignore the issue of defect anomalies. We derive the Ward identities in the next subsection. These involve a number of defect operators, one of which plays a pre-eminent role: the displacement operator defined in eq. \eqref{dispdef}. In subsection \ref{constraints}, we focus on the displacement operator and its properties, and derive some constraints on its appearance in the defect OPEs of a generic theory. In subsection \ref{subsec:2d}, we take a look at two dimensional interfaces, and prove unitarity bounds for the Zamolodchikov norm $C_\textup{D}$ of the displacement. Finally, in subsection \ref{examples}, we consider examples of free defect theories, in which specific identities can be given to the operators defined in subsection \ref{subsec:ward}, in terms of the elementary fields.

\subsection{The Ward identities for diffeomorphisms and Weyl transformations}
\label{subsec:ward}

  Let us consider a defect quantum field theory defined on a manifold $\mathcal{M}$ and coupled to a background metric $g$. We define the embedding of the defect sub-manifold $\mathcal{D}$ through
\beq
x^\m=X^\m(\si^a),
\eeq
where the coordinates $\si^a$, $a=1,\ldots,p$ provide a parametrization of $\mathcal{D}$.
We assume that a vacuum energy functional is defined through a functional integral of the form
\beq
W[g,X]= \log \int [D\p] \exp \left\{ - S[g,X,\p,\partial\p,\dots] \right\}.
\label{vacfun}
 \eeq
We ask the action functional to be invariant under diffeomorphisms on $\mathcal{M}$, once the embedding function is allowed to change accordingly:
\begin{align}
&S[g+\de_\xi g,\, X+\de_\xi X,\,\p+\de_\xi \p,\,\partial\p+\de_\xi\partial\p,\dots] = S[g,X,\p,\partial\p,\dots], \label{diffact} \\
&\de_\xi X^\m =\xi^\m, \label{diffemb}\\
&\de_\xi g_{\m\n}=-\nabla_\m \xi_\n-\nabla_\n \xi_\m. \label{diffmetric}
\end{align}
This can be achieved by coupling the defect degrees of freedom to local geometric quantities. The induced metric $\ga_{ab}$, the $q$-tuple of normal unit vectors $n_A^\m$, the extrinsic curvatures $K_{ab}^A$ and the spin connection on the normal bundle $\m_{a IJ}$ are defined in eqs. \eqref{indmetdef}, \eqref{normalvecdef}, \eqref{extrcurvdef} and \eqref{mutens} respectively.
Further available building blocks are given by components of the bulk Riemann tensor evaluated at the defect, etc. All of these quantities transform as tensors under a diffeomorphism in $\mathcal{M}$, up to a local rotation in the normal bundle - see in particular eqs. (\ref{vardif}). 
We find therefore convenient to ask for symmetry under local rotations in the normal bundle. Then the defect action may be constructed as the integral of a scalar function over $\mathcal{D}$. Let us emphasize that the set of counterterms needed to make the functional integral finite should be diffeomorphism invariant as well.
Of course, we also assume the action to be invariant under reparametrizations of the defect.

In this section, we shall first follow the procedure employed in \cite{McAvity:1993ue}, which consists in defining many defect operators, as the response of the partition function to the variations of each defect geometric quantity. Later we comment on the relation with a quicker approach, in which the only operators involved are the stress-tensor and the variation of the partition function with respect to the embedding coordinate $X(\si)$.

The Ward identities that follow from diffeomorphism invariance are obtained by the standard procedure. Given a set of bulk local operators $O_i$, let us also define the abbreviation
\beq
\mc{X}= O_1(x_1,z_1)\dots O_n(x_n,z_n).
\eeq
The following equation holds:
\begin{multline}
-\de_\xi \braket{\mc{X}}-\int_\mathcal{M} \nabla_\m \xi_\n
\braket{T^{\m\n}\, \mc{X}}
+\int_\mathcal{D}\left\langle \left( \frac{1}{2} B^{ab} \de_\xi \ga_{ab} +\eta_\m{}^a \de_\xi e^\m_a
+\la_\mu{}^I \de_\xi n_I^\m \right.\right. \\ 
+\frac{1}{2}C^{ab}_I\de_\xi K_{ab}^I 
+  D_\mu\, \de_\xi X^\m +j^{aIJ}\de_\xi \m_{aIJ}+\dots \Big)
\mc{X} \Big\rangle=0.
\label{wfirststep}
\end{multline}
Here we made use of the standard definition of the stress-tensor:
\beq
T^{\m\n}=-\frac{2}{\sqrt{g}}\frac{\de S}{\de g_{\m\n}};
\label{stressdef}
\eeq
It is important that this definition is adopted only for points in the bulk. The operators $B^{ab},\ \eta_\m{}^a,\ \la_\m{}^I,\ C^{ab}_I,\ D_\m,\, j^{aIJ}$ are local in local theories, and are defined by \emph{minus} the variation of the action with respect to the relevant quantities, in analogy with \eqref{stressdef}. In particular, $D_\m$ is obtained by the variation with respect to $X^\m$, keeping fixed the intrinsic and extrinsic geometry of the defect. Finally, the dots stand for higher dimension geometric quantities, that we disregard for simplicity.
Let us note that eq. (\ref{wfirststep}) holds for other symmetries of the vacuum functional, 
provided the variations under diffeomorphisms $\delta_\xi g_{\mu\nu}, \delta_\xi \gamma_{ab},\dots$ are appropriately replaced; we will use this fact repeatedly below. 

Taking into account eqs. (\ref{vardif}),
the relation \eqref{wfirststep} simplifies to
\beq
-\de_\xi \braket{\mc{X}}-\int_\mathcal{M} \nabla_\m \xi_\n
\braket{T^{\m\n}\, \mc{X}}
+\int_\mathcal{D}\braket{ \left(\pa_a\xi^\m \eta_\m{}^a+n^\m_I\pa_\mu\xi^\nu \la_\nu{}^I
+\xi^\mu D_\mu +\dots \right)
\mc{X} }=0.
\label{diffwardint}
\eeq
In deriving this, we used the invariance of the path integral under local rotations, which implies the following Ward identity, valid in the absence of defect operators:
\beq
\braket{\left(\la^{\m[I}n^{J]}_\m+\frac{1}{2}C^{ab[I} K^{J]}_{ab}
+ \nabla_a j^{aIJ}\right) \mc{X}}=0;
\label{wardrot}
\eeq
this can be seen by plugging the variations in eq. (\ref{rots}) into the analogue of eq. (\ref{wfirststep}).

The components of $D_\m$ parallel to the defect are further related to the other local operators defined above. The link is provided by the Ward identities associated to reparametrizations of the defect $\si'^a=\si^a+\zeta^a$, under which
\beq
\de_\zeta X^\mu = -e^\m_a\zeta^a,
\label{reparx}
\eeq
which in turn induces variations of all other quantities which are summarized in eq. \eqref{varrep}.
In the absence of operators localized at the defect, the following identity is then easily obtained:
\begin{multline}
\braket{D_a\mc{X}}=\langle\biggr(\nabla^b B_{ab}-K_{ab}^I\eta_I{}^b+\nabla_b\eta_a{}^b+\Gamma_{ab}^\m
\eta_\m{}^b+\m_{aIJ}\la^{IJ}
-\partial_a n^\m_I \la_\m{}^I \\
\left.+ \frac{1}{2}\left(-\nabla_a K^I_{bc}+2\nabla_cK_{ab}^I+2K_{ab}^I\nabla_c\right)
C_A^{bc}+\left(2K_{bc}^I K^c{}_a{}^J-\frac{1}{2} R_{abIJ}+ R_{a[IJ]b}\right)j^{bIJ}\right)\mc{X}\rangle.
\label{wardpar}
\end{multline}

We can work out the consequences of Weyl invariance on correlation functions in the presence of a defect in the same way. A Weyl transformation acts on the metric as
\beq
\de_\si g_{\m\n}= 2\si(x)\, g_{\m\n}
\label{weyl}
\eeq
and induces the variations given in eq. (\ref{varweyl}). Therefore, the Ward identity for Weyl transformations reads
\begin{equation}
-\de_\si 
\braket{\mc{X}}  + \int_\mathcal{M}\si\braket{T^\m_\m\, \mc{X}} 
+\int_\mathcal{D}\si \braket{ \big(B^a_a-n^\m_I \la_\m{}^I+\frac{1}{2}K_{ab}^I C^{ab}_I\big) \mc{X}}
-\frac{1}{2}\int_\mathcal{D}\pa_\m \si\, n^\m_I \braket{C^{aI}_a \mc{X}} =0.
\label{weylwardint}
\end{equation}

As usual, the transformation law of correlation functions in a conformal field theory can be inferred from the ones above, due to the fact that conformal transformations are a subgroup of diffeo$\times$Weyl transformations, for manifolds which possess conformal killing vectors. The variation of the partition function under conformal transformations vanishes in the absence of the defect, but this is not true any more in our set-up. Let us consider a conformal killing vector $\hat\xi$, for which
\begin{equation}
\label{ckv}
\nabla_{(\mu} \hat\xi_{\nu)} = - \frac{\nabla_\rho \hat\xi^\rho}{d} g_{\mu\nu} \equiv -\hat\sigma g_{\mu\nu} ~.
\end{equation}

Effecting a diffeomorphism of parameter $\hat\xi^\mu$ and a compensating Weyl rescaling of parameter
$\hat\sigma$
corresponds to a conformal transformation that leaves the 
metric invariant. However, no Weyl transformation can undo the action of the diffeomorphism on the embedding function $X^\m$, hence the non vanishing variation of the partition function.
Let us decompose the diffeomorphism in his tangent and normal components:
\beq
\hat\xi^\m (X)=e^\m_a \hat\xi^a+n^\m_I \hat\xi^I.
\eeq 
A diffeomorphism tangent to the defect can be undone by a reparametrization: in the end, we are left with the 
variation induced by the components of the conformal killing vector which do not preserve the defect. 
In order to see this, one starts from the Ward identities associated to the composition of the diffeomorphism $\hat\xi$ - eq. \eqref{diffwardint} - with the compensating Weyl transformation - eq. \eqref{weylwardint}. Accordingly, the stress-tensor cancels out in the sum. Then, one eliminates $\hat\xi^a D_a$ by use of the Ward identity for reparametrizations, eq. (\ref{wardpar}). Finally, a straightforward but lengthy manipulation 
shows that all remaining terms involving the parallel components $\xi^a$ correspond to transverse rotations 
and can be reabsorbed using eq. (\ref{wardrot}). The resulting formula is\footnote{For instance, if $\mc{X}$ contains a scalar field $O$, then $\de_\xi O=-\xi^\m\pa_\m O$ and $\de_{\hat\si} O=-\hat\si \Delta O$.}
\begin{equation}
\begin{aligned}
-(\de_{\hat\xi}+\de_{\hat\si})\braket{\mc{X}}
& =-\int_\mathcal{D}\hat\xi^I \braket{D_I\,\mc{X}} + \int_\mathcal{D}  
\hat\xi_{I} K_{ab}^I\braket{ B^{ab} \mc{X}}
+ \int_\mathcal{D} \left(\hat\xi^J \Ga^\n_{IJ}+ e^{\n a}\nabla_a\hat\xi_{I}\right) \braket{\la_\n{}^I \mc{X}} 
\\ 
& +\int_D\left(-\nabla_b\hat\xi^I n^\m_I+\hat\xi_{I}K^I_{ab} e^{a\m}+\hat\xi^I \Ga^\m_{Ib}\right)
\braket{\eta_\m{}^b\mc{X}}
\\
& +\frac{1}{2}\int_\mathcal{D} \Bigg( -  \nabla_{a}\nabla_{b}
\hat\xi^I +K_{ac}{}^I K_{b}{}^{cJ} 
\hat\xi_{J}  + \hat\xi^J R_{Jab}{}^I \Bigg)\braket{C^{ab}_I \mc{X}} \\
& +\int_\mathcal{D}  \left( 2 K_{ab}^{[I} \nabla^b \hat\xi^{J]} 
-\frac{1}{2} \hat\xi^K R_{KaIJ}+\hat\xi^K R_{K[IJ]a} \right) \braket{j^{a IJ}\mc{X}}.
\end{aligned}
\label{wardconfkill}
\end{equation}

Eq. \eqref{wardconfkill} says that the role of a certain subset of defect operators contained in the OPE of the stress-tensor is to implement a conformal transformation which does not fix the defect.

From now on, we would like to focus on a flat defect in flat space, and recover explicitly the contact terms of the stress-tensor with the defect, via the unintegrated form of equations \eqref{diffwardint} and \eqref{weylwardint}. Before doing it, it is maybe worth stressing that eq. \eqref{weylwardint} contains already all the information about the trace of the stress-tensor in the case of scale but not conformal invariant theories. We can choose the usual cartesian coordinates $\m=(a,i)$, pick an adapted basis of normal vector fields ($I\to i$), and place the defect in $X^i=0$. The Ward identities for diffeomorphism invariance read
\begin{multline}
\braket{\partial_\m T^{\m\n}(x)\, \mc{X}} =
- \de_\mc{D}(x) \de^\n_i \braket{\big(D^i(x^a)-\pa_a\eta^{ia}(x^a)\big)\mc{X}} 
- \de_\mc{D}(x) \de^\n_a\,\braket{ \pa_b B^{ab}(x^a)\, \mc{X}} \\
+\pa_i \de_\mc{D}(x)\,\braket{\la^{\n i}(x^a) \, \mc{X}} + \textup{contact terms associated with $\mc{X}$},
\label{wardflat}
\end{multline}
where $\de_\mc{D}(x)$ is a delta function with support on the defect, in particular $\de_\mc{D}(x)=\de^q(x^j)$ in this case. Eq. \eqref{wardpar} reduces to
\beq
\braket{\left(D_a-\pa_b\eta_a{}^b\right)\mc{X}}=\braket{\pa_b B_a^b\, \mc{X}}.
\label{wardparflat}
\eeq
The physical meaning of these equations is transparent: they regulate the exchange of energy and momentum between the bulk and the defect. In particular, when the defect is decoupled both $D_a$ and $\eta_\m{}^a$ generically vanish: eq. \eqref{wardparflat} tells us that in these cases the defect stress-tensor is separately conserved, as expected. Notice also that $C_{ab}^I$ does not appear in eqs. \eqref{wardflat} and \eqref{wardparflat}. This is expected, since a coupling with the extrinsic curvatures is never required in order to reformulate the theory in a diffeomorphism invariant way. On the contrary, this coupling can be necessary to enforce Weyl invariance - as can be foreseen by comparing the third line of eq. \eqref{varweyl} with the transformation law of the derivative of a bulk primary operator. It is therefore not surprising that $C_{ab}^I$ does appear in the Ward identities associated with the trace of the stress-tensor in a CFT:
\begin{multline}
\braket{T^\m_\m(x)\,\mc{X}} =
-\de_\mc{D}(x) 
\braket{ \big( B^a_a(x^a)-\la_i{}^i(x^a)\big) \mc{X}}
-\frac{1}{2}\pa_i\de_\mc{D}(x) \braket{ C^{a\,i}_a(x^a)\, \mc{X}}  \\
+ \textup{contact terms associated with $\mc{X}$}.
\label{weylward}
\end{multline}
One can summarize \eqref{wardflat} and \eqref{weylward} by adding contact terms to the definition of the stress-tensor itself. In other words, let us define
\beq
T^{\m\n}_\textup{tot}=T^{\m\n}+\de_\mc{D}(x) 
\left(\de^\m_a\de^\n_b B^{ab}
-2\de^{(\m}_a \de^{\n)}_i \left(\la^{ai}+\frac{1}{2}\pa_b C^{ab\,i}\right)-
\de^{(\m}_i\de^{\n)}_j \la^{ij}\right)
+\frac{1}{2}\pa_i\de_\mc{D}(x) \de^\m_a\de^\n_b C^{ab\,i}.
\label{ttot}
\eeq
In terms of $T_\textup{tot}$, the Ward identities take a simpler form:
\beq
\pa_\m T_\textup{tot}^{\m a} =0, \qquad
\pa_\m T_\textup{tot}^{\m i} =- \de_\mc{D} (x) \textup{D}^i
+\pa_k \de_\mc{D}(x) \la^{[ik]}, \qquad
(T_{\textup{tot}})^\m_\m=0,
\label{newward}
\eeq
where we defined the \emph{displacement operator}:
\beq
\textup{D}_i = D_i-\pa_a \eta_i{}^a+\pa_a\la^a{}_i+\frac{1}{2}\pa_a\pa_bC^{ab}_i.
\label{dispdef}
\eeq
The form \eqref{newward} of the Ward identities - which clearly still has to be interpreted as an operatorial equation - make it manifest that one can construct globally conserved currents for the space-time symmetries preserved by the defect.\footnote{Eq. \eqref{wardrot} implies that the second addend in $\pa_\m T^{\m i}_\textup{tot}$ is a total derivative. This allows to define a conserved current for rotations in transverse directions, which in general contains a contribution from internal degrees of freedom of the defect.}

By using eqs. \eqref{vargenim} to \eqref{decvarg}, it is easy to see that $T_\textup{tot}$ and $\textup{D}_i$ are defined by the total variations with respect to the bulk metric and the embedding function respectively. This provides a compact way of writing the consequences of diffeomorphism and Weyl invariance; on the other hand, it obscures the complete set of possible contact terms of the stress-tensor, which from the CFT point of view correspond to the presence of specific singularities in its defect OPE.

Finally, the flat space version of eq. \eqref{wardconfkill} is the following:
\beq
(\de_{\xi_\textup{c}}+\de_{\hat\si})\braket{\mc{X}}
=\int_\mathcal{D}\hat{\xi}^i
 \braket{\textup{D}_i\,\mc{X}}  
.
\label{confkillflat}
\eeq
We see that the displacement operator completely encodes the effect of conformal transformations on correlation functions in the presence of a flat defect.
Furthermore, eq. \eqref{newward} fixes both the scale dimension $\D_D=p+1$ and the normalization of this operator. It follows that its Zamolodchikov norm $C_\textup{D}$, defined by
\beq
\braket{\textup{D}_i(x) \textup{D}_j(0)} = C_\textup{D}\,\frac{\de_{ij}}{(x^2)^{p+1}},
\label{displtwopoint}
\eeq
is a piece of CFT data.

\subsection{Constraints on CFT data.}
\label{constraints}

The integrated Ward identity \eqref{confkillflat} provides information on the coupling of the displacement to bulk operators, analogously to what happens for the appearance of the stress-tensor in the bulk OPE. Indeed, when $\mc{X}$ comprises only one operator, both sides in eq. \eqref{confkillflat} are free of cross-ratios, and only depend on the CFT data. This generates constraints, of which we consider three examples, namely the defect OPE of a scalar, vector, and 2-index tensor primaries.

The easiest way of proceeding is to lift eq. \eqref{confkillflat} to the projective light-cone. On the l.h.s., we just need to apply the appropriate generator of the Lorentz group in $d+2$ dimensions. On the r.h.s., the corresponding killing vector is contracted with the displacement, and one should make sense of the integration by defining the correct measure on the light-cone. The issue of integration has be settled in \cite{SimmonsDuffin:2012uy}, whose results we borrow. We are going to assume that the operator $\textup{D}_i$ is a primary, and check that the functional form after integration matches the left hand side.

Let us start from a scalar operator $O_{\D,0}(P)$. Under the Lorentz group, the change of its one-point function coincides with the coordinate transformation, so the generator is the same as the killing vector:
\beq
\mc{J}_{AI}=2\, P_{[A} \frac{\pa}{\pa P^{I]}}.
\eeq
We can then write eq. \eqref{confkillflat} as follows:
\beq
P_A \frac{\pa}{\pa P^I} \braket{O_{\D,0}(P)} = - \int D^p Q\, Q_A \braket{O_{\D,0}(P)\textup{D}_I(Q)}.
\label{killembedsca}
\eeq
We just need to plug in the form of the correlators involved, that is:
\begin{align}
\braket{O_{\D,0}(P_1)\textup{D}(P_2,Z_2)} &=b_{O\textup{D}} \frac{Z_2\wbullet P_1}{(P_1\wbullet P_1)^{(\D-p)/2}(-2P_1\gbullet P_2)^{(p+1)}}, 
\label{Di-scalar}\\
\braket{O_{\D,0}(P_1)}&=\frac{a_O}{(P_1\wbullet P_1)^{\D/2}},
\end{align}
The integrand has dimension $p$, so that the integral is well defined and can be computed along the lines of \cite{SimmonsDuffin:2012uy}. We obtain the following scaling relation: 
\beq
\D\, a_O = \left(\frac{\pi}{4}\right)^{\frac{p}{2}} \frac{\sqrt{\pi}}{\Gamma\left(\frac{1}{2}(p+1)\right)}\, b_{O\textup{D}}.
\label{AoBdo}
\eeq
This equation states that the coefficient of the identity and the one of the displacement in the defect OPE of a scalar operator are linearly related. It is important to notice that, since the displacement is not canonically normalized, the coefficient appearing in the defect OPE is $b_{O\textup{D}}/C_\textup{D}$. Analogous considerations apply in the rest of this subsection. Eq. \eqref{AoBdo} is the generalization of a result found by Cardy in the codimension one case \cite{Eisenriegler}, where the normal component of the stress-tensor plays the role of the displacement \cite{McAvity:1993ue,Gliozzi:2015qsa}. In that case, the group of transverse rotations is trivial and the displacement does not carry indices, but the result can be obtained by directly plugging $p=d-1$ in eq. \eqref{AoBdo}.

Odd spin primaries do not acquire a one-point function in parity invariant theories - unless they are pseudo-tensors, a case which we do not consider here. The left hand side of eq. \eqref{confkillflat} vanishes, while the right hand side for a spin one primary $O_{\D,1}(P,Z)$ can be computed using\footnote{The relation with the conventions of section $3$ is $b_{1100}=b_{O\textup{D}}^1, b_{0010}=b_{O\textup{D}}^2$}
\begin{multline}
\braket{O_{\D,1}(P_1,Z_1)\textup{D}(P_2,Z_2) }
=\frac{1}{(P_1\wbullet P_1)^{(\D-p-1)/2} (-2 P_1\gbullet P_2)^{p+1}} \\
\times \left(b_{O\textup{D}}^1\frac{ P_1\wbullet Z_2 (P_1\gbullet P_2 \,
P_1\wbullet Z_1 - P_1\wbullet P_1\, P_2\gbullet Z_1)}{ P_1\wbullet P_1 \, P_1\gbullet P_2}\right.\\
\left.+b_{O\textup{D}}^2\frac{ P_1\wbullet P_1\,
   Z_1\wbullet Z_2-P_1\wbullet Z_2\, P_1\wbullet Z_1}{P_1\wbullet P_1}\right).
\end{multline}
The r.h.s. of eq. \eqref{confkillflat} does not vanish:
\beq
 \int D^p Q\, Q_A \braket{ O_{\D,1}(P,Z)\textup{D}_I(Q)}
\neq 0 ,
\label{killembedvec}
\eeq
therefore we conclude that $b_{O\textup{D}}^1=b_{O\textup{D}}^2=0$, that is, spin one primaries do not couple with the displacement. This conclusion holds true in the codimension one case, in which only the structure proportional to $b_{O\textup{D}}^1$ survives, and still does not vanish upon integration.

Let us turn to the spin 2 case. The action of a conformal transformation on the one-point function of a tensor is lifted simply as a Lorentz transformation involving both the coordinate of the field insertion and the auxiliary vector, so that eq. \eqref{confkillflat} reads in this case
\beq
2 \left(P_{[A} \frac{\pa}{\pa P^{I]}}+Z_{[A} \frac{\pa}{\pa Z^{I]}}\right)\braket{O_{\D,2}(P,Z)}
= - \int D^p Q\, Q_A \braket{O_{\D,2}(P,Z)\textup{D}_I(Q) }.
\label{killembedten}
\eeq
The one and two-point functions involved are written in eqs. \eqref{onepointboh}, \eqref{eq:Twopointfunctiondisplastresstensor} respectively. Eq. \eqref{killembedten} translates into the following relations:\footnote{The relation with the conventions of section $3$ is $b_{1200}=b_{O\textup{D}}^1, b_{0110}=b_{O\textup{D}}^3, b_{1001}=b_{O\textup{D}}^2$. 
}
\begin{subequations}
\begin{align}
b_{O\textup{D}}^2 &= \frac{1}{p+1} \left(\frac{\D}{2} b_{O\textup{D}}^3-b_{O\textup{D}}^1\right), \\
b_{O\textup{D}}^3 &=  2^{p+2} \pi^{-\frac{p+1}{2}} \Gamma\! \left(\frac{p+3}{2}\right) a_O.
\end{align}
\label{BtdAt}
\end{subequations}
Let us add a few comments. If the codimension is one, the structures multiplying $b_{O\textup{D}}^2$ and $b_{O\textup{D}}^3$ vanish, and so does the integral of the remaining one: this allows in particular the displacement to couple with the stress-tensor, which is expected since the first is the defect limit of a component of the second. Comparing the relations \eqref{BtdAt} with the ones which follow from conservation \eqref{eq:conservation}, one sees that compatibility is ensured by $\D=d$. Finally, when $O_{\m\n}=T_{\m\n}$ is the stress-tensor, another linear combination of the parameters in its correlator with the displacement can be seen to be equal to $C_\textup{D}$, defined in eq. \eqref{displtwopoint}. Thus the correlator is completely specified in terms of the latter, and of the coefficient of the one-point function $a_T$. To see this, let us look back at eq. \eqref{wardflat}, and specifically let us choose the free index in a direction orthogonal to the defect. By comparison with eq. \eqref{dispdef} we see that the $\de$-function contribution on the defect is given by the displacement plus descendants. The Ward-identity fixes the scale dimension of all the operators involved, so the displacement is certainly orthogonal to $\pa_a \la^a{}_i$ and $\pa_a\pa_bC^{ab}_i$. We can therefore write the following equality:
\beq
\braket{\partial_\m T^{\m i}(x)\, \textup{D}^j(0)} =
- \de_\mc{D} \braket{\textup{D}^i(x^a)\textup{D}^j(0)}.
\label{paTD}
\eeq
This equation is easily lifted to the projective light-cone by use of the Todorov operator:
\beq
\frac{1}{d}\,\pa_{P_1} \cdot D_{Z_1} \braket{ T(P_1,Z_1)\textup{D}(P_2,Z_2) }
= -\frac{C_\textup{D}\, \de^q(P_1^J) \ Z_1\wbullet Z_2}{(-2P_1\gbullet P_2)^{p+1}} .
\label{paTDcone}
\eeq
Notice that, although on the Poincar\'e section
\beq
-2 P_1\gbullet P_2=\abs{(x_1-x_2)^a}^2+\abs{x_1^i}^2,
\eeq
eq. \eqref{paTDcone} is correct, thanks to the $\de$-function which kills the dependence on the transverse coordinates. Upon integration against a test function, eq. \eqref{paTDcone} provides the following relation:
\beq
b_{T\textup{D}}^2=\frac{1}{p}\left((d+q-2)\frac{b_{T\textup{D}}^3}{2}+q\frac{C_\textup{D}}{\O_{q-1}}\right),
\label{b2b3cd}
\eeq
$\O_{q-1}=2\pi^{q/2}/\Ga (q/2)$ being the volume of $S_{q-1}$. As promised, this relation, together with eqs. \eqref{BtdAt}, fixes the bulk-to-defect coupling of the displacement with the stress-tensor in terms of the norm of the displacement ($C_\textup{D}$) and the coefficient of the one-point function of the stress-tensor ($a_T$).

\subsection{Displacement and reflection in two dimensional CFTs}
\label{subsec:2d}

The story of two dimensional defect conformal field theories is rich, and dates back to the essential work of Cardy \cite{Cardy:1984bb,Cardy:1989ir} and Cardy and Lewellen \cite{Cardy:1991tv}. Here we would like to comment on the role of the displacement operator in this case. It is not difficult to see that $C_\textup{D}$ is intimately related with the \emph{reflection coefficient} introduced in \cite{Quella:2006de}. A straightforward analysis of the defect operators present in the defect OPE of the stress-tensor also allows to prove unitarity bounds for this coefficient. This analysis is greatly simplified by some easy consequences of holomorphy. We put a $\textup{CFT}_1$ on the upper half and a $\textup{CFT}_2$ on the lower half of the complex plane, parametrized by $z=x+\ii y$ . Since both components of the stress-tensor of either theory, $T(z)$ and $\bT(\bz)$, are purely (anti)holomorphic, their defect OPE cannot be singular. It follows at once by dimensional analysis that the only defect operator surviving among the ones appearing in eqs. \eqref{wardflat}, \eqref{weylward} is the displacement operator. This means in particular that the component $T_{xy}$ is continuous across the defect, or in other words
\beq
T^1(x)-\bT^1(x)=T^2(x)-\bT^2(x), \quad x\in \mathbb{R}.
\label{eq:2d:T_cond_interf}
\eeq
In fact, in any dimension an interface CFT can be mapped to a boundary CFT by folding the system. In $2d$, mapping the $\textup{CFT}_2$ on the upper half plane corresponds to exchanging holomorphic and anti-holomorphic fields, so that the interface is equivalent to a boundary condition for the theory $\textup{CFT}_1\times\overline{\textup{CFT}_2}$.\footnote{Consequences of the folding trick are in fact not completely understood - see \cite{Fuchs:2010hk} for some comments in the case of topological defects - especially in non-rational theories, which lack a nice factorization of holomorphic and anti-holomorphic parts. However, related subtleties are not relevant for the considerations below. We thank Jurgen Fuchs for correspondence on this point.} The condition \eqref{eq:2d:T_cond_interf} is simply Cardy's condition for the stress-tensor of the folded theory $T^1+\bT^2$.
Holomorphy and translational and scale invariance fix all correlators of $T^1$ and $T^2$ up to constants. The gluing condition \eqref{eq:2d:T_cond_interf} relates many of the constants:
\begin{align}
\braket{T^1(z)T^2(z')}=&\frac{a/2}{(z-z')^4} \label{eq:2d:T1T2_a}  \\
\braket{\bT^1(\bz)\bT^2(\bz')}=&\frac{(a+\bar{c}_1-c_1)/2}{(\bz-\bz')^4} = \frac{(a+\bar{c}_2-c_2)/2}{(\bz-\bz')^4}  \\
\braket{T^1(z)\bT^1(\bz')}=&\frac{(c_1-b)/2}{(z-\bz')^4}   \\
\braket{T^2(z)\bT^2(\bz')}=&\frac{(c_2-b)/2}{(z-\bz')^4}.   \\
\end{align}
This imposes the condition
\beq
c_1-\bar{c}_1=c_2-\bar{c}_2,
\eeq
but we will assume
\beq
c_1=\bar{c}_1,\quad c_2=\bar{c}_2
\eeq
in the following. Also, the parameters $a$ and $b$ are related by the remaining correlator:
\beq
\braket{T^1(z)\bT^2(\bz')}=\frac{(a-b)/2}{(z-\bz')^4}.
\eeq
Here the minus sign is dictated by translational invariance along the real direction. However, this correlator doesn't fall off when the fields are far away on the opposite sides of the interface, therefore the boundary condition in this direction prompts us to fix
\beq
a=b.
\eeq
The displacement operator is (up to a sign)
\beq
\textup{D}(x)=(T^1+\bT^1-T^2-\bT^2)(x)=2 (T^1-T^2)(x)=2 (\bT^1-\bT^2)(x).
\label{eq:2d:displacement_def}
\eeq
Since there are four operators of dimension $2$ and one gluing condition, there are other two dimension $2$ operators on the interface. One of them might be taken to be $T=T^1+\bT^2$, the boundary stress-tensor for the folded theory, which is also, up to a factor $2$, the displacement for the folded theory. Notice that, unless $c_1=c_2$, this operator is not orthogonal to $\textup{D}$. The coefficient of the displacement two-point function is
\beq
C_\textup{D}=2(c_1+c_2-2a) \geq 0,
\label{eq:2d:C_D_a}
\eeq
where the inequality holds in a unitary theory. We can actually obtain a stronger lower bound and an upper bound for $C_\textup{D}$ by considering the matrix of two-point functions of all the dimension $2$ fields at our disposal. We go to the folded picture and denote:
\beq
\tT^2(z)=\bT^2(\bz'),\quad \bz=z',\ \Im{z'}<0.
\eeq
Then one can verify that, with respect to the total stress tensor,
\beq
T(z)= T^1(z)+\tT^2(z),
\eeq
the following two fields are primaries\footnote{In this section primaries are intended to be Virasoro primaries, as customary in 2d CFT.} (in the notation of \cite{Quella:2006de})
\begin{align}
W(z) =& c_2 T^1(z)-c_1 \tT^2(z), \\
\bar{W}(\bz) =& c_2 \bT^1(\bz)-c_1 \tilde{\bT}^2(\bz).
\end{align}
Being (anti)holomorphic, these fields have non-singular defect OPEs, and the coefficients of the two point functions of the boundary operators compute the overlap of the corresponding states in radial quantization. The matrix of inner products should have positive eigenvalues in a unitary theory:
\beq G=\frac{1}{2}
\begin{pmatrix}
c_1+c_2                    &       0                                                  &       0      \\
0                                &     c_1c_2(c_1+c_2)                          &   (c_1+c_2)(c_1c_2-a(c_1+c_2))   \\
0                                &     (c_1+c_2)(c_1c_2-a(c_1+c_2))   &     c_1c_2(c_1+c_2)   
\end{pmatrix}.
\eeq
The eigenvalues of $G$ are
\beq
\la_1   = \frac{1}{2}(c_1+c_2),  \qquad
\la_2  = \frac{a}{2} (c_1+c_2)^2, \qquad
\la_3   = -\frac{1}{2}(a (c_1+c_2)-2c_1c_2).
\eeq
From positivity of $\la_2$ and $\la_3$ it follows that
\beq
2\,\frac{(c_1-c_2)^2}{c_1+c_2} \leq C_\textup{D} \leq 2\,(c_1+c_2).
\label{eq:2d:displacement_bounds}
\eeq
Notice that the upper bound is saturated by the case of a boundary condition, for which $a=0$ in eq. \eqref{eq:2d:T1T2_a}. 

As already mentioned, a general definition of reflection and transmission coefficients in $2d$ CFT was put forward in \cite{Quella:2006de}. It is easy to see that the reflection coefficient $\mc{R}$ can be expressed in terms of the coefficient of the two-point function of the displacement operator:
\beq
\mc{R}=\frac{C_\textup{D}}{2(c_1+c_2)}.
\eeq
We find therefore that, in a unitary theory, reflection is less than unity, as it should, and transparency is bounded by the square of the difference of the central charges:
\beq
\left(\frac{c_1-c_2}{c_1+c_2}\right)^2 \leq \mc{R} \leq 1.
\eeq

 \subsection{Examples}
\label{examples}

When dealing with strongly coupled CFTs, usually the conformal data are all that matters, and the explicit expression of renormalized operators in terms of elementary fields is inaccessible. On the contrary, when perturbation theory makes sense, it is useful to consider the free field composite operators as starting point. Here we give a few free theory examples of the defect operators which appear in eqs. \eqref{wfirststep}-\eqref{dispdef} above. Dealing with free theories, we never need to worry about running couplings, so that conformal invariance of the defects that we consider simply follows from dimensional analysis. Most of the examples have already been considered elsewhere. In what follows, we will sometimes employ the notation $\sigma=\{\sigma^a\}$, along with the usual $x_\parallel = \{x^a\}$ for the parallel coordinates and $x_\perp = \{x^i\}$ for the orthogonal ones.

\paragraph{Minimal coupling to a free scalar}

Let us start by considering a single free scalar field $\phi$ in a flat $d$ dimensional Euclidean space. 
Rather than the usual QFT normalization of the field $\phi$ we will use the CFT one,\footnote{In the CFT normalization, the free action is
\begin{equation}
\label{freeacCFT}
\frac{1}{(d-2)\Omega_{d-1}}\int_{\mathcal{M}}\! d^dx\, \frac{1}{2}\, \partial_\mu \phi\, \partial^\mu\phi~,
\end{equation}
where $\Omega_{d-1}= 2\pi^{d/2}/\Gamma(d/2)$ is the volume of $S_{d-1}$.} in which the two-point
function is
\begin{equation}
\label{fsf2p}
G_0(x,y) \equiv \vev{\phi(x)\phi(y)}_0 = \frac{1}{|x-y|^{2\Delta}}~. 
\end{equation}

We place in the vacuum the following planar $p$-dimensional defect: 
\begin{equation}
\label{fsextop}
\mathcal{O}_\mc{D}= \exp\left(\lambda\int_{\mc{D}} 
d^p\sigma\phi(\sigma)\right)~.
\end{equation}
Notice that $\mc{O}_\mc{D}$ is the extended operator itself, according to the notation of eq. \eqref{genericcorr}. In what follows, we sometimes alternatively use the defect action $S_\mc{D}=-\log \mc{O}_\mc{D}$.
The defect is conformal when $p$ equals the dimension $\Delta$ of the scalar:
\begin{equation}
\label{fspq}
p = \Delta = \frac{d}{2} - 1~,~~~
q\equiv d - p = \frac{d}{2} +1~;
\end{equation}
the scalar Wilson line in four dimensions is one example \cite{Kapustin:2005py}.

The equation of motion 
reads
\begin{equation}
\label{fseom}
\Box \phi =- \lambda \,(d-2)\Omega_{d-1}\,\delta_\mc{D}~.
\end{equation}

In the presence of the defect, the one-point function of $\p$ can be computed in various ways. For instance, one can solve the equation of motion \eqref{fseom}, or use the properties of free exponentials to get 
\begin{equation}
\label{fsopd}
\vev{\phi(x)} = \frac{\vev{\phi(x)\mathcal{O}_{\mathcal{D}}}_0}{\vev{\mathcal{O}_{\mathcal{D}}}_0}
= \lambda \int d^p \sigma\, 
\vev{\phi(x)\phi(\sigma)}_0.
\end{equation}
Either way,
\begin{equation}
\label{fssol}
\vev{\phi(x)} = 
\frac{a_\phi}{|x_\perp|^\Delta}~,~~~
\mbox{with}~~~
a_\phi = 2 \lambda\frac{\Omega_{d-1}}{\Omega_{q-1}}~.
\end{equation}
For completeness, let us also write down the two-point function:
\begin{equation}
\label{fs2p}
\begin{aligned}
\vev{\phi(x)\phi(y)} & = 
\frac{\vev{\phi(x)\phi(x)\mathcal{O}_{\mathcal{D}}}_0}{\vev{\mathcal{O}_{\mathcal{D}}}_0}
= \vev{\phi(x)\phi(x)}_0 + \vev{\phi(x)} \vev{\phi(y)}\\
& = \frac{1}{|x-y|^{2\Delta}} + 
\frac{a_\phi^2}{|x_\perp|^\Delta |y_\perp|^\Delta}
= \frac{1}{|x_\perp|^\Delta |y_\perp|^\Delta }
\left(\frac{1}{\xi^\Delta} + a_\phi^2\right)~.
\end{aligned}
\end{equation}
The form of this correlator means in particular that the defect OPE of the fundamental field is only mildly deformed from the case of a trivial defect: the identity appears, but the other couplings remain untouched.

The (improved) stress-tensor for the free scalar field $\phi$ reads
\begin{equation}
\label{fsTmunu}
T_{\mu\nu} = 
\frac{1}{(d-2)\Omega_{d-1}}
\left(
\partial_\mu\phi\,\partial_\nu\phi - \frac 12 \delta_{\mu\nu} \partial\phi\cdot\partial\phi -
\frac 14\frac{d-2}{d-1}\left(\partial_\mu\partial_\nu - \delta_{\mu\nu}\square\right)\phi^2
\right)~.
\end{equation}

Contact terms in the Ward identities of the stress-tensor follow from the definition (\ref{fsTmunu}):
\begin{equation}
\pa_\m T^{\m\n} = \frac{1}{(d-2)\Omega_{d-1}} \pa^\n\p\, \Box \p,
 \qquad T_\mu^{~\mu} = \frac{1}{(d-2)\Omega_{d-1}}\frac{d-2}{2}\,\phi\square\phi~,
\end{equation} 
and from the use of the e.o.m. \eqref{fseom}:
\begin{equation}
\label{Tmumudope}
\braket{\pa_\m T^{\m\n}\mc{X}}=-\la \braket{\pa^\n\p\, \mc{X}}\,\delta_\mc{D}(x),
\qquad \vev{T_\mu^{~\mu}\mc{X}} =-\lambda
\,\frac{d-2}{2}\,\vev{\phi\, \mc{X}}\,\delta_\mc{D}(x)~.
\end{equation}  
This is in agreement with the Ward identities \eqref{wardflat}, \eqref{weylward}. Indeed, the defect in curved space-time becomes
\begin{equation}
\label{fsdefcurved}
O_\mathcal{D} = \lambda \int_\mathcal{D} d^p\sigma \sqrt{\gamma}\, \phi\left(X(\sigma)\right)~,
\end{equation}
and
\begin{equation}
\label{fsBab}
B^{ab} \equiv \left. \frac{2}{\sqrt{\gamma}} \frac{\delta O_\mathcal{D}}{\delta\gamma_{ab}}
\right|_{\textup{flat}} = 
\lambda\, \phi\,\de_{ab}~,
\end{equation}
We also have, in the flat space limit,
\begin{equation}
\label{fsDmuflat}
D_a = \lambda\,\partial_a\phi~,~~~
\textup{D}_i=D_i = \lambda\,\partial_i\phi~. 
\end{equation}
Eq.s (\ref{fsBab}) and (\ref{fsDmuflat}) are consistent with the Ward identity for reparametrization symmetry, eq. (\ref{wardparflat}). 
The norm $C_\textup{D}$ of the displacement follows from the two-point function%
\begin{equation}
\label{fsCD}
\vev{\mathrm{D}_i(x_\parallel) \mathrm{D}_j(y_\parallel)} = \lambda^2
\left.\frac{\partial~}{\partial x^i }\frac{\partial~}{\partial y^j }\vev{\phi(x)\phi(y)}\right|_{x_\perp=y_\perp=0}, \qquad
C_{\mathrm{D}} = 2\lambda^2 \Delta.
\end{equation}

The coupling of $\phi$ to the displacement is also 
easily computed:
\begin{equation}
\label{fsBD}
\vev{\mathrm{D}_i(y_\parallel) \phi(x)} = \lambda\left.\frac{\partial~}{\partial y^i } \vev{\phi(y)\phi(x)}\right|_{y_\perp=0}
= 2\Delta\lambda \frac{x_i}{((x-y)_\parallel^2 + x_\perp^2)^{\Delta+1}}~,
\end{equation}
and since 
\begin{equation}
\label{fsBphiD}
b_{\phi \mathrm{D}} = 2\Delta\lambda
= \Delta \left(\frac{\pi}{4}\right)^{-\frac{p}{2}}\frac{\Gamma\left(\frac{p+1}{2}\right)}{\sqrt{\pi}}\, a_\phi~, 
\end{equation}
this explicitly verifies the general relation (\ref{AoBdo}) between the coefficient of the identity and the one of the displacement in the defect OPE of $\phi$. 

The one-point function of the stress tensor in presence of the defect can be computed by diagrammatic methods or by evaluating $T_{\mu\nu}$ on the classical solution, i.e., on the one-point function eq. \eqref{fssol}. The result agrees with the expected form eq. \eqref{tensonept} with
\begin{equation}
\label{fsaT}
a_T =- \frac{(d-2)}{4(d-1)} \frac{a_\phi^2}{\Omega_{d-1}}~. 
\end{equation}
Similarly, a direct computation shows that the coupling between the displacement operator and the stress-tensor is of the form \eqref{bulkdefc}, with $\D=d$, $\wh{\D}=p+1$ and with the following defect OPE coefficients:
\begin{gather}
b^1_{\textup{D}T}=- \lambda ^2\,\frac{d (d-2)}{2
   (d-1)}\, \pi
   ^{-\frac{d+2}{4}}
    \Gamma \left(\frac{d+6}{4}\right), \quad 
b^2_{\textup{D}T}= -\lambda ^2\, \frac{(d-2)^3}{16
   (d-1)}\,
   \pi
   ^{-\frac{d+2}{4}}
    \Gamma
   \left(\frac{d-2}{4}\right), \notag \\
b^3_{\textup{D}T}= -  \lambda ^2 \frac{d(d-2)}{2
   (d-1)}\,
    \pi
   ^{-\frac{d+2}{4}}\,
   \Gamma
   \left(\frac{d+2}{4}\right).
   \label{TDscalar}
\end{gather}
Equations \eqref{fsCD} \eqref{fsaT} \eqref{TDscalar} verify the constraints \eqref{BtdAt} and \eqref{b2b3cd}.


\paragraph{Minimal coupling to a $p$-form} Let us consider an (Abelian) $p$-form
$A$ minimally coupled to a $p$-dimensional object $\mathcal{D}$. The action reads
\begin{equation}
\label{pfaction}
\frac{1}{2}\int_{\mathcal{M}} F\wedge {}^*F - \lambda \int_\mathcal{D} A =
\frac{1}{2(p+1)!}\int_{\mathcal{M}} d^d x\, F_{\mu_1\ldots \mu_{p+1}} 
F^{\mu_1\ldots \mu_{p+1}}\, - \lambda \int_\mathcal{D} d^px_\parallel A_{1\ldots p}~, 
\end{equation}
where $F = dA$. This system is scale invariant when
\begin{equation}
\label{pfp}
p = \frac{d}{2}-1~.
\end{equation}
In fact, there is full conformal invariance, as can be checked from the trace of the stress-tensor eq. \eqref{pfTmunu}. This is a simple generalization of a charged particle in four dimensions, which corresponds to the case $p=1$ in $d=4$.
The equations of motion read
\begin{equation}
\label{pfeom}
\partial^\mu F_{\mu\, 1 \ldots p} = -\lambda\, \delta_\mc{D}~,  
\end{equation} 
while $\partial^\mu F_{_\mu\nu_1\ldots \nu_p}=0$ for all other sets of indices 
$\nu_1,\ldots \nu_p$.
The bulk stress-energy tensor for the $p$-form field is given by 
\begin{equation}
\label{pfTmunu}
T_{\mu\nu} = \frac{1}{p!} 
\left(F_{\mu\rho_1\ldots \rho_p} F_\nu^{~\rho_1\ldots\rho_p} - \de_{\mu\nu}
\frac{1}{2(p+1)} F_{\rho_1\ldots\rho_{p+1}}F^{\rho_1\ldots\rho_{p+1}}
\right)~.
\end{equation} 
Direct computation of the one-point function yields the expected form eq. \eqref{tensonept}, with
\begin{equation}
\label{fsaTvector}
a_T =- \left(\frac{\lambda}{\Omega_{q-1}}\right)^2~.
\end{equation}
The trace of the stress tensor vanishes identically, therefore the only contact terms appear in its derivative:
\begin{equation}
\label{pfTmumu}
\partial_\mu T^{\mu i}= -\lambda \delta_\mc{D}\, F_{i\,1\ldots p}~,\qquad T_\mu^{~\mu}(x) = 0~.
\end{equation}

The defect action in eq. (\ref{pfaction}) is directly generalized to a curved case, since it is written as the integral of a form:
\begin{equation}
\label{pfSdef}
S_\mathcal{D} = -\lambda \int_\mc{D} A = -\frac{\lambda}{p!} \int_\mc{D} A_{\mu_1\ldots\mu_p}
e^{\mu_1}_{a_1}\ldots e^{\mu_p}_{a_p} d\sigma^{a_1}\wedge \ldots d\sigma^{a_p}~.
\end{equation} 
Since the induced metric $\gamma_{ab}$, the normal vectors $n^\mu_I$ and the extrinsic curvature $K^I_{ab}$ do not appear, the operator $B_{ab}$, $\lambda_\mu{}^I$ and $C_I^{ab}$ vanish, so that \eqref{pfTmumu} is consistent with the Ward identity for Weyl rescalings, eq. (\ref{weylward}).

We have instead non-trivial operators
\begin{equation}
\label{pfetaamu}
\eta_\mu{}^a = -\frac{\delta S_\mc{D}}{\delta e^\mu_a} = \frac{\lambda}{(p-1)!} A_{\mu a_2 \ldots a_p} \epsilon^{a a_2\ldots a_p}
\end{equation}
and
\begin{equation}
\label{pfDmu}
D_\mu = -\frac{\delta S_\mc{D}}{\delta X^\mu} = \frac{\lambda}{p!} \partial_\mu A_{a_1\ldots a_p} \epsilon^{a_1\ldots a_p}~.
\end{equation}
These are not primary operators, since they are not gauge invariant. The gauge invariant combination is the one appearing in the Ward identities \eqref{wardflat}, which coincides with the displacement operator, i.e.
\begin{equation}
\label{pfdisp}
\mathrm{D}_i = D_i - \partial_a \eta_i{}^a = 
\lambda F_{i0\ldots p-1}~.  
\end{equation}
This is in agreement with eq. \eqref{pfTmumu}.


\paragraph{Vector coupled with lower dimensional matter.}

We now go back to a free vector field in a generic dimension $d$, but this time we add degrees of freedom on a $p$ dimensional subspace. If we choose these additional fields to possess a global $SO(q)_I$ symmetry, we can couple the two theories by a symmetry breaking term that only preserves the diagonal of $SO(q)_I \times SO(q)$, the latter being the usual transverse rotational symmetry. It is convenient to start from the Stuckelberg Lagrangian, so that we have enough fields at disposal to make the defect coupling gauge invariant. The redundant description contains a vector $A_\m$ and a scalar $B$, and a symmetry is imposed under the following local transformations:
\beq
A_\m \to A_\m + \pa_\m \La, \qquad B \to B+m \La.
\eeq
We couple the usual action with a defect in the following way:
\beq
S= \int_\mathcal{M} d^d x \left( \frac{1}{4} F_{\m\n}^2 + 
\frac{m^2}{2}\bigg(A_\m-\frac{1}{m}\pa_\m B\bigg)^{\!2} \right)
+\frac{\la}{2} \int_\mathcal{D} d^p x \left(
(\pa_a \p^i)^2+2 \m \left(A_i-\frac{1}{m}\pa_i B\right) \p^i\right).
\label{sstuck}
\eeq
While at low energy the bulk theory contains a massive vector, at high energy the longitudinal degree of freedom decouples. We will come back later to the massless limit. 
The equations of motion are
\begin{align}
\pa_\m F^{\m\n}- m^2 (A^\n-\pa^\n B/m)&= \de^\n_i\, \delta_\mc{D}(x)\, \la\m\, \p^i, \\
m^2 \pa^\m(A_\m-\pa_\m B/m) &= -\pa_i \delta_\mc{D}(x)\,\la\m\, \p^i, \label{Beom}\\
 \m (A^i-\pa^i B/m)  &= \Box \p^i.
\end{align}
Notice that the e.o.m. are compatible with the antisymmetry of the field strength, in the sense that $\pa_\m\pa_\n F^{\m\n} =0$.
The stress-tensor reads
\beq
T_{\m\n} = F_{\m\r} F_\n{}^\r
+ m^2 \left(A_\m-\pa_\m B/m\right) \left(A_\n-\pa_\n B/m\right)
-\de_{\m\n} \left(\frac{1}{4} F_{\r\si}^2+ 
\frac{m^2}{2}\left(A_\m-\pa_\m B/m\right)^{\!2}  \right).
\eeq
Let us consider its divergence:
\beq
\pa^\m T_{\m\n} = \delta_\mc{D}\,  \la\m \p^i F_{\n i}-\pa_i \delta_\mc{D}\,\la\m \p^i\left(A_\n-\pa_\n B/m\right).
\label{warddiffvecstep}
\eeq
It is important to notice that the fields appearing in the contact terms preserve their full space-time dependence. In other words, upon integration by parts the derivative acting on the delta function in the second addend ends up acting on $A$ and $B$. On the contrary, the Ward identities \eqref{wardflat} were written assuming that defect operators only depend on the parallel coordinates. With this in mind, we can rewrite eq. \eqref{warddiffvecstep} as follows:
\beq
\pa^\m T_{\m\n} = \delta_\mc{D}\,\la\m \p^i \pa_\n (A_i-\pa_i B/m)-\pa_i \delta_\mc{D}\, \la\m \p^i\left(A_\n-\pa_\n B/m\right),
\label{warddiffvec}
\eeq
where now all fields and their derivatives are evaluated at the defect and do not carry dependence on transverse coordinates any more.

When coupling the theory to a background metric, we also need to upgrade the $SO(q)_I$ symmetry to a local one, so that the bulk-to-boundary mixing term couples the rotations in the normal bundle to the internal symmetry. For the same reason, the background gauge field for the internal symmetry has to coincide with the spin connection $\m_{a\,IJ}$ as defined in \eqref{mutens}.
Altogether, the defect action in curved space time - without any improvement, since we cannot achieve conformal invariance in the bulk anyway - is
\beq
S_\mathcal{D}= \frac{\la}{2} \int_\mathcal{D} \left(
\left(D_a \p^I\right)^2+2 \m \left(A_\m-\pa_\m B/m \right)n^\m_I\p^I\right), \qquad 
D_a \p^I= \pa_a\p^I+\m_{a}{}^I{}_J \p^J.
\label{Sveccurved}
\eeq

Let us identify the defect operators appearing in the contact terms coming from the action \eqref{Sveccurved}:
\begin{align}
B_{ab} &= \frac{\la}{2} \left(2\pa_a \p^i\pa_b\p^i
-\left((\pa_c\p^i)^2+2\m \left(A_i-\pa_i B/m\right)\p^i\right)\de_{ab}\right), \\
D_\m &= -\la\m \p^i\,\pa_\m\! \left(A_i-\pa_i B/m\right), \\
\la_\m{}^i &=- \la \m \left(A_\m-\pa_\m B/m\right) \p^i,\\
j_a{}^{ij}&= -\la \pa_a \p^{[i} \p^{j]}.
\end{align}
On shell, one finds in particular
\begin{align}
\pa^a B_{ab} &= \la \Box \p^i \pa_b \p^i
-\la\m \pa_b\left(\left(A_i-\pa_i B/m\right)\p^i\right) = D_b, \\
\pa^a j_a{}^{ij} &= -\la\m \left(A^{[i}-\pa^{[i}B/m\right)\p^{j]} = \la^{[ij]},
\end{align}
in agreement with eqs. \eqref{wardparflat} and \eqref{wardrot}.
Plugging this into \eqref{wardflat}, we find perfect agreement with eq. \eqref{warddiffvec}.

Let us now look for a fixed point of the action \eqref{sstuck}. In order to have a non singular massless limit on the defect, $\m$ needs to vanish with $m$. This means that the transverse degrees of freedom of the photon decouple from the lower dimensional matter, and we end up with the coupling of a bulk with a defect scalar field:
\beq
S_\textup{massless}= \int_\mathcal{M} d^d x 
\,\frac{1}{2}\left(\pa_\m B\right)^2
+\frac{\la}{2} \int_\mathcal{D} d^p x \left(
(\pa_a \p^i)^2-2 \kappa \pa_i B \p^i\right), \quad \kappa=\frac{\m}{m}.
\label{sstuckmless}
\eeq
All couplings are dimensionless for a codimension two defect. In this case one can achieve Weyl invariance by improving both the bulk and the defect action with the usual coupling to the Ricci scalar, and furthermore by supplementing the defect action with a linear coupling to the extrinsic curvatures:
\beq
S_\mathcal{D}= \frac{\la}{2} \int_\mathcal{D} d^p x \sqrt{\ga} \left(
\big(D_a \p^I\big)^2+\frac{p-2}{4(p-1)}\wh{R}\,(\p^I)^2-2 \kappa \pa_I B \p^I+\kappa B \p_I K^I\right).
\eeq
By means of the massless limit of eq. \eqref{Beom}, and repeating the step which led us from \eqref{warddiffvecstep} to \eqref{warddiffvec}, we obtain the trace of the stress-tensor in flat space:
\beq
T_\m{}^\m=\frac{d-2}{2}\la \kappa \left(-\delta_\mc{D}\, \pa_i B \p^i
+\pa_i \delta_\mc{D}\, B\p^i \right).
\eeq
Comparing with the operators $B_{ab},\,\la_\m{}^I$ and $C^I_{ab}$ extracted from $S_\mathcal{D}$, we find that these are the right contact terms for a conformal defect when $p=d-2$, as expected.


\section{Conclusions and outlook}

In this work, we have used the embedding formalism to analyze correlation functions in a generic CFT endowed with a conformal defect. The tensor structures appearing in a two-point function of symmetric traceless (bulk or defect) primaries have been classified. We also started the exploration of the crossing constraints for defects of generic codimension, by deriving the conformal blocks for the scalar two-point function in the defect channel and setting up the light-cone expansion for the bulk channel. Moreover, for codimension two defects, we found that the bulk channel blocks for identical external scalars are equal to the blocks of the four-point function of a homogeneous CFT (i.e. without defects), up to a change of variables. This means in particular that the bulk blocks are now known in closed form for codimension two defects in even dimensional CFTs, and of course all the results available in odd dimensions apply as well (see \cite{Rychkov:2015lca} for a recent progress in $d=3$.).
Finally, we described the possible protected defect operators appearing in the OPE of the stress-tensor with the defect, and derived constraints on the CFT data starting from the appropriate Ward identities. In the two dimensional case, we pointed out that the Zamolodchikov norm of the displacement operator contains the same information as the reflection coefficient defined in \cite{Quella:2006de}, and we derived unitarity bounds for the latter.

These results can be extended in multiple directions. It should be possible to find an algorithmic way of constructing correlation functions involving an arbitrary number of primaries. Also, the number of independent structures depend on the dimension of space and of the defect, so one may inquire the case of low dimensional CFTs. We did not consider mixed symmetry tensors or spinors. The latter might be done generalizing the results contained in \cite{Costa:2014rya} and \cite{Weinberg:2010fx} respectively. Much in the same way, we expect that many of the techniques that have been developed for the four-point function could be bent to the purpose of computing conformal blocks for a scalar two-point function in the presence of a defect. This is work in progress \cite{defblocks}. It is also worth mentioning that the simplest bootstrap equation, the one we considered, does not exhaust the constraints coming from crossing symmetry. Indeed, it is not difficult to realize that all correlation functions automatically obey crossing only when the three-point functions $\braket{O_1 O_2 \wh{O}}$ do.

The results presented here are sufficient to set up a bootstrap analysis of codimension two defects. Upon completion of some of the formal developments outlined above, it will be possible to do the same for generic flat extended operators. Two-point functions of bulk operators can be bootstrapped with the method of determinants \cite{Gliozzi:2013ysa}, while the linear functional method \cite{Rattazzi:2008pe} meets the obstruction of lack of positivity in the bulk channel.\footnote{See for instance the introduction of \cite{Rychkov:2015lca} for a comparison between the two techniques.} On the contrary, four-point functions of defect operators are clearly amenable to the latter technique \cite{Gaiotto:2013nva}. Finally, it is also possible to study the bootstrap equations analytically in the presence of defects, and obtain asymptotic information about the spectrum of defect primaries \cite{analyticdef}.

\section*{Acknowledgements}
The authors would like to thank J. Fuchs, D. Gaiotto, F. Gliozzi, J. Penedones and E. Trevisani for discussions and comments on the draft.
The hospitality offered by the Perimeter Institute for Theoretical Physics, where part of the project was carried out, is gratefully acknowledged.
Research at Perimeter Institute
is supported by the Government of Canada through
Industry Canada and by the Province of Ontario through
the Ministry of Research \& Innovation.
M.B. thanks M. Frau, A. Lerda, C. Maccaferri and I. Pesando for discussions. The work of M.B. is partially supported  by the Compagnia di San Paolo contract ``MAST: Modern Applications of String Theory'' TO-Call3-2012-0088.
E.L. is grateful to A. Bernamonti, F. Galli, A. Gnecchi, G. Tartaglino-Mazzucchelli and in particular to N. Bobev and A. Van Proeyen for the support and many illuminating discussions and suggestions. The work of E.L. is in part supported by COST Action MP1210 The String Theory Universe, the Interuniversity Attraction Poles Programme
initiated by the Belgian Science Policy (P7/37), the European Research Council grant no. ERC-2013-CoG 616732 HoloQosmos and the  National  Science Foundation  of Belgium (FWO) Odysseus grants G.001.12 and G.0.E52.14N.
V.G. thanks M. Costa and J. Penedones for discussions. V.G. would also like to thank FAPESP grant 2015/14796-7 and CERN/FIS-NUC/0045/2015. Centro de Fisica do Porto is partially
funded by the Foundation for Science and Technology of Portugal (FCT). 
M.M. would like to thank L. Bianchi, R. Myers and M. Smolkin for discussions on related topics.

\appendix

\section{Notations and conventions}\label{notations}

$\mathcal{M}$ is the $d$-dimensional space-time with coordinates $x^\mu$, $\mu=1,...,d$. $\mc{M}=\mathbb{R}^d$ throughout the paper, except for section \ref{sec:ward}. The coordinates in the embedding space are $P^M$, with $M=(+,-,\mu)$, except in subsection \ref{subsec:spherical}, where we use Cartesian coordinates $M=(0,\mu,d+1)$, with 
$P^\pm=P^0\pm P^{d+1}$.

The defect is a $p$ dimensional sub manifold $\mc{D}$ with coordinates $\sigma^a$, $a=1,...,p$, embedded into $\mathcal{M}$ by $x^\mu=X^\mu(\sigma)$. We denote the codimension with $q$: $p+q=d$. $\mc{D}$ is a plane or a sphere throughout the paper, except in section \ref{sec:ward}. For a flat defect in $\mathbb{R}^d$, we use an adapted system of coordinates $x^\mu= (x^a, x^i)$, where the defect $\mc{D}$ lies at $x^i=0$. In the embedding light-cone, the defect is a $(p+2)$-dimensional sub-manifold, and we use an adapted system of coordinates $P^M=(P^A,P^I)$, where the defect lies at $P^I=0$. In subsection \ref{subsec:spherical} and in section \ref{sec:ward}, we also denote by $I,J=1,...,q$ an index in the normal bundle to $\mc{D}$.  

The stability group of the defect is $SO(p+1,1)\times SO(q)$, where $q$ is the codimension. Bulk operators considered in this paper are rank-$J$ symmetric, traceless, tensors of $SO(d)$. Defect operators are rank-$j$ symmetric, traceless, tensors of $SO(p)$ and rank-$s$ symmetric, traceless, tensors of their flavour symmetry group $SO(q)$. We denote them as follows: 
\begin{equation}
O^{\mu_1,...\mu_J}_{\Delta}(x^\m)\quad\mbox{and}\quad
\widehat{O}^{a_1,...a_j; i_1,...i_s}_{\widehat{\Delta}}(x^a).
\end{equation}
When several operators are present, we distinguish them with a further label $n$ (and not $i$).
Operators uplifted to the embedding space are encoded in homogeneus polynomials of an auxiliary variable $Z$ (or $Z$ and $W$ for defect operators), so we use notations like
\begin{equation}
O_{\Delta,J}(P,Z)\quad\mbox{and}\quad \widehat{O}_{\widehat{\Delta},j,s}(P,Z,W)
\end{equation}
 In certain cases, like with the stress tensor, the labels $\Delta,J$ (or $\widehat{\Delta},j,s$) are redundant and we omit them.

We denote the conformal blocks as follows:
\begin{equation}
\begin{array}{ll}
\displaystyle{f_{\Delta,J}} &\displaystyle{\mbox{two-point function, bulk channel,}}\\
\displaystyle{\widehat{f}_{\widehat{\Delta},j,s}}&\displaystyle{\mbox{two-point function, defect channel,}}\\
\displaystyle{f_{\Delta,J}^{4pt}}&\displaystyle{\mbox{four-point function without defect}}.
\end{array}
\end{equation}

\section{OPE channels of scalar primaries and the conformal blocks}
\label{scalapp}
This appendix collects some facts related to bulk scalar primaries. In particular, we describe the most general defect OPE of a free scalar in \ref{subsub:freesca}, and we give a formula for the scalar block in the bulk-channel, eq. \eqref{scalar_res_real}. All other results are not new.

\subsection{Bulk and defect OPEs}

The regular OPE in the bulk can be expressed as
\begin{equation}\label{OPEI}
{O}_{1}(x_1){O}_{2}(x_2)=\sum_{k} \frac{c_{12k}}{(x_{12}^2)^{\frac{1}{2}(\Delta_1+\Delta_2-\Delta_k+J)}}{\mathcal{C}}^{(J)}(x_{12},\partial_2)_{\mu_1...\mu_J}{O}^{\mu_1...\mu_J}_{k}(x_2),
\end{equation}
where $c_{12k}$ are the structure constants appearing in the three-points functions in the absence of a defect and ${\mathcal{C}}^{(J)}$ are differential operators fixed by $SO(d+1,1)$ symmetry \cite{Ferrara:1973yt}. For the exchange of a scalar, one can derive the following expression \cite{Osborn:1993cr}:
\begin{equation}\label{Fe_Gr_Ga_diff_op}
\begin{array}{ll}
\displaystyle{\mathcal{C}^{(0)}(s,\partial_2)}&\displaystyle{ =\frac{1}{B(a,b)}\int_0^1 \mathrm{d}\alpha\mbox{ } \alpha^{a-1}(1-\alpha)^{b-1}e^{\alpha s\cdot \partial_2}}\\
&\displaystyle{\times\sum_{m=0}\frac{1}{m!}\frac{1}{(\Delta_k+1-d/2)_m}[-\frac{1}{4}s^2 \alpha (1-\alpha)\partial_2^2]^m },
\end{array}
\end{equation}
where $(k)_m=\Gamma(k+m)/\Gamma(k)$, $B(a,b)$ is the Euler Beta function, $a=(\Delta_k+\Delta_{12})/2$, $b=(\Delta_k-\Delta_{12})/2$ and we assume $[s,\partial]=0$.
\vspace{10pt}

Similarly, the defect OPE can be written as follows:\footnote{With respects to eq. \eqref{eq:Structures2PtBulkdefect} we have $b_{s,0,0,0}= b_{O\widehat{O}}$ .}
\begin{equation}\label{compBOPE}
{O}(x)=\sum_{\widehat{O}}\frac{b_{O\widehat{O}}}{|x^i|^{\Delta-\widehat{\Delta}}}\, \mathcal{C}_{\widehat{{O}}}\!\left(|x^i|^2\pa_\parallel^2\right)\frac{x_{i_1}...x_{i_s}}{|x^i|^s}\,\widehat{O}^{i_1,...i_s}(x^a),
\end{equation}
where $\pa_\parallel^2\equiv \partial_a \partial^a$. The explicit form of the differential operator $\mathcal{C}_{\widehat{{O}}}$ is fixed by the $SO(p+~1,1)\times SO(q)$ symmetry:
\begin{equation}\label{diff_spinl_op}
\mathcal{C}_{\widehat{O}}\left(|x^i|^2\pa_\parallel^2\right)=\sum_{m=0}c_m\left(|x^i|^2\pa_\parallel^2\right)^m,\quad
c_m=\frac{1}{m!}\frac{(-4)^{-m}}{\left({\widehat{\Delta}}+1-\frac{p}{2}\right)_{m}}.
\end{equation}

\subsubsection{Defect OPE of a free scalar}
\label{subsub:freesca}

The spectrum of defect primaries can be complicated, no matter how simple the bulk theory is. This simply follows from the fact that new degrees of freedom with intricate dynamics might be present on the defect. However, bulk-to-defect couplings are constrained by the properties of the bulk. In a free theory, a differential operator exists that annihilates all correlators of the elementary field, and as a consequence it annihilates all its OPEs \cite{Dimofte:2012pd}. Consider for instance the theory of a free boson $\varphi$. Let us focus on the contribution of a defect primary to the OPE \eqref{compBOPE}. When applying the Laplace operator $\pa^2=\pa^2_{\perp}+\pa^2_{\parallel}$ to the right hand side, we can disregard the parallel part, which acts on the primary to give descendants. Denoting the transverse twist $\widehat{\tau}\equiv \widehat{\Delta}-s$, we then get
\beq
0 \sim b_{O\widehat{O}} \left(\Delta-\widehat{\tau} \right)\left(\Delta-\widehat{\tau} +2-q-2s\right) \frac{x_{i_1}...x_{i_s}}{|x^i|^{\Delta-\widehat{\tau}+2}}\,\widehat{O}^{i_1,...i_s}(x^a)+\dots
\eeq
where of course $\D=(d-2)/2$. It follows that $b_{O\widehat{O}}=0$ unless
\begin{equation}\label{shortprim}
\widehat{\tau}=\D, \quad \textup{or} \quad \widehat{\tau}=\Delta+2-q-2s.
\end{equation}
We see that primaries with protected dimension are generically induced on the defect. The first equation isolates a tower of primaries that we may denote $\wh{\pa_\perp^s\varphi}$. The second equation has solutions only for finitely many values of the transverse spin $s$ in a unitary theory, precisely $s\leq (4-q)/2$. The inequality is saturated for a defect scalar at the unitarity bound, and in particular, at $s=1$ this is possible only for codimension two defects. This is precisely the case of the last example discussed in subsection \ref{examples}. Finally, the identity can appear in the defect OPE of $\varphi$ only when $q=d/2+1$, which matches the first of the examples in the same subsection.

\subsection{The Casimir equations for the two-point function}\label{Derivation_casimir}

For the sake of completeness, we give here a derivation of the Casimir equations that appear in section \ref{sec:blocks}. In a defect CFT, there are two ways of doing radial quantization: we can define a state on a sphere $\Sigma$ centered in some point in the bulk, or on a sphere $\wh{\Sigma}$ centered in a point on the defect. In the first case, in the absence of insertions and when the radius of the sphere is smaller than the distance from the defect, the state on the sphere is the conformal invariant vacuum $\ket{0}$. The partition function in the presence of the defect is given by the superposition of this vacuum with a state $\bra{\mc{D}}$ defined by evaluating the path integral on the exterior of $\Si$ - see fig. \ref{fig:bulkchannel}. If we center the quantization on the defect, the sphere $\wh{\Si}$ is decorated by its intersection with the defect - see fig. \ref{fig:channels} - and a new vacuum $\ket{\wh{\mc{D}}}$ is thus defined. Of course, we can write a two-point function in the two schemes:
\begin{figure}[t]
\begin{center}
\includegraphics[scale=.4]{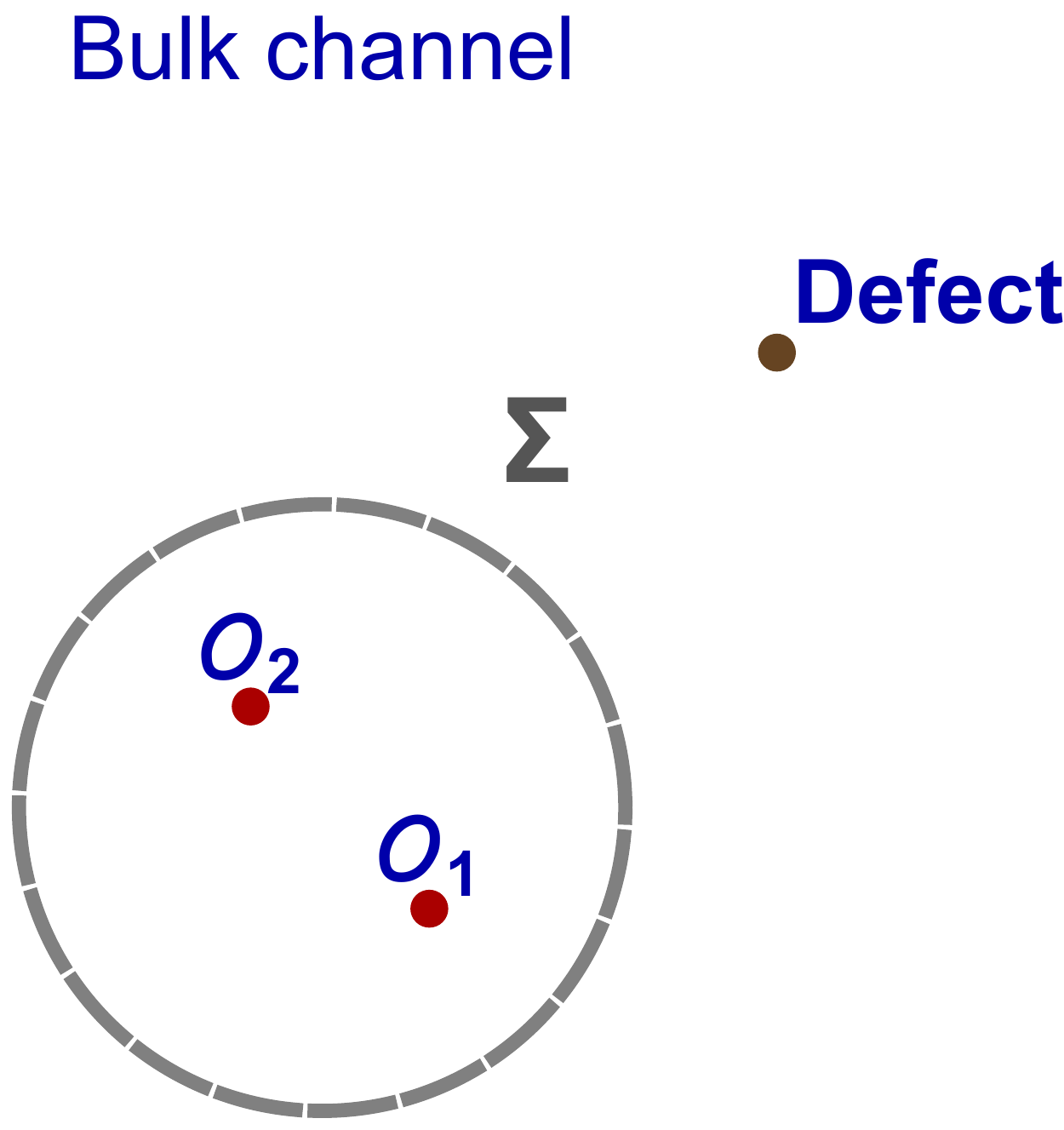}
\end{center}
\caption{ One way to do radial quantization is to quantize around a bulk point. We can always choose a sphere which does not intersect the defect and such that both insertions lie in its interior.}\label{fig:bulkchannel}
\end{figure}
\begin{align}
\braket{O_1(x_1)O_2(x_2)} &= \bra{\mc{D}} O_1(x_1)O_2(x_2) \ket{0} \notag\\
                                  &= \bra{\mc{\wh{D}}} O_1(x_1)O_2(x_2) \ket{\mc{\wh{D}}}.
                                  \label{quantizations}
\end{align}
The conformal block decompositions in the bulk and defect channels correspond to the insertion of a complete set of states in the two lines of eq. \eqref{quantizations} respectively:
\begin{figure}[t]
\begin{center}
\includegraphics[scale=.4]{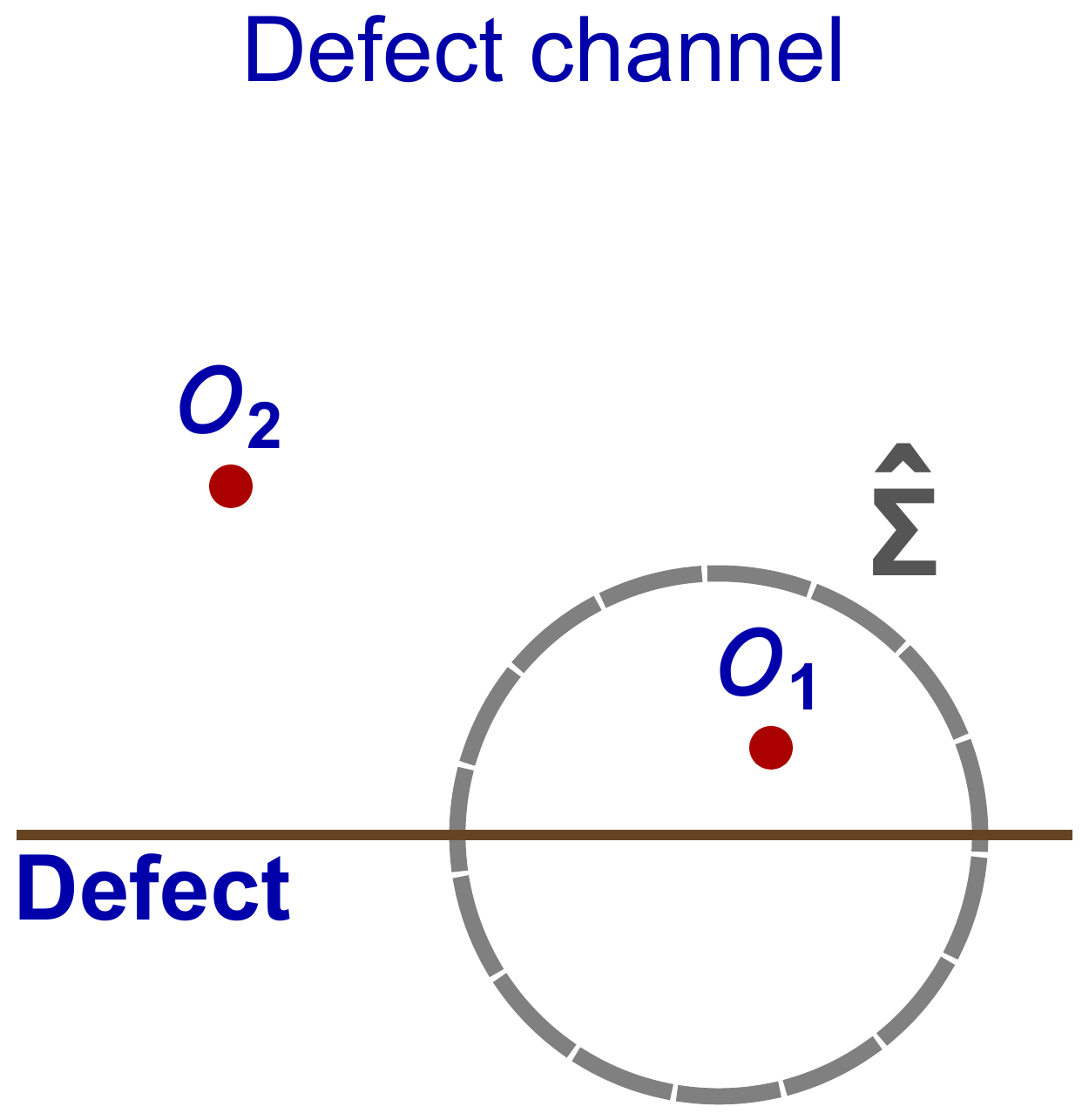}
\end{center}
\caption{ The theory can also be quantized around a point on the defect. This is usually associated with the defect OPE. }\label{fig:channels}
\end{figure}
\begin{align}
\braket{O_1(x_1)O_2(x_2)} &= 
      \sum_{\al = O,P_\m O, \dots}\braket{\mc{D}|\al}\frac{1}{\braket{\al|\al}}
      \bra{\al} O_1(x_1)O_2(x_2) \ket{0} +\textup{other families}, \label{bulkidentity}\\
       &= \sum_{\wh{\al} = \wh{O},P_a \wh{O}, \dots}
       \bra{\mc{\wh{D}}}O_1(x_1)\ket{\wh{\al}}\frac{1}{\braket{\wh{\al}|\wh{\al}}}
      \bra{\wh{\al}} O_2(x_2) \ket{\mc{\wh{D}}} +\textup{other families}.\label{defidentity}
\end{align}
We will focus on a single conformal family from here on. The generators of the full conformal group $J_{MN}$ act on scalar primaries as follows
\begin{equation}\label{diff_comm}
\ii [J_{MN},O_i(x_i)] =\mathcal{J}_{MN}^{(i)}O_i (x_i),\qquad 
\mathcal{J}_{MN}=\left(P_M\frac{\partial}{\partial P^N}-P_N\frac{\partial}{\partial P^M}\right).
\end{equation}
Bulk and defect vacua are invariant under a different set of generators:
\beq
J_{MN} \ket{0} = 0, \quad J_{AB}\ket{\mc{\wh{D}}}= J_{IJ}\ket{\mc{\wh{D}}}=0.
\label{genvacua}   
\eeq
If we define the Casimir operators of $SO(d+1,1)$, $SO(p+1,1)$ and $SO(q)$ respectively:
\beq
J^2=\frac{1}{2} J_{MN} J^{MN}, \quad L^2=\frac{1}{2} J_{AB} J^{AB},
\quad S^2=\frac{1}{2} J_{IJ} J^{IJ},
\eeq
their action on any state belonging to the family of $O$ or $\wh{O}$ is given by
\beq
J^2 \ket{\al} = C_{\D,J} \ket{\al}, 
\quad L^2\ket{\wh{\al}}=\wh{C}_{\wh{\D},0}\ket{\wh{\al}},
\quad S^2\ket{\wh{\al}}=\wh{C}_{0,s}\ket{\wh{\al}},
\label{casstate}
\eeq
with $C_{\D,J}=\D (\D-d)+J(J+d-2)$ and $\wh{C}_{\wh{\D},s}=\wh{\D} (\wh{\D}-p)+s(s+q-2)$. The Casimir equations are now easily obtained. For the bulk channel, we apply the operator $J^2$ in eq. \eqref{bulkidentity}, we use alternatively eq. \eqref{diff_comm} and eqs. \eqref{genvacua}, \eqref{casstate}, and we obtain
\beq
\begin{split}
\sum_{\al = O,P_\m O, \dots}\braket{\mc{D}|\al}\frac{1}{\braket{\al|\al}}
      \bra{\al}\frac{1}{2} [J_{MN},[J^{MN}, &O_1(x_1)O_2(x_2)]] \ket{0}  \\
      &= -\frac{1}{2}\left(\mathcal{J}_{MN}^{(1)}+\mathcal{J}_{MN}^{(2)}\right)^2
              \braket{O_1(x_1)O_2(x_2)}_O\\
              &=C_{\D,J} \braket{O_1(x_1)O_2(x_2)}_O.
\end{split}
\eeq
Here the subscript means restriction to the contribution of a single conformal family:
\beq
\braket{O_1(x_1)O_2(x_2)}_O = \sum_{\al = O,P_\m O, \dots}\braket{\mc{D}|\al}\frac{1}{\braket{\al|\al}}
      \bra{\al} O_1(x_1)O_2(x_2) \ket{0}.
\eeq
The Casimir equation in the defect channel is obtained similarly. In this case the Casimir operator is only made to act on one of the operators. For instance, starting from
\beq
\sum_{\wh{\al} = \wh{O},P_a \wh{O}, \dots}
       \bra{\mc{\wh{D}}}\frac{1}{2}[J_{AB}[J^{AB},O_1(x_1)]]\ket{\wh{\al}}\frac{1}{\braket{\wh{\al}|\wh{\al}}}
      \bra{\wh{\al}} O_2(x_2) \ket{\mc{\wh{D}}},
\eeq
one immediately obtains the first of the eqs. \eqref{def_eigval}. The second one can be derived analogously.

\subsection{The bulk-channel block for an exchanged scalar}\label{app_scalar}

Here, we obtain a formula for the scalar conformal block starting from the OPE \eqref{OPEI}. The result follows from manipulating the r.h.s. of the following equation:
\begin{equation}
a_k \xi^{-(\D_1+\D_2)/2}f_{\Delta_k,0}(\xi,\theta)=\frac{|x_{1}^i|^{\Delta_1}|x_{2}^i|^{\Delta_2}}{(x_{12}^2)^{\frac{1}{2}(\Delta_1+\Delta_2-\Delta_k)}}\mathcal{C}^{(0)}(x_{12},\partial_2)\langle{O}_k(x_2)\rangle,
\end{equation}
where the differential operator $\mathcal{C}^{(0)}$ is defined in \eqref{Fe_Gr_Ga_diff_op} and
\begin{equation}\label{onpta}
\langle{O}_k(x_2)\rangle=\frac{a_{{k}}}{|x_{2}^i|^{\Delta_k}}.
\end{equation}
For ease of notation, in this appendix we use $s^\mu=x_{12}^\mu$. Let us compute the expression $\mathcal{C}^{(0)}(s,\partial_2){|x_{2}^i|^{-\Delta_k}}$. Since $[s,\partial_2]=0$ by definition, $e^{\alpha s\cdot \partial_2}$ acts as a translation: 
\begin{equation}
e^{\alpha s\cdot \partial_2}{|x_{2}^i|^{-\Delta_k}}={|x_{2}^i+\alpha s^i|^{-\Delta_k}},
\end{equation}
Furthermore:
\begin{equation}
(\partial_{2\perp}^2)^m |x_{2}^i|^{-\Delta_k}=4^m \left(\frac{\Delta_k}{2}\right)_m \left(\frac{\Delta_k}{2}+1-\frac{q}{2}\right)_m |x_{2}^i|^{-(\Delta_k+2m)},
\end{equation}
where $\partial_{2\perp}\equiv \partial_2^i\partial_2^i$. Hence
\begin{equation}
\begin{array}{ll}
\displaystyle{\mathcal{C}^{(0)}(s,\partial_2){|x_{2}^i|^{-\Delta_k}}}&\displaystyle{ =\frac{1}{B(a,b)}\int_0^1 \mathrm{d}\alpha\mbox{ } \alpha^{a-1}(1-\alpha)^{b-1}}\\
&\displaystyle{\times\sum_{m=0}\frac{1}{m!}\frac{\left(\frac{\Delta_k}{2}\right)_m \left(\frac{\Delta_k}{2}+1-\frac{q}{2}\right)_m}{(\Delta_k+1-d/2)_m}\frac{[-s^2 \alpha (1-\alpha)]^m}{|x_{2}^i|^{(\Delta_k+2m)}} }\\
\\
\displaystyle{}&\displaystyle{ =\frac{1}{B(a,b)}\int_0^1 \mathrm{d}\alpha\mbox{ } \alpha^{a+m-1}(1-\alpha)^{b+m-1}}\\
&\displaystyle{\times\sum_{m=0}\frac{1}{m!}\frac{\left(\frac{\Delta_k}{2}\right)_m \left(\frac{\Delta_k}{2}+1-\frac{q}{2}\right)_m}{(\Delta_k+1-d/2)_m}\frac{(-s^2)^m |x_{2}^i|^{-(\Delta_k+2m)}}{(1+\alpha^2\frac{|s^i|^2}{|x_{2}^i|^2}+2\alpha \frac{s^i x_{2}^i}{|x_{2}^i|})^{(\Delta_k+2m)}}}.
\end{array}
\end{equation}
Recall that $a=(\Delta_k+\Delta_{12})/2$, $b=(\Delta_k-\Delta_{12})/2$. Using now 
\begin{equation}
\int_0^1 \mathrm{d}\alpha\mbox{ } \alpha^{w-1}(1-\alpha)^{r-1}(1-\alpha x')^{-\rho} (1-\alpha y')^{-\sigma}=B(w,r)F_1 (w,\rho,\sigma,w+r; x';y'),
\end{equation}
we find
\begin{equation}\label{serul}
\begin{array}{ll}
\displaystyle{\mathcal{C}^{(0)}(s,\partial_2){|x_{2}^i|^{-\Delta_k}}}&\displaystyle{ =\frac{1}{B(a,b)}\int_0^1 \mathrm{d}\alpha\mbox{ } \alpha^{a+m-1}(1-\alpha)^{b+m-1}}\\
&\displaystyle{\times\sum_{m=0}\frac{1}{m!}\frac{\left(\frac{\Delta_k}{2}\right)_m \left(\frac{\Delta_k}{2}+1-\frac{q}{2}\right)_m}{(\Delta_k+1-d/2)_m}B(a+m,b+m)}\\
&\displaystyle{\times \frac{(-s^2)^m}{ |x_{2}^i|^{(\Delta_k+2m)}}F_1 \left(a+m, m+\frac{\Delta_k}{2}, m+\frac{\Delta_k}{2},2m+\Delta_k;x',y'\right)},
\end{array}
\end{equation}
where $x',y'$ are solutions of
\begin{equation}\left\{
\begin{array}{ll}
\displaystyle{x'+y'=-2 \frac{s^i x_2^i}{|x_{2}^i|^2}},\\
\displaystyle{x'y' =\frac{|s^i|^2}{|x_{2}^i|^2}}.
\end{array}
\right.
\end{equation}
The Appell function in \eqref{serul} is secretly a hypergeometric:
\begin{equation}
F_1(w,\rho,\sigma,\rho+\sigma; x';y')=(1-y')^w {}_2 F_1 \left(w,\rho,\rho+\sigma; \frac{x+y'}{1-y'}\right).
\end{equation}
This formula also allows to reconstruct the explicit dependence on the cross-ratios:
\begin{multline}
\displaystyle{
f_{\Delta,0}(\xi,\phi)=\frac{1}{B(a,b)}{\xi^{\frac{\Delta_k}{2}}}\sum_{m}\frac{1}{m!}\frac{(\frac{\Delta_k}{2})_m (\frac{\Delta_k}{2}+1-\frac{q}{2})_m}{(\Delta_k+1-\frac{d}{2})_m}(-\xi e^{- i \phi})^m\times}\\
\displaystyle{B(m+a,m+b)e^{- i a \phi} {}_2F_1\left(m+a,m+\frac{\Delta_k}{2},\Delta_k+2m; 2 i \sin\phi e^{- i\phi}\right)}.
\label{scalar_res}
\end{multline}
\eqref{scalar_res} can be finally recast in a form which agrees with the light-cone expansion \eqref{lightconexp} by means of the quadratic transformation \eqref{rel_hyp}:
\begin{multline}\label{scalar_res_real}
f_{\Delta_k,0}(\xi,\phi)=\frac{1}{B(a,b)}{\xi^{\frac{\Delta_k}{2}}}\sum_{m}\frac{1}{m!}\frac{(\frac{\Delta_k}{2})_m (\frac{\Delta_k}{2}+1-\frac{q}{2})_m}{(\Delta_k+1-\frac{d}{2})_m}(-\xi)^m\\
\times B(m+a,m+b){}_2F_1\left(\frac{m+a}{2},\frac{m+b}{2},\frac{\Delta_k+1}{2}+m; \sin^2\phi\right).
\end{multline}
As a special case, \eqref{scalar_res} reduces to a nice closed form when $\phi=0$:
\begin{equation}\label{scala_res0}
\begin{array}{ll}
\displaystyle{
f_{\Delta_k,0}(\xi,0)={\xi^{\frac{\Delta_k}{2}}} {}_3F_2\left(1-\frac{q}{2}+\frac{\Delta_k}{2},a,b;\frac{1}{2}+\frac{\Delta_k}{2},1-\frac{d}{2}+\Delta_k;-\frac{\xi}{4}\right)},
\end{array}
\end{equation}
and reproduces the boundary CFT result \cite{McAvity:1995zd,Liendo:2012hy} when $q=1$:
\begin{equation}
\begin{array}{ll}
\displaystyle{ f_{\Delta_k,0}(\xi)=\xi^{\frac{\Delta_k}{2}}{}_2F_1\left(a,b,\Delta_k+1-\frac{d}{2};-\frac{\xi}{4}\right)}.
\end{array}
\end{equation}

\section{Differential geometry of sub-manifolds}
\label{diffgeom}

In this appendix we collect some definitions and elementary results in differential geometry of sub-manifolds
which are useful in section \ref{sec:ward}. We refer, for instance, to \cite{Eisenhart} and \cite{Aminov} for more details.

The induced metric on a sub-manifold $\mathcal{D}$ is defined as follows:
\beq
\gamma_{ab}=e^\m_a\, e^\n_b\, g_{\m\n},\qquad e^\m_a=\frac{\partial X^\m}{\partial \sigma^a}~.
\label{indmetdef}
\eeq
We raise and lower latin indices with $\ga$, so that the inverse induced metric can be written as
\beq
\ga^{ab}=e_\m^a e_\n^b g^{\m\n}~.
\label{indinvdef}
\eeq
We choose a set of $q$ unit vector fields $n_I^\m$ normal to the defect through the following conditions:
\beq
n_\m^I e^\m_a =0,\qquad n_\m^I n^{\m J}=\de^{IJ}~.
\label{normalvecdef}
\eeq
When restricted to the defect, the bulk metric enjoys the following decomposition, which is just the completeness relation for the basis in the tangent bundle of $\mathcal{M}$:
\beq
g_{\m\n}(X)=e_\m{}^a e_{\n a}+n_\m^I n_\n^I~.
\label{metricdecomp}
\eeq

The extrinsic curvatures are defined as follows:
\beq
\nabla_a e_b{}^\m =\pa_a e_b{}^\m-\hat{\Ga}_{ab}^c e_c{}^\m+\Ga_{\la\si}^\m e_a{}^\la e_b{}^\si 
=n^\m_I K^I_{ab}~, \qquad K_{ab}^I=K_{(ab)}^I~.
\label{extrcurvdef}
\eeq
We denoted with a hat the Christoffel symbols of the induced metric. The Weingarten decomposition holds:
\beq
\nabla_a n^\m_I=\pa_a n^\m_I+ \Ga_{\la\n}^\m e_a\!{}^\la\, n_I{}^\n - \m_a{}^J\!{}_I\, n^{\m}_J =  -K_{ab}^Ie^{\m b}~, \qquad \m_{aIJ}=\m_{a[IJ]}~.
\label{mutens}
\eeq
$\m_{aIJ}$ is sometimes called the torsion tensor, and provides a spin connection for the normal bundle. 
In this way $\nabla_a$, which we defined to act on both flat and curved indices, is covariant both under diffeomorphisms in $\mathcal{D}$ and local orthogonal transformations in the normal bundle. In other terms, 
an infinitesimal local rotation in the normal bundle, parametrized by an antisymmetric matrix $\omega$, 
acts as
\begin{equation}
\label{rots}
\delta_\o n^\mu_I = \omega_{IJ} n^{\mu J}~,~~~
\delta_\o K_{Iab} = \omega_{IJ} K^J_{ab}~,~~~
\delta_\o \mu_{aIJ} = -\nabla_a \omega_{IJ}~.
\end{equation}

The bulk and defect Riemann tensors and the extrinsic curvatures are related by the Gauss-Codazzi equations:
\begin{align}
&\hat{R}_{abcd} = K_{ac}^I K_{bd}^I-K_{ad}^I K_{bc}^I
+R_{abcd}~, \label{gausscodazzi1} \\
&\nabla_c K_{ab}^I - \nabla_b K_{ac}^I = R_{a}{}^I{}_{bc}~. \label{gausscodazzi2}
\end{align}
Here we defined $R_{abcd}\equiv R_{\m\n\r\si} e^\m_a e^\n_b e^\r_c e^\si_d$, and similarly in the second line, where the contraction with a normal vector appears.

Hereafter, we report some notions in variational calculus on sub-manifolds, which are needed in section \ref{sec:ward}.
The variations of the normal vectors, of the extrinsic curvatures 
and of the torsion tensor are induced,
through their definition in eq.s (\ref{indmetdef}-\ref{mutens}), from the variations of 
the metric $g_{\mu\nu}$ (and of its Christoffel connection) evaluated on $\mathcal{D}$ and of the tangent vectors $e^\mu_a$. From eq. (\ref{indmetdef}) we find
\begin{equation}
\label{vargenim}
\delta\gamma_{ab} = 2 e_{\m (a} \delta e^\m_{b)} + e^\m_a e^\nu_b \delta g_{\m\n}~.
\end{equation}
Decomposing the variation of the normal vectors on the basis given by $e^\mu_a$ and $n^\mu_I$ and 
considering the variation of eq.s (\ref{normalvecdef}) one finds%
\footnote{It is useful to write down also the variation of $n_{\mu I} \equiv g_{\mu\nu} n^\nu_I$:
\begin{equation}
\label{vargen21}
\delta n_{\m I} =
\left(\frac 12 n^\rho_I n^\sigma_J \delta g_{\rho\sigma} + \o_{IJ} \right) n_\m^J 
- n_{I\rho} \delta e^\rho_a e^a_\mu~. 
\end{equation}
}
\begin{equation}
\label{vargenn1}
\delta n^\m_I = \left(-\frac 12 n^\rho_I n^\sigma_J \delta g_{\rho\sigma} + \omega_{IJ} \right) n^{\m J} 
-\left(n^\rho_I e^\sigma_a \delta g_{\rho\sigma} + n_{I\rho} \delta e^\rho_a\right) e^{a\mu}~, 
\end{equation}
where the antisymmetric matrix $\omega_{IJ}$ is undetermined. 
The variation of the extrinsic curvatures can be derived from eq. (\ref{extrcurvdef}), and it involves also the variation of the Christoffel connection:
\begin{equation}
\label{vargenK}
\delta K^I_{ab}  = \nabla_{(a} \delta e^\m_{b)} n^I_\m + n^I_\m \Gamma^\m_{\la\si} \delta e^\la_{(a} e^\si_{b)} 
+ n^I_\m \delta \Gamma^\m_{\la\si} e^\la_a e^\si_b + n^\m_J K^J_{ab} \delta n^I_\m~.
\end{equation}
Finally, from eq. (\ref{mutens}) we obtain the variation of the torsion tensor $\mu$:
\begin{equation}
\label{vargenmu}
\delta\mu_a^{IJ} = - \nabla_a\delta n^{[I\m} n^{J]}_\m - n^{[J}_\m \Gamma^\m_{\la\nu}\delta e^\la_a n^{I]\nu}
- n^{[J}_\m \delta \Gamma^\m_{\la\nu} e^\la_a n^{I]\nu} + K^{[I}_{ab} e^{b\mu} \delta n^{J]}_\m~.
\end{equation}
From these formul\ae\, we can obtain the expressions pertaining to the various symmetries by specializing
the form of $\delta e^\mu_a$, $\delta g_{\mu\nu}$ and $\delta\Gamma^\mu_{\nu\rho}$. 
Note that in general the total variation
$\delta g_{\mu\nu}(X) = g^\prime_{\mu\nu}(X+\delta X)-g_{\mu\nu}(X)$
can be decomposed into a variation in form, $\delta_g g_{\mu\nu}$, and a part due to the change in the argument:
\begin{equation}
\label{decvarg}
\delta g_{\mu\nu} = \delta_g g_{\mu\nu} +  \delta_X g_{\mu\nu}~,~~~\mbox{with}~
\delta_X  g_{\mu\nu} = \delta X^\lambda \partial_\lambda g_{\mu\nu}~.
\end{equation}
The same applies to the connection $\Gamma^\mu_{\nu\rho}$ (and to any bulk quantity evaluated on $\mathcal{D}$).
The variation of the tangent vectors is instead only due to the change in the embedding function:
$\delta e^\mu_a = \delta_X e^\mu_a$ 

\paragraph{Diffeomorphisms}
 Under an infinitesimal diffeomorphism $x^\mu \to x^{\prime\mu} = x^\mu + \xi^\mu(x)$, the embedding coordinates of the sub-manifold change by $\delta_\xi X^\mu = \xi^\mu$, so that
\begin{equation}
\label{vardiffe}
\delta_\xi e^\m_a = \partial_a \xi^\m~.
\end{equation}
The in form variations of the metric and the Christoffel symbols are
\begin{equation}
\label{varformgG}
\begin{aligned}
\delta_{\xi,g} g_{\mu\nu} & = -2 \nabla_{(\mu} \xi_{\nu)}~,\\
\delta_{\xi,g} \Gamma^\mu_{\nu\rho} & = \frac{1}{2}g^{\mu\tau} 
\left(\nabla_\nu \delta_g g_{\rho\tau} + \nabla_\rho \delta_g g_{\nu\tau} 
- \nabla_\tau \delta_g g_{\nu\rho}\right) = 
\xi^\sigma R_{\sigma (\nu\rho)}^{\phantom{\sigma (\nu\rho)}\mu} 
- \nabla_{(\nu} \nabla_{\rho)} \xi^\mu~.
\end{aligned}
\end{equation}
The total variations of the same quantities follow 
the usual law:%
\begin{equation}
\label{vardiffgG}
\begin{aligned}
\delta_\xi g_{\mu\nu} & = -2\partial_{(\mu}\xi^\lambda g_{\nu)\lambda}~,\\
\delta_\xi \Gamma^\mu_{\nu\rho} & = \partial_\sigma\xi^\mu \Gamma^\sigma_{\nu\rho} 
- \partial_\nu\xi^\sigma \Gamma^\mu_{\sigma\rho} - \partial_\rho\xi^\sigma \Gamma^\mu_{\sigma\nu}
-\partial^2_{\nu\rho}\xi^\mu~. 
\end{aligned}
\end{equation}
Inserting this into eq.s (\ref{vargenim}-\ref{vargenmu}) we find in the end
\begin{equation}
\label{vardif}
\begin{aligned}
\delta_\xi \gamma_{ab} & = 0~,\\
\delta_\xi n^\mu_I & = n^\lambda_I \partial_\lambda\xi^\mu + \omega^\prime_{IJ} n^{\mu J}~,\\
\delta_\xi K^I_{ab} & = \omega^\prime_{IJ} K^J_{ab}~,\\
\delta_\xi \mu_{aIJ} & = -\nabla_a \omega^\prime_{IJ}~,
\end{aligned}
\end{equation}
with $\omega^\prime_{IJ}$ an arbitrary antisymmetric matrix.%
\footnote{With respect to eq. (\ref{vargenn1}) we have $\omega^\prime_{IJ} = -\partial_\lambda\xi^\rho n^\lambda_{[I} n_{J]\rho} + \omega_{IJ}$.}
Up to transverse rotations, these are just the expected tensorial transformations: 
$n^\mu_I$ is a vector, the other quantities are scalars.

\paragraph{Re-parametrizations}
Under an infinitesimal redefinition $\sigma^a\to \sigma^a + \zeta^a(\sigma)$ of the coordinates on $\mathcal{D}$, the embedding functions change by $\delta_\zeta X^\mu = -e^\mu_a \zeta^a$. The bulk metric varies just because of this shift in its argument:
\begin{equation}
\label{varrepg}
\delta_\zeta g_{\mu\nu} = 
- \zeta^a e^\lambda_a \partial_{\lambda} g_{\mu\nu}~.
\end{equation}
The various quantities describing the geometry of 
$\mathcal{D}$ vary in form according to their tensor structure in the indices $a,b\ldots$, so in particular
\begin{equation}
\label{varrepe}
\delta_\zeta e^\mu_a = -\partial_a\zeta^b \e^\mu_b - \zeta^b\partial_b e^\mu_a = 
-\nabla_a\zeta^b e^\mu_b - K^I_{ab} \zeta^b n^\mu_I + \Gamma^{\mu}_{b\lambda} e^\lambda_a \zeta^b~,
\end{equation}    
where in the second step we introduced covariant derivatives and made use of eq. \ref{extrcurvdef}. 
Inserting eq.s (\ref{varrepg}),\ref{varrepe}) into eq.s (\ref{vargenim}-\ref{vargenmu}) we find then
\begin{equation}
\label{varrep}
\begin{aligned}
\delta_\zeta \gamma_{ab} & = -2 \nabla_{(a} \zeta_{b)}~,\\
\delta_\zeta n^\mu_I & = K^I_{ab} e^{\mu a} \zeta^b + \Gamma^\mu_{\lambda b} n^\lambda_I \zeta^b + 
\omega^{\prime\prime}_{IJ} n^{\mu J}~,\\
\delta_\zeta K^I_{ab} & = -2 \nabla_{(a} \zeta^c K_{b)c}^I - \zeta^c \nabla_c K^I_{ab} + 
\omega^{\prime\prime}_{IJ} K^{J}_{ab}~,\\
\delta_\zeta \mu_{aIJ} & = K^{[Ib}_{a} K^{J]}_{bc} \zeta^c - \frac 12 \zeta^c R_{caIJ} - \zeta^c R_{c[IaJ]} - 
\nabla_a \omega^{\prime\prime}_{IJ}~,
\end{aligned}
\end{equation}
where $\omega^{\prime\prime}_{IJ}$ is an arbitrary antisymmetric matrix.\footnote{With respect to eq. (\ref{vargenn1}) we have $\omega^{\prime\prime}_{IJ} = 
+ \zeta^c \Gamma^\lambda_{c\rho} n_{\lambda[I} n^\rho_{J]} + \omega_{IJ}$.}

\paragraph{Weyl rescalings}
If we perform an arbitrary infinitesimal  rescaling of the metric 
\begin{equation}
\label{varweylg}
\delta_\sigma g_{\mu\nu}(X) = 2\sigma(X) g_{\mu\nu}(X)~,
\end{equation}
which does not affect the position $X$ and the tangent vectors $e^\mu_a$, from eqs. (\ref{vargenim}-\ref{vargenmu}) 
we find
\begin{equation}
\label{varweyl}
\begin{aligned}
\delta_\sigma \gamma_{ab} & = 2 \sigma \gamma_{ab}~,\\
\delta_\sigma n^\mu_I & = -\sigma n^\mu_I + \omega_{IJ} n^{\mu J}~,\\
\delta_\sigma K^I_{ab} & = \sigma K^I_{ab} - \gamma_{ab}n^{\mu I} \partial_\mu \sigma +\omega^{IJ} K_{Jab}~,\\
\delta_\sigma \mu_{aIJ} & = -\nabla_a \omega_{IJ}~.
\end{aligned}
\end{equation}

\vfill
\newpage


\bibliographystyle{JHEP-2}

\providecommand{\href}[2]{#2}\begingroup\raggedright\endgroup

\end{document}